\documentclass[12pt]{article}
\usepackage{graphicx,color}
\usepackage{chicago}
\usepackage{subfigure}
\usepackage{amsmath, amsfonts, amssymb, amstext}

\newtheorem{thm}{Theorem}%[section]
\newtheorem{lemma}{Lemma}%[section]
\newtheorem{remark}{Remark}%[section]
\newtheorem{proposition}{Proposition}%[section]
\definecolor{DarkBlue}{rgb}{0.1,0.1,0.5}
\definecolor{Red}{rgb}{0.9,0.0,0.1}

\usepackage{setspace}
\def\0{{\bf 0}}
\title{Partial Correlation Estimation by Joint Sparse Regression Models}
\author{Jie Peng  \footnote{Equal contributors} \footnote{Correspondence author: jie@wald.ucdavis.edu}, Pei
  Wang$^{\ast \ddag}$, Nengfeng Zhou$^\S$, Ji Zhu$^\S$\\
  $^\dag$ Department of Statistics, University of California, Davis, CA 95616. \\
  $^\ddag$ Division of Public Health Science, Fred Hutchinson Cancer Research Center,\\ Seattle, WA 98109.\\
  $^\S$ Department of Statistics, University of Michigan, Ann Arbor, MI 48109.\\
}

\date{}

\begin{document}
\maketitle
\clearpage
\begin{abstract}
In this paper, we propose a computationally efficient approach
---\texttt{space}(Sparse PArtial Correlation Estimation)--- for selecting non-zero partial correlations under the
high-dimension-low-sample-size setting. This method assumes the
overall sparsity of the partial correlation matrix and employs
sparse regression techniques for model fitting. We illustrate the
performance of \texttt{space} by extensive simulation studies. It
is shown that \texttt{space} performs well in both non-zero
partial correlation selection and the identification of hub
variables, and also outperforms two existing methods. We then
apply \texttt{space} to a microarray breast cancer data set and
identify a set of \textit{hub genes} which may provide important
insights on genetic regulatory networks. Finally, we prove that,
under a set of suitable assumptions, the proposed procedure is
asymptotically consistent in terms of model selection and
parameter estimation.

{\bf key words:} concentration network,
high-dimension-low-sample-size, lasso, shooting, genetic regulatory network

\end{abstract}
\newpage
{\centering \section{INTRODUCTION}}
There has been a large amount of literature on \textit{covariance
selection}: the identification and estimation of non-zero entries
in the inverse covariance matrix (a.k.a. \textit{concentration
matrix} or \textit{precision matrix}) starting from the seminal
paper by \shortciteN{Dempster1972}. Covariance selection is very
useful in elucidating associations among a set of random
variables, as it is well known that non-zero entries of the
concentration matrix correspond to non-zero partial correlations.
Moreover, under Gaussianity, non-zero entries of the concentration
matrix imply conditional dependency  between corresponding
variable pairs conditional on the rest of the variables \shortcite{edward}. Traditional methods does not work
unless the sample size ($n$) is larger than the number of variables
($p$) \shortcite{Whittaker1990,edward}. Recently, a number of
methods have been introduced to perform covariance selection for
data sets with $p>n$, for example, see \shortciteN{meinshausen},
\shortciteN{Yuan}, \shortciteN{Li},
\shortciteN{SchaferStrimmer07}.

In this paper, we propose a novel approach using sparse regression
techniques for covariance selection.  Our work is partly motivated
by the construction of \textit{genetic regulatory networks (GRN)}
based on high dimensional gene expression data.  Denote the
expression levels of $p$ genes as $y_1,\cdots,y_p$. A
\textit{concentration network} is defined as an undirected graph,
in which the $p$ vertices represent the $p$ genes and an edge
connects gene $i$ and gene $j$ if and only if the partial
correlation $\rho^{ij}$ between $y_i$ and $y_j$ is non-zero. Note
that, under the assumption that $y_1,\cdots,y_p$ are jointly normal, the partial correlation
$\rho^{ij}$ equals to ${\rm Corr}(y_i,y_j|y_{-(i,j)})$, where $y_{-(i,j)}=\{y_k: 1 \leq k \not= i,j \leq p\}$.
Therefore,
$\rho^{ij}$ being nonzero is equivalent to $y_i$ and $y_j$ being
conditionally dependent given all other variables
$y_{-(i,j)}$. The proposed
method is specifically designed for the
high-dimension-low-sample-size scenario. It relies on the
assumption that the partial correlation matrix is sparse (under
normality assumption, this means that most variable pairs are
conditionally independent), which is reasonable for many real life
problems. For instance, it has been shown that most genetic
networks are intrinsically sparse
\shortcite{gardner,jeong,tegner}. The proposed method is also
particularly powerful in the identification of \textit{hubs}:
vertices (variables) that are connected to (have nonzero partial
correlations with) many other vertices (variables). The existence
of hubs is a well known phenomenon for many large networks, such
as the internet, citation networks, and protein interaction
networks \shortcite{newman}. In particular, it is widely believed
that genetic pathways consist of many genes with few interactions
and a few hub genes with many interactions \shortcite{barabasi}.

Another contribution of this paper is to propose a novel algorithm
\texttt{active-shooting} for solving penalized optimization problems
such as lasso \shortcite{Ti96}. This algorithm is computationally
more efficient than the original \texttt{shooting} algorithm, which
was first proposed by \shortciteN{Fu1998shooting} and then extended
by many others including \shortciteN{Genkin07} and
\shortciteN{Tibshirani2007PathCO}. It enables us to implement the
proposed procedure efficiently, such that we can conduct extensive
simulation studies involving $\sim 1000$ variables and hundreds of
samples. To our knowledge, this is the first set of intensive
simulation studies for covariance selection with such high
dimensions.

A few methods have also been proposed recently to perform covariance
selection in the context of $p \gg n$. Similar to the method
proposed in this paper, they all assume sparsity of the partial
correlation matrix. \shortciteN{meinshausen} introduced a
variable-by-variable approach for neighborhood selection via the
lasso regression. They proved that neighborhoods can be consistently
selected under a set of suitable assumptions. However, as regression
models are fitted for each variable separately, this method has two
major limitations. First, it does not take into account the
intrinsic symmetry of the problem (i.e., $\rho^{ij}=\rho^{ji}$).
This could result in loss of efficiency, as well as contradictory
neighborhoods. Secondly, if the same penalty parameter is used for
all $p$ lasso regressions as suggested by their paper, more or less
equal effort is placed on building each neighborhood. This
apparently is not the most efficient way to address the problem,
unless the degree distribution of the network is nearly uniform.
However, most real life networks have skewed degree distributions,
such as the \textit{power-law networks}. As observed by
\shortciteN{SchaferStrimmer07}, the neighborhood selection approach
limits the number of edges connecting to each node. Therefore, it is
not very effective in hub detection.
%\textcolor{red}{On the other hand, if different penalty parameters
%are used in model fitting, the method suffers from great
%variability. (??? maybe delete this sentence???)}
On the contrary, the proposed method is based on a joint sparse
regression model, which simultaneously performs neighborhood
selection for all variables. It also preserves the symmetry of the
problem and thus utilizes data more efficiently. We show by
intensive simulation studies that our method performs better in
both model selection and hub identification. Moreover, as a joint
model is used, it is easier to incorporate prior knowledge such as
network topology into the model. This is discussed in Section
\ref{subsec:model}.

Besides the regression approach mentioned above, another class of
methods employ the maximum likelihood framework. \shortciteN{Yuan}
proposed a penalized maximum likelihood approach which performs
model selection and estimation simultaneously and ensures the
positive definiteness of the estimated concentration matrix.
However, their algorithm can not handle high dimensional data. The
largest dimension considered by them is $p = 10$ in simulation and
$p = 5$ in real data. \shortciteN{TiSparse07} proposed an
efficient algorithm \texttt{glasso} to implement this method, such
that it can be applied to problems with high dimensions. We show
by simulation studies that, the proposed method
performs better than \texttt{glasso} in both model selection and
hub identification.   \shortciteN{rothman08} proposed another algorithm to implement the method of
 \shortciteN{Yuan}.  The computational cost is on the same order of
 \texttt{glasso}, but in general not as efficient as \texttt{glasso}.
\shortciteN{Li} introduced a threshold
gradient descent (TGD) regularization procedure.
\shortciteN{SchaferStrimmer07} proposed a shrinkage covariance
estimation procedure to overcome the ill-conditioned problem of
sample covariance matrix when $p>n$.
%\shortciteN{BickelLevina08}
%proposed to regularize the covariance matrix by hard thresholding
%for families of covariance matrices satisfying suitable sparsity
%assumptions. However, the latter two methods are
%  for estimating the covariance matrix, rather than the
%  concentration matrix and hence they do not result in sparse estimator of the concentration matrix.
  There are also a large class of methods covering the situation where
variables have a natural ordering, e.g., longitudinal data, time series,
spatial data, or spectroscopy.  See \shortciteN{wu03}, \shortciteN{BickelLevina08},
\shortciteN{huang06} and \shortciteN{levina_zhu06}, which are all based on the
modified Cholesky decomposition of the concentration matrix.  In this
 paper, we, however, focus on the general case where an ordering of the
 variables is not available.

The rest of the paper is organized as follows. In Section 2, we
describe the joint sparse regression model, its implementation and
the \texttt{active-shooting} algorithm. In Section 3, the
performance of the proposed method is illustrated through
simulation studies and compared with that of the neighborhood
selection approach and the likelihood  based approach \texttt{glasso}. In
Section 4, the proposed method is applied to a microarray
expression data set of $n=244$ breast cancer tumor samples and
$p=1217$ genes. In Section 5, we study the asymptotic properties
of this procedure. A summary of the main results are given in
Section 6. Technique details are provided in the Supplemental
Material.

{\centering \section{METHOD}}
\subsection{Model}
\label{subsec:model}

In this section, we describe a novel method for detecting pairs of
variables having nonzero partial correlations among a large number
of random variables based on i.i.d. samples. Suppose that, $(y_1,\cdots,y_p)^{T}$ has a joint distribution with mean $0$ and
covariance  $\mathbf{\Sigma}$, where $\mathbf{\Sigma}$ is a $p$ by $p$ positive
definite matrix. Denote the partial correlation between $y_i$ and
$y_j$ by $\rho^{ij}$ ($1 \leq i<j \leq p$). It is defined as ${\rm Corr}(\epsilon_i,
\epsilon_j)$, where $\epsilon_i$ and $\epsilon_j$ are the
prediction errors of the best linear predictors of $y_i$ and $y_j$
based on $y_{-(i,j)}=\{y_k: 1 \leq k \not= i,j \leq p\}$,
respectively.
%%%%%%%%%%%%%%%
%%{\color{blue} where $\epsilon_i$ is the residue of the regression
%%model $y_i \sim y_{-(i,j)}=\{y_k: 1 \leq k \not= i,j \leq p\}$ and
%%$\epsilon_j$ is the residue of the regression model $y_j \sim
%%y_{-(i,j)}$.}
%%%%%%%%%%%%%%%%%
Denote the \textit{concentration matrix} $\mathbf{\Sigma}^{-1}$ by
$(\sigma^{ij})_{p\times p}$. It is known that,
$\rho^{ij}=-\frac{\sigma^{ij}}{\sqrt{\sigma^{ii}\sigma^{jj}}}$.
Let $y_{-i}:=\{y_k: 1 \leq k \not=i \leq p\}$. The following
well-known result (Lemma \ref{lemma:reg}) relates the estimation
of partial correlations to a regression problem.  {\lemma:
\label{lemma:reg}
 For $1 \leq i \leq p$, $y_i$ is expressed as $y_i=\sum_{j\not=i}\beta_{ij}y_j
+\epsilon_i$, such that $\epsilon_i$ is uncorrelated with $y_{-i}$
if and only if $\beta_{ij} = -\frac{\sigma^{ij}}{\sigma^{ii}} =
\rho^{ij} \sqrt{\frac{\sigma^{jj}}{\sigma^{ii}}}$. Moreover, for
such defined $\beta_{ij}$, ${\rm
  Var}(\epsilon_i)=\frac{1}{\sigma^{ii}}, {\rm
Cov}(\epsilon_i,
\epsilon_j)=\frac{\sigma^{ij}}{\sigma^{ii}\sigma^{jj}}$. }

Note that, under the normality assumption, $\rho^{ij}={\rm
Corr}(y_i,y_j|y_{-(i,j)})$ and in Lemma \ref{lemma:reg}, we can
replace ``uncorrelated" with ``independent". Since $\rho^{ij}={\rm
sign}(\beta_{ij})\sqrt{\beta_{ij}\beta_{ji}}$, the search for
non-zero partial correlations can be viewed as a model selection
problem under the regression setting. In this paper, we are mainly
interested in the case where the dimension $p$ is larger than the
sample size $n$. This is a typical scenario for many real life
problems. For example, high throughput genomic experiments usually
result in data sets of thousands of genes for tens or at most
hundreds of samples. However, many high-dimensional problems are
intrinsically sparse. In the case of genetic regulatory networks, it
is widely believed that most gene pairs are not directly interacting
with each other. Sparsity suggests that even if the number of
variables is much larger than the sample size, the effective
dimensionality of the problem might still be within a tractable
range. Therefore, we propose to employ sparse regression techniques
by imposing the $\ell_1$ penalty on a suitable loss function to
tackle the high-dimension-low-sample-size problem.

Suppose $\mathbf{Y}^{k}=(y^{k}_1,\cdots,y^k_p)^T$ are i.i.d. observations
from $(0, \mathbf{\Sigma})$, for $k=1,\cdots,n$. Denote the sample of the
$i$th variable as $\mathbf{Y}_i=(y^1_{i},\cdots,y^n_{i})^{T}$. Based on
Lemma \ref{lemma:reg}, we propose the following joint loss
function
\begin{eqnarray}
\label{eqn:loss_l2}
L_n(\mathbf{\theta},\mathbf{\sigma},\mathbf{Y})&=&\frac{1}{2}\Bigl(\sum_{i=1}^p w_i
||\mathbf{Y}_i-\sum_{j \not=
i}\beta_{ij}\mathbf{Y}_j||^2 \Bigr)\nonumber\\
&=&\frac{1}{2}\Bigl(\sum_{i=1}^p w_i||\mathbf{Y}_i-\sum_{j \not=
i}\rho^{ij}\sqrt{\frac{\sigma^{jj}}{\sigma^{ii}}}\mathbf{Y}_j||^2 \Bigr),
\end{eqnarray}
where $\mathbf{\theta}=(\rho^{12},\cdots, \rho^{(p-1)p})^T$,
$\mathbf{\sigma}=\{\sigma^{ii}\}_{i=1}^p$; $\mathbf{Y}=\{\mathbf{Y}^k\}_{k=1}^n$; and
$w=\{w_i\}_{i=1}^p$ are nonnegative weights. For example, we can
choose $w_i=1/{\rm Var}(\epsilon_i)=\sigma^{ii}$ to weigh
individual regressions in the joint loss function according to their
residual variances, as is done in regression with heteroscedastic noise. We propose
to estimate the partial correlations $\mathbf{\theta}$ by minimizing a
penalized loss function
\begin{eqnarray}
\label{eqn:loss_pen}
\mathcal{L}_n (\mathbf{\theta}, \mathbf{\sigma}, \mathbf{Y}) = L_n(\mathbf{\theta}, \mathbf{\sigma},
\mathbf{Y}) + \mathcal{J}(\mathbf{\theta}),
\end{eqnarray}
where the penalty term $\mathcal{J}(\mathbf{\theta})$ controls the overall
sparsity of the final estimation of $\theta$. In this paper, we
focus on the $\ell_1$ penalty \shortcite{Ti96}:
\begin{eqnarray}
\label{eqn:l1_pen} {\mathcal
J}(\mathbf{\theta})=\lambda||\mathbf{\theta}||_1=\lambda\sum_{1 \leq i<j\leq
p}|\rho^{ij}|.
\end{eqnarray}

The proposed joint method is referred to as \texttt{space} (Sparse
PArtial Correlation Estimation) hereafter. It is related to the
\textit{neighborhood selection approach} by \shortciteN{meinshausen}
(referred to as \texttt{MB} hereafter), where a lasso regression is
performed separately for each variable on the rest of the variables.
However, \texttt{space} has several important advantages.
\begin{enumerate}

\item[(i)] In \texttt{space}, sparsity is utilized for the partial
correlations $\theta$ as a whole view. However, in the
neighborhood selection approach, sparsity is imposed on each
neighborhood. The former treatment is more natural and utilizes
the data more efficiently, especially for networks with hubs. A
prominent example is the genetic regulatory network, where master
regulators are believed to exist and are of great
interest.

\item[(ii)] According to Lemma \ref{lemma:reg}, $\beta_{ij}$ and
$\beta_{ji}$ have the same sign. The proposed method assures this
sign consistency as it estimates $\{\rho^{ij}\}$ directly.
However, when fitting $p$ separate (lasso) regressions, it is
possible that ${\rm sign}(\widehat{\beta}_{ij})$ is different from
${\rm sign}(\widehat{\beta}_{ji})$, which may lead to
contradictory neighborhoods.

\item[(iii)] Furthermore, the utility of the symmetric nature of
the problem allows us to reduce the number of unknown parameters
in the model by almost half ($p(p+1)/2$ for \texttt{space} vs.
$(p-1)^2$ for \texttt{MB}), and thus improves the  efficiency.

\item[(iv)] Finally, prior knowledge of the network structure are
  often available. The joint model is more flexible in incorporating
  such prior knowledge. For example, we may assign different weights $w_i$ to
different nodes according to their ``importance''. We have already
discussed the residual variance weights, where
$w_{i}=\sigma^{ii}$. We can also consider the weight that is
proportional to the (estimated) degree of each variable, i.e., the
estimated number of edges connecting with each node in the
network. This would result in a preferential attachment effect
which explains the cumulative advantage phenomena observed in many
real life networks including GRNs \shortcite{barabasiAlbert}.
\end{enumerate}

These advantages help enhance the performance of \texttt{space}.
As illustrated by the simulation study in Section
\ref{sec:simulation}, the proposed joint method performs better
than the neighborhood selection approach in both non-zero partial
correlation selection and hub detection.

As compared to the penalized maximum likelihood approach
\texttt{glasso} \shortcite{TiSparse07}, the simulation study in
Section \ref{sec:simulation} shows that \texttt{space} also
outperforms \texttt{glasso} in both edge detection and hub
identification under all settings that we have considered. In
addition, \texttt{space} has the following advantages.
\begin{enumerate}
\item[(i)] The complexity of \texttt{glasso} is $O(p^3)$, while as
discussed in Section 2.2, the \texttt{space} algorithm has the
complexity of $\min(O(np^2),O(p^3))$,
  which is much faster than the algorithm of \shortciteN{Yuan}
  and in general should also be faster than \texttt{glasso} when $n<p$,
  which is the case in many real studies.
\item [(ii)] As discussed in Section 6, \texttt{space} allows for
trivial
  generalizations to other penalties of the form of
  $|\rho^{ij}|^q$ rather than simply $|\rho^{ij}|$, which
  includes ridge and bridge \cite{Fu1998shooting} or other more complicated penalties
  like SCAD \cite{Fan2001SCAD}. The \texttt{glasso} algorithm, on the other hand, is
  tied to the lasso formulation and cannot be extended to other
  penalties in a natural manner.

  \item [(iii)]  In Section 5, we prove that our method
  consistently identifies the correct network neighborhood when
  {\it both $n$ and $p$} go to $\infty$.  As far as
  we are aware, no such theoretical results have been developed
  for the penalized maximum likelihood approach.

%\item [(iv)] \textcolor{red}{As a regression based method,
%\texttt{space} is more robust to distributional assumptions.}
\end{enumerate}

 Note that, in the penalized loss function (\ref{eqn:loss_pen}),
$\mathbf{\sigma}$ needs to be specified. We propose to estimate $\mathbf{\theta}$
and $\mathbf{\sigma}$ by a two-step iterative procedure. Given an initial
estimate $\mathbf{\sigma}^{(0)}$ of $\mathbf{\sigma}$, $\mathbf{\theta}$ is estimated by
minimizing the penalized loss function (\ref{eqn:loss_pen}), whose
implementation is discussed in Section \ref{sec:implementation}.
Then given the current estimates $\mathbf{\theta}^{(c)}$ and
$\mathbf{\sigma}^{(c)}$, $\mathbf{\sigma}$ is updated based on Lemma
\ref{lemma:reg}: $1/\widehat{\sigma}^{ii}=\frac{1}{n}||\mathbf{Y}_i-\sum_{j
\not=i}\widehat{\beta}^{(c)}_{ij}\mathbf{Y}_j||^2$, where
$\widehat{\beta}^{(c)}_{ij} = (\rho^{ij})^{(c)}
\sqrt{\frac{(\sigma^{jj})^{(c)}}{(\sigma^{ii})^{(c)}}}$. We then
iterate between these two steps until convergence. Since
$1/\sigma^{ii}\leq {\rm
Var}(y_i)=\sigma_{ii}$, we can use $1/\widehat{\sigma}_{ii}$ as
the initial estimate of $\sigma^{ii}$, where
$\widehat{\sigma}_{ii}=\frac{1}{n-1}\sum_{k=1}^n
(y^k_i-\bar{y}_i)^2$ is the sample variance of $y_i$. Our
simulation study shows that, it usually takes no more than three
iterations for this procedure to stabilize.

\subsection{Implementation}
\label{sec:implementation} In this section, we discuss the
implementation of the \texttt{space} procedure: that is,
minimizing (\ref{eqn:loss_pen}) under the $\ell_1$ penalty
(\ref{eqn:l1_pen}). We first re-formulate the problem, such that
the loss function (\ref{eqn:loss_l2}) corresponds to the $\ell_2$
loss of a ``regression problem.'' We then use the
\texttt{active-shooting} algorithm proposed in Section
\ref{subsec:adaptive} to solve this lasso regression problem
efficiently.

Given $\mathbf{\sigma}$ and positive weights $w$, let
$\mathcal{Y}=(\tilde{\mathbf{Y}}_1^T, ..., \tilde{\mathbf{Y}}_p^T)^T$ be a $np \times
1$ column vector, where $\tilde{\mathbf{Y}}_i=\sqrt{w_i}\mathbf{Y}_i$
($i=1,\cdots,p$); and let
$\mathcal{X}=(\tilde{\mathcal{X}}_{(1,2)},\cdots,\tilde{\mathcal{X}}_{(p-1,p)})$
be a $np$ by $p(p-1)/2$ matrix, with
$$
\begin{array}{ccccc}
\tilde{\mathcal{X}}_{(i,j)}=(0, ..., 0,& \sqrt{\frac{
\tilde{\sigma}^{jj}}{\tilde{\sigma}^{ii}}}\tilde{\mathbf{Y}}_j^T, & 0, ...,
0, & \sqrt{\frac{
\tilde{\sigma}^{ii}}{\tilde{\sigma}^{jj}}}\tilde{\mathbf{Y}}_i^T, & 0,
..., 0)^T\\
&\uparrow & & \uparrow&\\
 & i^{th} \textrm{block} & & j^{th} \textrm{block} &
\end{array},
$$
where $\tilde{\sigma}^{ii}=\sigma^{ii}/w_i$ ($i=1,\cdots,p$). Then
it is easy to see that the loss function (\ref{eqn:loss_l2})
equals to $\frac{1}{2}||\mathcal{Y}-\mathcal{X}\mathbf{\theta}||_2^2$, and
the corresponding $\ell_1$ minimization problem is equivalent to:
$\min_{\mathbf{\theta}} \frac{1}{2}||\mathcal{Y}-\mathcal{X}\mathbf{\theta}||_2^2
+\lambda ||\mathbf{\theta}||_1$. Note that, the current dimension
$\tilde{n}=np$ and $\tilde{p}=p(p-1)/2$ are of a much higher order
than the original $n$ and $p$. This could cause serious
computational problems. Fortunately, $\mathcal{X}$ is a block
matrix with many zero blocks. Thus, algorithms for lasso
regressions can be efficiently implemented by taking into
consideration this structure (see the Supplemental Material for
the detailed implementation). To further decrease the
computational cost, we develop a new algorithm
\texttt{active-shooting} (Section \ref{subsec:adaptive}) for the
\texttt{space} model fitting. \texttt{Active-shooting} is a
modification of the \texttt{shooting} algorithm, which was first proposed by
\shortciteN{Fu1998shooting} and then extended by many others including \shortciteN{Genkin07} and \shortciteN{Tibshirani2007PathCO}. \texttt{Active-shooting} exploits
the sparse nature of sparse penalization problems in a more
efficient way, and is therefore computationally much faster. This
is crucial for applying \texttt{space} for large $p$ and/or
$n$. It can be shown that the computational cost of \texttt{space}
is $\min(O(np^2),O(p^3))$, which is the same as applying $p$
individual lasso regressions as in the neighborhood selection
approach. We want to point out that, the proposed method can also
be implemented by \texttt{lars} \shortcite{lars}. However, unless the
exact whole solution path is needed, compared with
\texttt{shooting} type algorithms, \texttt{lars} is
computationally less appealing \shortcite{Tibshirani2007PathCO}.
(Remark by the authors: after this paper was submitted, recently the \texttt{active-shooting} idea was also proposed by \shortciteN{Friedman08}.)

Finally, note that the concentration matrix should be positive
definite. In principle, the proposed method (or more generally,
the regression based methods) does not guarantee the positive
definiteness of the resulting estimator, while the likelihood
based method by \shortciteN{Yuan} and \shortciteN{TiSparse07}
assures the positive definiteness. While admitting that this is
one limitation of the proposed method, we argue that, since we are
more interested in model selection than parameter estimation in
this paper, we are less concerned with this issue.  Moreover, in
Section \ref{sec:theory}, we show that the proposed estimator is
consistent under a set of suitable assumptions. Therefore, it is
asymptotically positive definite. Indeed, the \texttt{space}
estimators are rarely non-positive-definite under the high
dimensional sparse settings that we are interested in. More
discussions on this issue can be found in
Section~\ref{sec:simulation}.

\subsection{Active Shooting}
\label{subsec:adaptive} In this section, we propose a
computationally very efficient algorithm \texttt{active-shooting}
for solving lasso regression problems.
\texttt{Active-shooting} is motivated by the \texttt{shooting}
algorithm \shortcite{Fu1998shooting}, which solves the lasso
regression by updating each coordinate iteratively until
convergence. \texttt{Shooting} is computationally very competitive
compared with the well known \texttt{lars} procedure
\shortcite{lars}. Suppose that we want to minimize an $\ell_1$
penalized loss function with respect to $\beta$
$$
f(\beta)=\frac{1}{2}||\mathbf{Y}-\mathbf{X}\mathbf{\beta}||_2^2+\gamma \sum_{j}|\beta_j|,
$$
where $\mathbf{Y}=(y_1,\cdots,y_n)^T$, $\mathbf{X}=(x_{ij})_{n\times
p}=(\mathbf{X}_1:\cdots:\mathbf{X}_p)$ and $\mathbf{\beta}=(\beta_1,\cdots,\beta_p)^T$. The
\texttt{shooting} algorithm proceeds as follows:
\begin{itemize}
\item [1.] Initial step: for $j=1,\cdots,
p,$
\begin{equation}
\label{eqn:initial}
\begin{array}{rl}
\beta_j^{(0)}
&=~ \textrm{arg}\min_{\beta_j}\{\frac{1}{2}||\mathbf{Y}-\beta_j\mathbf{X}_j||^2+\gamma
|\beta_j|\}\\
 &=~ \textrm{sign}(\mathbf{Y}^{T}\mathbf{X}_j)\frac{(|\mathbf{Y}^{T}\mathbf{X}_j|-\gamma)_{+}}{\mathbf{X}_j^{T}\mathbf{X}_j},
\end{array}
\end{equation}
where $(x)_{+}=xI_{(x>0)}$.
\item[2.] For $j=1, ..., p$, update
$\beta^{(old)}\longrightarrow\beta^{(new)}:$
\begin{equation}
\label{eqn:shoot_update}
\begin{array}{rl}
\beta^{(new)}_i &=~ \beta^{(old)}_i, i\neq j;\\
\beta^{(new)}_j &=~
\textrm{arg}\min_{\beta_j}{\frac{1}{2}\left\|\mathbf{Y}-\sum_{i\neq
j}\beta_i^{(old)}\mathbf{X}_i-\beta_j \mathbf{X}_j\right\|^2+\gamma |\beta_j|}\\
&=~ \textrm{sign}\left(\frac{(\mathbf{\epsilon}^{(old)})^T\mathbf{X}_j}{\mathbf{X}^T_j\mathbf{X}_j}+\beta_j^{(old)}\right)
\left( \left| \frac{(\mathbf{\epsilon}^{(old)})^T\mathbf{X}_j}{\mathbf{X}^T_j\mathbf{X}_j} + \beta^{(old)}_j
\right| - \frac{\gamma}{\mathbf{X}^T_j\mathbf{X}_j}\right)_+,
\end{array}
\end{equation}
where $\mathbf{\epsilon}^{(old)}=\mathbf{Y}-\mathbf{X}\mathbf{\beta}^{(old)}$. \item[3.] Repeat step $2$ until
convergence.
\end{itemize}
At each updating step of the \texttt{shooting} algorithm, we define
the set of currently non-zero coefficients as the \textit{active
  set}. Since under sparse models, the active set
should
remain small, we propose to first update the coefficients within the
active set until convergence is achieved before moving on to update
other coefficients. The \texttt{active-shooting} algorithm proceeds
as follows:
\begin{itemize}\itemsep=-3pt
\item[1.] Initial step: same as the initial step of
\texttt{shooting}.
\item[2.] Define the current active set $\Lambda=\{k:
\hbox{current}~ \beta_k\neq 0 \}$.
\vspace{-5pt}\begin{itemize}\itemsep=-3pt\vspace{-3pt}
\item[$\textrm{ }(2.1)$] For each $k \in \Lambda$, update
$\beta_k$ with all other coefficients fixed at the current value
as in equation (\ref{eqn:shoot_update});
 \item[$\textrm{ }(2.2)$] Repeat (2.1)
until convergence is achieved on the active set.
\end{itemize}\vspace{-5pt}
\item[3.] For $j=1$ to $p$, update $\beta_j$ with all other
coefficients fixed at the current value as in equation
(\ref{eqn:shoot_update}). If no $\beta_j$ changes during this
process, return the current $\beta$ as the final estimate.
Otherwise, go back to step 2.
\end{itemize}
\begin{table}[h]
 \centering \caption{The numbers of
iterations required by the \texttt{shooting} algorithm and the
\texttt{active-shooting} algorithm to achieve
convergence ($n=100$, $\lambda=2$). ``coef. $\#$'' is the number
of non-zero coefficients}

\begin{tabular}{c|c|c c}
 \hline\hline {$p$} & {coef. $\#$}& {\texttt{shooting}} &
{\texttt{active-shooting}} \\
\hline {200}& {14}& {29600} & {4216} \\
\hline {500}& {25} & {154000} & {10570} \\
\hline {1000}& {28} & {291000} & {17029}\\
\hline\hline
\end{tabular}
\label{table:adaptive}
\end{table}

The idea of \texttt{active-shooting} is to focus on the set of
variables that is more likely to be in the model, and thus it
improves the computational efficiency by achieving a faster
convergence. We illustrate the improvement of the
\texttt{active-shooting} over the \texttt{shooting} algorithm by a
small simulation study of the lasso regression (generated in the
same way as in Section 5.1 of \shortciteN{Tibshirani2007PathCO}).
The two algorithms result in exact same solutions. However, as can
be seen from Table \ref{table:adaptive}, \texttt{active-shooting} takes much fewer
iterations to converge (where one iteration is counted whenever an
attempt to update a $\beta_j$ is made).
%The convergence of \texttt{active-shooting} is
%about $10$ times faster than that of the \texttt{shooting}
%algorithm.
In particular, it takes less than $30$ seconds (on average) to fit
the \texttt{space} model by \texttt{active-shooting} (implemented
in \texttt{c} code) for cases with $1000$ variables, $200$ samples
and when the resulting model has  around $1000$ non-zero partial
correlations on a server with two Dual/Core, CPU 3 GHz and 4 GB
RAM. This great computational advantage enables us to conduct
large scale simulation studies to examine the performance of the
proposed method (Section \ref{sec:simulation}).

{\remark : In the initial step, instead of using the univariate
soft-shrinkage estimate, we can use a previous estimate as the
initial estimate if such a thing is available. For example, when
iterating between $\{\rho^{ij}\}$ and $\{\sigma^{ii}\}$, we can
use the previous estimate  of $\{\rho^{ij}\}$  in the current
iteration as the initial value. This can further improve the
computational efficiency of the proposed method, as a better initial
value implies a faster convergence. Moreover, in practice, often
estimates are desired for a series of tuning parameters $\lambda$,
whether it is for data exploration or for the selection of
$\lambda$. When this is the case, a \textit{decreasing-lambda}
approach can be used to facilitate computation.  That is, we start
with the largest $\lambda$ (which results in the smallest model),
then use the resulting estimate as the initial value when fitting
the model under the second largest $\lambda$ and continue in this
manner until all estimates are obtained.}

\subsection{Tuning \label{sec:tuning}}
 The choice of the tuning parameter $\lambda$ is
of great importance.  Since the \texttt{space} method uses a lasso
criterion, methods that have been developed for selecting the
tuning parameter for lasso can also be applied to \texttt{space},
such as the GCV in \shortciteN{Ti96}, the CV in
\shortciteN{Fan2001SCAD}, the AIC in \shortciteN{Buhlmann06} and
the BIC in \shortciteN{ZouEtAl07}. Several methods have also been
proposed for selecting the tuning parameter in the setting of
covariance estimation, for example, the MSE based criterion in
\shortciteN{SchaferStrimmer07}, the likelihood based method in
\shortciteN{huang06} and the cross-validation and bootstrap
methods in \shortciteN{Li}. In this paper, we propose to use a
``BIC-type" criterion for selecting the tuning parameter mainly
due to its simplicity and computational easiness. For a given
$\lambda$, denote the \texttt{space} estimator by
$\widehat{\theta}_\lambda=\{\widehat{\rho}_\lambda^{ij}: 1 \leq i
< j \leq p\}$ and
$\widehat{\sigma}_\lambda=\{\widehat{\sigma}_\lambda^{ii}: 1 \leq
i \leq p\}$. The
 corresponding residual sum of squares for  the $i$-th regression:
$y_i=\sum_{j\not=i}\beta_{ij}y_j +\epsilon_i$ is
\begin{eqnarray*}
RSS_i(\lambda)=\sum_{k=1}^n \left(y^k_i-\sum_{j \not=
i}\widehat{\rho}_\lambda^{ij}\sqrt{\frac{\widehat{\sigma}_\lambda^{jj}}{\widehat{\sigma}_\lambda^{ii}}}y^k_j\right)^2.
\end{eqnarray*}
 We then define a ``BIC-type" criterion for the $i$-th
regression as
\begin{eqnarray}
\label{eqn:bic_sep} BIC_i(\lambda)=n \times
\log(RSS_i(\lambda))+\log n \times \#\{j: j \not =i,
\widehat{\rho}_\lambda^{ij} \not = 0\}.
\end{eqnarray}
Finally, we define $BIC(\lambda):=\sum_{i=1}^p BIC_i(\lambda)$ and
select $\lambda$ by minimizing $BIC(\lambda)$. This method is
referred to as \texttt{space.joint} hereafter.

In  \shortciteN{Yuan}, a BIC criterion is proposed for the
penalized maximum likelihood approach. Namely
\begin{equation}
\label{eqn:bic} BIC(\lambda):=n \times
\left[-\log|\widehat{\mathbf{\Sigma}}_\lambda^{-1}|+\rm{trace}(\widehat{\mathbf{\Sigma}}_\lambda^{-1}\mathbf{S})\right]+\log
n \times  \#\{(i,j): 1 \leq i \leq j \leq p, \widehat{\sigma}_\lambda^{ij}
\not= 0\},
\end{equation}
where $\mathbf{S}$ is the sample covariance matrix, and
$\widehat{\mathbf{\Sigma}}_\lambda^{-1}=(\widehat{\mathbf{\sigma}}_\lambda^{ij})$
is the estimator under $\lambda$. In this paper, we refer to this
method as \texttt{glasso.like}. For the purpose of comparison, we
also consider the selection of the tuning parameter for \texttt{MB}.
Since \texttt{MB} essentially performs $p$ individual lasso
regressions, the tuning parameter can be selected for each of them
separately. Specifically, we use criterion (\ref{eqn:bic_sep})
(evaluated at the corresponding \texttt{MB} estimators) to select
the tuning parameter $\lambda_i$ for the $i$-th regression. We
denote this method as \texttt{MB.sep}. Alternatively, as suggested
by \shortciteN{meinshausen}, when all $Y_i$ are standardized to have
sample standard deviation one, the same
$\lambda(\alpha)=\sqrt{n}\Phi^{-1}(1-\frac{\alpha}{2p^2})$ is
applied to all regressions. Here, $\Phi$ is the standard normal
c.d.f.; $\alpha$ is used to control the false discovery rate and is
usually taken as  $0.05$ or $0.1$. We denote this method as
$\texttt{MB.alpha}$. These methods are examined by the simulation
studies in the next section.

%%%%%%%%%%%%%%%%%%%%%%%%%%%%%%%%%%%%%%%%%%%%%%%%%%%%%%%%%%%%%%%%%%%%%
{\centering \section{SIMULATION}\label{sec:simulation}}

%%%%%%%%%%%%%%%%%%%%%%%%%%%%%%%%%%%%%%%%%%%%%%%%%%%%%%%%%%%%%%%%%%%%%
In this section, we conduct a series of simulation experiments to
examine the performance of the proposed method \texttt{space} and
compare it with the neighborhood selection approach \texttt{MB} as
well as the penalized likelihood method \texttt{glasso}. For all
three methods, variables are first standardized to have sample
mean zero and sample standard deviation one before model
fitting. For \texttt{space}, we consider three different types of
weights: (1) uniform weights: $w_i=1$; (2) residual variance based
weights: $w_i=\widehat{\sigma}^{ii}$; and (3) degree based
weights: $w_i$ is proportional to the estimated degree of $y_i$,
i.e., $\#\{j: \widehat{\rho}^{ij} \not= 0, j \not=i\}$. The
corresponding methods are referred as \texttt{\textbf{space}},
\texttt{\textbf{space.sw}} and \texttt{\textbf{space.dew}},
respectively. For all three \texttt{space} methods, the initial
value of $\sigma^{ii}$ is set to be one. Iterations are used for
these \texttt{space} methods as discussed in Section
\ref{subsec:model}. For \texttt{space.dew} and \texttt{space.sw},
the initial weights are taken to be one (i.e., equal weights). In
each subsequent iteration, new weights are calculated based on the
estimated residual variances (for \texttt{\textbf{space.sw}}) or
the estimated degrees (for \texttt{\textbf{space.dew}}) of the
previous iteration. For all three \texttt{space} methods, three
iterations (that is updating between $\{\sigma^{ii}\}$ and
$\{\rho^{ij}\}$) are used since the procedure converges very fast
and more iterations result in essentially the same estimator. For
\texttt{glasso}, the diagonal of the concentration matrix is not
penalized.
%(penalize.diagonal=FALSE).}

%A lot of real life large networks exhibit a modular structure
%comprised of many disjoint or loosely connected components of
%relatively small size. For example, experiments on model organisms
%like yeast or bacteria suggest that the transcriptional regulatory
%networks have modular structures \shortcite{Lee2002}. Therefore,
%we simulate networks consisting of disjoint modules.
We simulate networks consisting of disjointed modules. This is
done because many real life large networks exhibit a modular
structure comprised of many disjointed or loosely connected
components of relatively small size. For example, experiments on
model organisms like yeast or bacteria suggest that the
transcriptional regulatory networks have modular structures
\shortcite{Lee2002}. Each of our network modules is set to have
$100$ nodes and generated according to a given degree
distribution, where the \textit{degree} of a node is defined as
the number of edges connecting to it. We mainly consider two
different types of degree distributions and denote their
corresponding networks by \texttt{Hub network} and
\texttt{Power-law network} (details are given later). Given an
undirected network with $p$ nodes, the initial ``concentration
matrix" $(\tilde{\sigma}^{ij})_{p\times
  p}$
 is generated by
\begin{eqnarray}
\label{eqn:concen} \tilde{\sigma}^{ij}= \left\{\begin{array}{cc}
1,
& i=j;\\
0,& i\neq j \ \hbox{and no edge between nodes $i$ and $j$};\\
\sim Uniform([-1, -0.5]\cup[0.5, 1]), & i\neq j \ \hbox{ and an
edge connecting
nodes $i$ and $j$}.\\
\end{array} \right.
\end{eqnarray}
We then rescale the non-zero elements in the above matrix to
assure positive definiteness. Specifically, for each row, we first
sum the absolute values of the off-diagonal entries, and then
divide each off-diagonal entry by $1.5$ fold of the sum. We then
average this re-scaled matrix with its transpose to ensure
symmetry. Finally the diagonal entries are all set to be one. This
process results in diagonal dominance. Denote the final matrix as
$\mathbf{A}$. The covariance matrix $\mathbf{\Sigma}$ is then determined by
$$\mathbf{\Sigma}(i,j)=\mathbf{A}^{-1}(i,j)/\sqrt{\mathbf{A}^{-1}(i,i)\mathbf{A}^{-1}(j,j)}.$$
Finally, i.i.d. samples $\{\mathbf{Y}^k\}_{k=1}^n$ are generated from
${\rm Normal}(0,\mathbf{\Sigma})$. Note that, $\mathbf{\Sigma}(i,i)=1$, and
$\mathbf{\Sigma}^{-1}(i,i)=\sigma^{ii} \geq 1$.

%\vspace{5pt}

\subsection*{Simulation Study I}
\noindent{\textbf{Hub networks}\hspace{3pt} In the first set of
simulations, module networks are generated by inserting a few hub
nodes into a very sparse graph. Specifically, each module consists
of three hubs with degrees around $15$, and the other $97$ nodes
with degrees at most four. This setting is designed to mimic the
genetic regulatory networks, which usually contains a few hub genes
plus many other genes with only a few edges. A network consisting of
five such modules is shown in Figure \ref{Fig:HubNet}. In this
network, there are $p=500$ nodes and $568$ edges. The simulated
non-zero partial correlations fall in $(-0.67, -0.1]\cup[0.1,
0.67)$, with two modes around -0.28 and 0.28. Based on this network
and the partial correlation matrix, we generate $50$ independent
data sets each consisting of $n=250$ i.i.d. samples. }

We then evaluate each method at a series of different values of
the tuning parameter $\lambda$. The number of total detected edges
($N_t$) decreases as $\lambda$ increases. Figure \ref{Fig:HubDetectEdge} shows the
number of correctly detected edges ($N_c$) vs. the number of total
detected edges ($N_t$) averaged across the $50$ independent data
sets for each method. We observe that all three \texttt{space}
methods (\texttt{space}, \texttt{space.sw} and \texttt{space.dew})
consistently detect more correct edges than the neighborhood
selection method \texttt{MB} (except for \texttt{space.sw} when
$N_t<470$) and  the likelihood based method \texttt{glasso}.
\texttt{MB} performs favorably over \texttt{glasso}
 when $N_t$ is relatively small (say less than $530$), but
 performs worse than \texttt{glasso} when $N_t$ is large.
Overall, \texttt{space.dew} is the best among all methods.
Specifically, when $N_t=568$ (which is the number of true edges),
\texttt{space.dew} detects $501$ correct edges on average with a
standard deviation $4.5$ edges. The corresponding sensitivity and
specificity are both $88\%$. Here, sensitivity is defined as the
ratio of the number of correctly detected edges to the total number
of true edges; and specificity is defined as the ratio of the number
of correctly detected edges to the number of total detected edges.
On the other hand, \texttt{MB} and \texttt{glasso} detect $472$ and
$480$ correct edges on average, respectively, when the number of
total detected edges $N_t$ is $568$.

In terms of hub detection, for a given $N_t$, a rank  is assigned to each variable $y_i$ based
on its estimated degree (the larger the estimated degree, the
smaller the rank value). We then calculate the average rank of the
$15$ true hub nodes for each method. The results are shown in
Figure \ref{Fig:HubRank}. This average rank would achieve the minimum value $8$
(indicated by the grey horizontal line), if the $15$ true hubs
have larger estimated degrees than all other non-hub nodes. As can
be seen from the figure, the average rank curves (as a function of
$N_t$) for the three \texttt{space} methods are very close to the
optimal minimum value $8$ for a large range of $N_t$. This
suggests that these methods can successfully identify most of the
true hubs. Indeed, for \texttt{space.dew}, when $N_t$ equals to
the number of true edges ($568$), the top $15$ nodes with the
highest estimated degrees contain at least $14$ out of the $15$
true hub nodes in all replicates. On the other hand, both
\texttt{MB} and \texttt{glasso} identify far fewer hub nodes, as
their corresponding average rank curves are much higher than the
grey horizontal line.

%\textcolor{red}{In terms of computational cost of these methods,
%as can be seen from Table 3, \texttt{MB} is the least
%computational intensive. The three \texttt{space} methods are
%similar in terms of computation, and require about $1.5$ as much
%time as \texttt{glasso} when three iterations are used.}

\begin{table}[h]
\centering \caption{Power (sensitivity) of  \texttt{space.dew} , \texttt{MB} and
\texttt{glasso} in identifying correct
edges when FDR is controlled at 0.05.}
\begin{tabular}{c|c|c||ccc}
\hline\hline Network&{$p$} & {$n$} & \texttt{space.dew}&
\texttt{MB} &
\texttt{glasso} \\
\hline\hline Hub-network&{500}&
{250} &  {0.844} & {0.784} &{0.655} \\
\hline
& & {200} &{0.707} & {0.656}&   {0.559}  \\
Hub-network& {1000} & {300} &  {0.856}  & {0.790}&  {0.690} \\
& & {500}  & {0.963} & {0.894}& { 0.826} \\\hline Power-law
network& 500&250 &{0.704} &{0.667}&{0.580}\\\hline\hline
\end{tabular}\label{table:power}
\end{table}

To investigate the impact of dimensionality $p$ and sample size
$n$, we perform simulation studies for a larger dimension with
$p=1000$ and various sample sizes with $n=200, 300$ and $500$. The
simulated network includes ten disjointed modules of size $100$ each
and has $1163$ edges in total. Non-zero partial correlations form
a similar distribution as that of the $p=500$ network discussed
above.  The ROC curves for \texttt{space.dew}, \texttt{MB} and
\texttt{glasso} resulted from these simulations are shown in
Figure \ref{Fig:HubNet1000}. When false discovery rate (=1-specificity) is controlled
at $0.05$, the power (=sensitivity) for detecting correct edges
is given in Table \ref{table:power}. From the figure and the table, we observe
that the sample size has a big impact on the performance of all
methods. For $p=1000$, when the sample size increases from $200$
to $300$, the power of \texttt{space.dew} increases  more than
$20\%$; when the sample size is $500$, \texttt{space.dew} achieves
an impressive power of $96\%$. On the other hand, the
dimensionality seems to have relatively less influence. When the
total number of variables is doubled from $500$ to $1000$, with
only $20\%$ more samples (that is $p=500, n=250$ vs. $p=1000, n=300$), all three methods achieve similar
powers. This is presumably because the larger network ($p=1000$)
is sparser than the smaller network ($p=500$) and also the
complexity of the modules remains unchanged. Finally, it is
obvious from Figure \ref{Fig:HubNet1000} that, \texttt{space.dew} performs best
among the three methods.

\begin{table}[h]
\small \centering \caption{Edge detection under the selected tuning
parameter $\lambda$. For \textit{average rank}, the optimal value is
$15.5$. For \texttt{MB.alpha},
 $\alpha=0.05$ is used.}

\begin{tabular}{c|ccccc}\hline\hline Sample
size &Method& Total edge detected & Sensitivity &
Specificity&Average rank\\\hline\hline

&\texttt{space.joint}& 1357& 0.821  & 0.703&28.6\\
$n=200$&\texttt{MB.sep}& 1240 &    0.751 & 0.703&57.5\\
& \texttt{MB.alpha}&404& 0.347&1.00&175.8\\
 &\texttt{glasso.like}&  1542&     0.821 &0.619&35.4\\\hline
&\texttt{space.joint}& 1481    &0.921&0.724&18.2 \\

$n=300$ &\texttt{MB.sep}&1456& 0.867&0.692&30.4\\
& \texttt{MB.alpha}&562& 0.483&1.00&128.9\\
&\texttt{glasso.like}&    1743  &0.920&0.614&21
\\\hline
&\texttt{space.joint}   &1525 &0.980&0.747&16.0 \\
$n=500$&\texttt{MB.sep}&  1555 & 0.940&0.706&16.9  \\
& \texttt{MB.alpha}&788& 0.678&1.00&52.1\\
&\texttt{glasso.like}&    1942 & 0.978 &0.586&16.5\\\hline\hline
\end{tabular}\label{Table:BIC}
\end{table}

We then investigate the performance of these methods at the
selected tuning parameters (see Section~\ref{sec:tuning} for details). For the
above Hub network with $p=1000$ nodes and $n=200, 300, 500$, the
results are reported in Table \ref{Table:BIC}. As can be seen from the table,
BIC based approaches tend to select large models (compared to the
true model which has $1163$ edges). \texttt{space.joint} and \texttt{MB.sep} perform
similarly in terms of specificity, and \texttt{glasso.like} works
considerably worse than the other two in this regard. On the other
hand, \texttt{space.joint} and \texttt{glasso.like} performs
similarly in terms of sensitivity, and are better than
\texttt{MB.sep} on this aspect. In contrast, $\texttt{MB.alpha}$
selects very small models and thus results in very high
specificity, but very low sensitivity. In terms of hub
identification, \texttt{space.joint} apparently performs better
than other methods (indicated by a smaller average rank over $30$ true hub nodes).
%Overall, \texttt{space.joint} performs well in
%edge detection and hub identification.
Moreover, the performances
of all methods improve with sample size.

 %\vspace{10pt}

\noindent{\textbf{Power-law networks} \hspace{3pt} Many real world
networks have a \textit{power-law (also a.k.a scale-free)} degree
distribution with an estimated power parameter $\alpha=2 \sim 3$
\shortcite{newman}. Thus, in the second set of simulations, the
module networks are generated according to a power-law degree
distribution with the power-law parameter $\alpha=2.3$, as this
value is close to the estimated power parameters for biological
networks \shortcite{newman}. Figure \ref{Fig:PowerLawNet} illustrates a network
formed by five such modules with each having $100$ nodes.
%Figure 5(a) shows the degree distribution of this network.
It can be seen that there are three obvious hub nodes in this
network with degrees of at least $20$. The simulated non-zero
partial correlations fall in the range $(-0.51, -0.08]\cup[0.08,
0.51)$, with two modes around -0.22 and 0.22. Similar to the
simulation done for Hub networks, we generate $50$ independent
data sets each consisting of $n=250$ i.i.d. samples. We then
compare the number of correctly detected edges by various methods.
The result is shown in Figure \ref{Fig:PowerLawCurve}. On average, when the number of
total detected edges equals to the number of true edges which is
$495$, \texttt{space.dew} detects $406$ correct edges, while
\texttt{MB} detects only $378$ and \texttt{glasso} detects only
$381$ edges. In terms of hub detection, all methods can correctly
identify the three hub nodes for this network.}

\noindent{\textbf{Summary} \hspace{3pt} These simulation results
suggest that when the (concentration) networks are reasonably
sparse, we should be able to characterize their structures with only
a couple-of-hundreds of samples when there are a couple of thousands
of nodes. In addition, \texttt{space.dew} outperforms \texttt{MB} by
at least $6\%$ on the power of edge detection under all simulation
settings above when FDR is controlled at $0.05$, and the
improvements are even larger when FDR is controlled at a higher
level say $0.1$ (see Figure \ref{Fig:HubNet1000}). Also, compared to
\texttt{glasso}, the improvement of \texttt{space.dew} is at least
$15\%$ when FDR is controlled at $0.05$, and the advantages  become
smaller when FDR is controlled at a higher level (see Figure
\ref{Fig:HubNet1000}). Moreover, the \texttt{space} methods perform
much better in hub identification than both \texttt{MB} and
\texttt{glasso}.

\subsection*{Simulation Study II}
In the second simulation study, we apply \texttt{space}, \texttt{MB} and \texttt{glasso} on networks with nearly
uniform degree distributions generated by following the simulation
procedures in \shortciteN{meinshausen}; as well as on the AR network
discussed in \shortciteN{Yuan} and \shortciteN{TiSparse07}. For
these cases, \texttt{space} performs comparably, if not
better than, the other two methods. However, for these networks
without hubs, the advantages of \texttt{space} become smaller
compared to the results on the networks with hubs. The results are
summarized below.

\noindent{\textbf{Uniform networks} \hspace{3pt} In this set of
simulation, we generate similar networks as the ones used in
\shortciteN{meinshausen}. These networks have uniform degree
distribution with degrees ranging from zero to four. Figure
\ref{Fig:Uniform} illustrates a network formed by five such modules
with each having $100$ nodes. There are in total $447$ edges. Figure
\ref{Fig:Uniform-curve} illustrates the performance of $\texttt{MB},
\texttt{space}$ and $\texttt{glasso}$ over $50$ independent data
sets each having $n=250$ i.i.d. samples. As can be been from this
figure, all three methods perform similarly. When the total number
of detected edges equals to the total number of true edges ($447$),
\texttt{space} detects $372$ true edges, \texttt{MB} detects $369$
true edges and $\texttt{glasso}$ $371$ true edges.

\noindent{\textbf{AR networks} \hspace{3pt} In this simulation, we
consider the so called AR network used in \shortciteN{Yuan} and
\shortciteN{TiSparse07}. Specifically, we have $\sigma^{ii}=1$ for
$i=1,\cdots p$ and  $\sigma^{i-1,i}= \sigma^{i,i-1}=0.25$ for
$i=2,\cdots,p$. Figure \ref{Fig:AR} illustrates such a network with
$p=500$ nodes and thus $499$ edges. Figure \ref{Fig:AR-curve}
illustrates the performance of $\texttt{MB}, \texttt{space}$ and
$\texttt{glasso}$ over $50$ independent data sets each having
$n=250$ i.i.d. samples. As can be been from this figure, all three
methods again perform similarly. When the total number of detected
edges equals to the total number of true edges ($499$),
\texttt{space} detects $416$ true edges, \texttt{MB} detects $417$
true edges and $\texttt{glasso}$ $411$ true edges. As a slight
modification of the AR network, we also consider a big circle
network with: $\sigma^{ii}=1$ for $i=1,\cdots p$; $\sigma^{i-1,i}=
\sigma^{i,i-1}=0.3$  for $i=2,\cdots,p$ and $\sigma^{1,p}=
\sigma^{p,1}=0.3$. Figure \ref{Fig:Cicle} illustrates such a network
with $p=500$ nodes and thus $500$ edges. Figure
\ref{Fig:Cicle-result} compares the performance of the three
methods. When the total number of detected edges equals to the total
number of true edges ($500$), \texttt{space}, \texttt{MB} and
\texttt{glasso} detect $478$, $478$ and $475$ true edges,
respectively.

We also compare the mean squared error (MSE) of estimation of
$\{\sigma^{ii}\}$. For the uniform network, the median (across all
samples and $\lambda$) of the square-root MSE is $0.108, 0.113,
0.178$ for $\texttt{MB}$, $\texttt{space}$ and $\texttt{glasso}$.
These numbers are $0.085, 0.089, 0.142$ for the AR network and
$0.128, 0.138, 0.233$ for the circle network. It seems that
\texttt{MB} and \texttt{space} work considerably better than
$\texttt{glasso}$ on this aspect.

\subsection*{Comments}
We conjecture that, under the sparse and high dimensional setting,
the superior performance in model selection of the regression based
method \texttt{space} over the penalized likelihood method \texttt{glasso} is partly
due to its simpler quadratic loss function. Moreover, since
\texttt{space} ignores the correlation structure of the regression
residuals, it amounts to a greater degree of regularization, which
may render additional benefits under the sparse and high dimensional
setting.

In terms of parameter estimation, we compare the entropy loss of
the three methods. We find that, they perform similarly when the
estimated models are of small or moderate size.
When the estimated models are large, \texttt{glasso} generally performs
better in this regard than the other two methods.
Since the interest of this paper lies in model selection, detailed
results of parameter estimation are not reported here.

As discussed earlier, one limitation of \texttt{space} is
its lack of assurance of positive definiteness.
However, for simulations reported above, the
corresponding estimators we have examined (over $3000$ in total)
are all positive definite.  To further investigate this issue, we design a few additional
simulations. We first consider
a case with a similar network structure as the Hub network,
however having a nearly singular concentration matrix (the
condition number is $16,240$; as a comparison, the condition
number for the original Hub network is $62$). For
this case, the estimate of \texttt{space} remains positive
definite until the number of total detected edges increases to
$50,000$; while the estimate of \texttt{MB} remains positive
definite until the number of total detected edges is more than
$23,000$. Note that, the total number of true edges of this model
is only $568$, and the model selected by \texttt{space.joint} has
$791$ edges. In the second simulation, we consider a
denser network ($p=500$ and the number of true edges is
$6,188$) with a nearly singular concentration matrix (condition
number is $3,669$). Again, we observe that, the \texttt{space} estimate only
becomes non-positive-definite when the estimated models are huge
(the number of detected edges is more than
$45,000$). This suggests that, for the regime we are
interested in in this paper (the sparse and high dimensional
setting), non-positive-definiteness does not seem to be a big
issue for the proposed method, as it only occurs when the
resulting model is huge and thus very far away from the true
model. As long as the estimated models are reasonably sparse, the
corresponding estimators by \texttt{space} remain positive
definite. We believe that this is partly due to the heavy
shrinkage imposed on the off-diagonal entries in order to ensure
sparsity.

Finally, we investigate the performance of these methods when the observations come from a non-normal distribution. Particularly, we consider the multivariate $t_{df}$-distribution with $df=3, 6,
  10$. The performances of all three methods deteriorate compared to the normal case, however the overall picture in terms of relative performance among these methods remains essentially unchanged (Table \ref{table:tdf}).
  \begin{table}[h]
  \small \centering
  \caption{{Sensitivity of different methods under different
      $t_{df}$-distributions when FDR is controlled at 0.05}}
  \begin{tabular}{c|c|c|c}
    \hline \hline
    & & \multicolumn{2}{c}{Sensitivity} \\
    \cline{3-4}
    df & Method & Hub & Power-law \\
    \hline
    & \texttt{space} & 0.369 & 0.286 \\
    3 & \texttt{MB} & 0.388 & 0.276 \\
    & \texttt{glasso} & 0.334 & 0.188 \\
    \hline
    & \texttt{space} & 0.551 & 0.392 \\
    6 & \texttt{MB} & 0.535 & 0.390 \\
    & \texttt{glasso} & 0.471 & 0.293 \\
    \hline
    & \texttt{space} & 0.682 & 0.512 \\
    10 & \texttt{MB} & 0.639 & 0.518 \\
    & \texttt{glasso} & 0.598 & 0.345 \\
    \hline \hline
  \end{tabular}\label{table:tdf}
  \end{table}

%\textcolor{red}{more discussion on computational: R/C code, cost,
%etc.; also implication of GRN in discussion section}

{\centering \section{APPLICATION}}

More than 500,000 women die annually of breast cancer world wide.
Great efforts are being made to improve the prevention, diagnosis
and treatment for breast cancer. Specifically, in the past couple of
years, molecular diagnostics of breast cancer have been
revolutionized by high throughput genomics technologies. A large
number of gene expression signatures have been identified (or even
validated) to have potential clinical usage. However, since breast
cancer is a complex disease, the tumor process cannot be understood
by only analyzing individual genes. There is a pressing need to
study the interactions between genes, which may well lead to better
understanding of the disease pathologies.

In a recent breast cancer study, microarray expression experiments
were conducted for $295$ primary invasive breast carcinoma samples
\shortcite{Chang2005,VanDe2002}. Raw array data and patient clinical
outcomes for $244$ of these samples are available on-line and are
used in this paper. Data can be  downloaded at
\texttt{http://microarray-pubs.stanford.edu/wound\_NKI/explore.html
}. To globally characterize the association among thousands of mRNA
expression levels in this group of patients, we apply the
\texttt{space} method on this data set as follows. First, for each
expression array, we perform the global normalization by centering
the mean to zero and scaling the median absolute deviation to one.
Then we focus on a subset of $p=1217$ genes/clones whose expression
levels are significantly associated with tumor progression
($p$-values from univariate Cox models $<$ 0.0008, corresponding FDR
$=0.01$). We estimate the partial correlation matrix of these $1217$
genes with \texttt{space.dew} for a series of $\lambda$ values. The
degree distribution of the inferred network is heavily skewed to the
right. Specifically, when $629$ edges are detected, $598$ out of the
$1217$ genes do not connect to any other genes, while five genes
have degrees of at least 10. The power-law parameter of this degree
distribution is $\alpha = 2.56$
%[Figure \ref{Fig:RealDataDegree}]
, which is consistent with
the findings in the literature for GRNs ~\shortcite{newman}. The
topology of the inferred network is shown in Figure
\ref{Fig:RealNet}, which supports the statement that genetic
pathways consist of many genes with few interactions and a few hub
genes with many interactions.

We then search for potential hub genes by ranking nodes according
to their degrees. There are 11 candidate hub genes whose degrees
consistently rank the highest under various $\lambda$ [see Figure
\ref{Fig:RealDegreeHub}]. Among these 11 genes, five are important
known regulators in breast cancer. For example, \textit{HNF3A}
(also known as \textit{FOXA1}) is a transcription factor expressed
predominantly in a subtype of breast cancer, which regulates the
expression of the cell cycle inhibitor $p27kip1$ and the cell
adhesion molecule E-cadherin. This gene is essential for the
expression of approximately $50\%$ of estrogene-regulated genes
and has the potential to serve as a therapeutic target
\shortcite{Nakshatri2007}. Except for \textit{HNF3A}, all the
other 10 hub genes fall in the same big network component related
to cell cycle/proliferation. This is not surprising as it is
well-agreed that cell cycle/proliferation signature is prognostic
for breast cancer. Specifically, \textit{KNSL6, STK12, RAD54L} and
\textit{BUB1} have been previously reported to play a role in
breast cancer: \textit{KNSL6} (also known as \textit{KIF2C}) is
important for anaphase chromosome segregation and centromere
separation, which is overexpressed in breast cancer cells but
expressed undetectably in other human tissues except testis
\shortcite{Shimo2007}; \textit{STK12} (also known as
\textit{AURKB}) regulates chromosomal segregation during mitosis
as well as meiosis, whose LOH contributes to an increased breast
cancer risk and may influence the therapy outcome
\shortcite{Tchatchou2007}; RAD54L is a recombinational repair
protein associated with tumor suppressors BRCA1 and BRCA2, whose
mutation leads to defect in repair processes involving homologous
recombination and triggers the tumor development
\shortcite{Matsuda1999}; in the end, BUB1 is a spindle checkpoint
gene and belongs to the BML-1 oncogene-driven pathway, whose
activation contributes to the survival life cycle of cancer stem
cells and promotes tumor progression. The roles of the other six
hub genes in breast cancer are worth of further investigation. The
functions of all hub genes are briefly summarized in Table \ref{Table:Annotation}.
%Finally, we want to mention that we have also applied \texttt{MB}
%and \texttt{glasso} on this data set, but neither method
%identifies any hubs.

\begin{table}[h]
\centering
%\scriptsize
\caption{Annotation of hub genes}
\begin{tabular}{c|c|l}
\hline\hline Index & Gene Symbol & Summary Function
(GO)\\\hline\hline
1 & CENPA & Encodes a centromere protein (nucleosome assembly) \\
2 & \textit{NA.} & \textit{Annotation not available}\\
3 & KNSL6 & Anaphase chromosome segregation (cell proliferation)\\
4 & STK12 & Regulation of chromosomal segregation (cell cycle) \\
5 & \textit{NA.} & \textit{Annotation not available}\\
6 & URLC9 & \textit{Annotation not available} (up-regulated in lung cancer)\\
7 & HNF3A & Transcriptional factor activity (epithelial cell
differentiation)\\
8 & TPX2 & Spindle formation (cell proliferation)\\
9 & RAD54L & Homologous recombination related DNA repair (meiosis)\\
10 & ID-GAP & Stimulate GTP hydrolysis (cell cycle)\\
11 & BUB1 & Spindle checkpoint (cell cycle)\\\hline\hline
\end{tabular}
\label{Table:Annotation}

\end{table}

%%%%%%%%%

{\centering \section{ASYMPTOTICS} \label{sec:theory}}
 In this section, we show that under appropriate
conditions, the \texttt{space} procedure achieves both model
selection consistency and estimation consistency. Use $\overline{\theta}$ and $\overline{\sigma}$ to denote
the true parameters of $\theta$ and $\sigma$. As discussed in
Section \ref{subsec:model}, when $\sigma$ is given, $\theta$ is
estimated by solving the following $\ell_1$ penalization problem:
\begin{eqnarray}
\label{eqn:l1} \widehat{\theta}^{\lambda_n}(\sigma)={\rm
arg}\min_{\theta} L_n(\theta,\sigma,
\mathbf{Y})+\lambda_n||\theta||_1,
\end{eqnarray}
where the \textit{loss function} $
L_n(\theta,\sigma,\mathbf{Y}):=\frac{1}{n}\sum_{k=1}^n
L(\theta,\sigma,\mathbf{Y}^k),$ with, for $k=1,\cdots,n$
\begin{eqnarray}
\label{eqn:loss}
 L(\theta,\sigma,\mathbf{Y}^k):=\frac{1}{2}\sum_{i=1}^p w_i(y_i^k-\sum_{j \not=
 i}\sqrt{\sigma^{jj}/\sigma{ii}}\rho^{ij}y_j^k)^2.
 %=\frac{1}{2}\sum_{i=1}^p \tilde{w}_i(\tilde{y}_i^k-\sum_{j \not=
 %i}\rho^{ij}\tilde{y}_j^k)^2,
\end{eqnarray}
%where $ \tilde{y}_i^k=\sqrt{\sigma^{ii}}y_i^k$,
%$\tilde{w}_i=w_i/\sigma^{ii}$.

%\subsection{Consistency results}
Throughout this section, we assume $\mathbf{Y}^1,\cdots,\mathbf{Y}^n$ are i.i.d.
samples from $N_p(0,\mathbf{\Sigma})$. The Gaussianity assumption here can
be relaxed by assuming appropriate tail behaviors of the
observations. The assumption of zero mean is simply for exposition
simplicity. In practice, in the loss function (\ref{eqn:l1}),
$\mathbf{Y}^k$ can be replaced by $\mathbf{Y}^k-\overline{\mathbf{Y}}$ where
$\overline{\mathbf{Y}}=\frac{1}{n}\sum_{k=1}^n \mathbf{Y}^k$ is the sample mean. All
results stated in this section still hold under that case.

 We first state regularity conditions that are
needed for the proof. Define $\mathcal{A}=\{(i,j):
\overline{\rho}^{ij}\not=0\}$.

\begin{itemize}
\item [\bf C0:] The weights satisfy $0<w_0\leq \min_i\{w_i\} \leq
\max_i\{w_i\} \leq w_{\infty}<\infty$

 \item [\bf C1:] There exist constants
$0<\Lambda_{\min} (\overline{\theta}) \leq \Lambda_{\max}(\overline{\theta})
<\infty $, such that the true covariance
$\overline{\mathbf{\Sigma}}=\overline{\mathbf{\Sigma}}(\overline{\theta},\overline{\sigma})$
satisfies: $ 0<\Lambda_{\min}(\overline{\theta}) \leq
\lambda_{\min}(\overline{\mathbf{\Sigma}}) \leq
\lambda_{\max}(\overline{\mathbf{\Sigma}}) \leq
\Lambda_{\max}(\overline{\theta}) < \infty,$ where $\lambda_{\min}$ and $\lambda_{\max}$ denote the smallest and largest eigenvalues of a matrix, respectively.

\item [\bf C2:] There exist a constant $\delta<1$ such that for
all $(i,j) \notin \mathcal{A}$
$$
\left|\overline{L}^{\prime\prime}_{ij,\mathcal{A}}(\overline{\theta},\overline{\sigma})\left[\overline{L}^{\prime\prime}_{\mathcal{A},\mathcal{A}}(\overline{\theta},\overline{\sigma})\right]^{-1}{\rm
sign}(\overline{\theta}_{\mathcal{A}})\right| \leq \delta (<1),
$$
where  for $1 \leq i<j\leq p, 1 \leq t<s \leq p$,
$$\overline{L}^{\prime\prime}_{ij,ts}(\overline{\theta},\overline{\sigma}):=E_{(\overline{\theta},\overline{\sigma})}\left(\frac{\partial ^2 L(\theta,\sigma,Y)}{\partial \rho^{ij} \partial \rho^{ts}}\Bigl|_{\theta=\overline{\theta},
\sigma=\overline{\sigma}}\right).$$

\end{itemize}
Condition C0 says that the weights are bounded away from zero and
infinity. Condition C1 assumes that the eigenvalues of the true
covariance matrix $\overline{\mathbf{\Sigma}}$ are bounded away from zero
and infinity. Condition C2 corresponds to the \textit{incoherence
condition} in \shortciteN{meinshausen}, which plays a crucial role
in proving model selection consistency of $\ell_1$ penalization
problems.

Furthermore, since $\overline{\sigma}$ is usually unknown, it needs to
be estimated. Use
$\widehat{\sigma}=\widehat{\sigma}_n=\{\widehat{\sigma}^{ii}\}_{i=1}^p$ to
denote one estimator. The following condition says
\begin{itemize}
\item[{\bf D}]: For any $\eta>0$, there exists a constant $C>0$,
such that for sufficiently large $n$, $\max_{1 \leq i \leq p}
|\widehat{\sigma}^{ii}-\overline{\sigma}^{ii}|\leq C (\sqrt{\frac{\log
n}{n}})$ holds with probability at least $1-O(n^{-\eta})$.
\end{itemize}
Note that, the theorems below hold even when $\widehat{\sigma}$ is
obtained based on the same data set  from which $\theta$ is
estimated as long as condition D is satisfied. The following
proposition says that, when $p<n$, we can get an estimator of $\sigma$ satisfying condition D by simply using the residuals of the
ordinary least square fitting.

{\proposition \label{prop2} Suppose $\mathbf {Y}=[\mathbf{Y}^1:\cdots:\mathbf{Y}^n]$
is a $p \times n$ data matrix with i.i.d. columns $\mathbf{Y}^i \sim
N_p(0,\mathbf{\Sigma})$. Further suppose that $p=p_n$ such that $p/n \leq
1-\delta$ for some $\delta>0$; and $\mathbf{\Sigma}$ has a bounded
condition number (that is assuming condition C1). Let
$\overline{\sigma}^{ii}$ denote the $(i,i)$-th element of
$\mathbf{\Sigma}^{-1}$; and let $\mathbf{e}_i$ denote the residual from regressing
$\mathbf{Y}^i$ on to
$\mathbf{Y}_{(-i)}:=[\mathbf{Y}^1:\cdots:\mathbf{Y}^{i-1}:\mathbf{Y}^{i+1}:\cdots: \mathbf{Y}^n]$,
that is
$$
\mathbf{e}_i=\mathbf{Y}^i-\mathbf{Y}_{(-i)}^T (\mathbf{Y}_{(-i)}
\mathbf{Y}_{(-i)}^T)^{-1} \mathbf{Y}_{(-i)} \mathbf{Y}^i.
$$
Define $ \widehat{\sigma}^{ii}=1/\widehat{\sigma}_{ii,-(i)}$,
where
$$
\widehat{\sigma}_{ii,-(i)}=\frac{1}{n-p-1}\mathbf{e}_i^T \mathbf{e}_i,
$$
then condition D holds for $\{\widehat{\sigma}^{ii}\}_{i=1}^p$.

 }
The proof of this proposition is omitted due to space limitation.

 We now state notations used in the main results. Let
$q_n=|\mathcal{A}|$ denote the number of nonzero partial
correlations (of the underlying true model) and let $\{s_n\}$ be a positive sequence of real
numbers such that for any $(i,j) \in \mathcal{A}$: $
|\overline{\rho}^{ij}| \geq s_n. $ Note that, $s_n$ can be viewed
as the signal size. We follow the similar strategy as in
\shortciteN{meinshausen} and \shortciteN{Massam2007} in deriving
the asymptotic result: (i) First prove estimation consistency and
sign consistency for the restricted penalization problem with
$\theta_{\mathcal{A}^c}=0$ (Theorem \ref{thm:one}). We employ the
method of the proof of Theorem 1 in \shortciteN{Fan2004}; (ii)
Then we prove that with probability tending to one, no wrong edge is
selected (Theorem \ref{thm:two}); (iii) The final consistency
result then follows (Theorem \ref{thm:three}).

{\thm \label{thm:one} (consistency of the restricted problem)
Suppose that  conditions C0-C1 and D are satisfied. Suppose
further that $q_n \sim o(\sqrt{\frac{n}{\log n}}), \ \lambda_n
\sqrt{\frac{n}{\log n}} \rightarrow \infty$ and
$\sqrt{q_n}\lambda_n \sim o(1)$, as $n \rightarrow \infty$. Then
there exists a constant $C(\overline{\theta})>0$, such that for
any $\eta>0$, the following events hold with probability at least
$1-O(n^{-\eta})$:
\begin{itemize}
\item there exists a solution
$\widehat{\theta}^{\mathcal{A},\lambda_n}=\widehat{\theta}^{\mathcal{A},\lambda_n}(\widehat{\sigma})$
of the restricted problem: \begin{eqnarray} \label{eqn:restrict}
 \min_{\theta: \theta_{\mathcal{A}^c}=0}
L_n(\theta,\widehat{\sigma},\mathbf{Y})+\lambda_n||\theta||_1,
\end{eqnarray}
where the loss function $L_n$ is defined via (\ref{eqn:loss}).

 \item (estimation consistency) any solution
$\widehat{\theta}^{\mathcal{A},\lambda_n}$ of the restricted
problem (\ref{eqn:restrict}) satisfies:
$$
||\widehat{\theta}^{\mathcal{A},\lambda_n}-\overline{\theta}_{\mathcal{A}}||_2
\leq C(\overline{\theta})\sqrt{q_n} \lambda_n.
$$

\item (sign consistency) if further assume that  the signal
 sequence satisfies:
 $
\frac{s_n}{\sqrt{q_n}\lambda_n} \rightarrow \infty, \ n
\rightarrow \infty$, then ${\rm
sign}(\widehat{\theta}^{\mathcal{A},\lambda_n}_{ij})={\rm
sign}(\overline{\theta}_{ij})$, for all $1 \leq i <j \leq p$.
%The
%same result holds if we instead assume: $ s_n\sqrt{n} \rightarrow
%\infty, $ and there exists a constant
%$\vartheta(\overline{\theta}) <\infty$, such that for all $(i,j)
%\in \mathcal{A}$,
%$$
%||(\overline{L}^{\prime\prime}_{\mathcal{A}_{ij}\mathcal{A}_{ij}}(\overline{\theta}))^{-1}\overline{L}^{\prime\prime}_{{\mathcal{A}_{ij},ij}}(\overline{\theta})||_1
%\leq \vartheta(\overline{\theta}),
%$$
%where $\mathcal{A}_{ij} :=\mathcal{A}/\{(i,j)\}$.
%\medskip

%\textit {Remark:} the quantity
%$||(\overline{L}^{\prime\prime}_{\mathcal{A}_{ij}\mathcal{A}_{ij}}(\overline{\theta}))^{-1}\overline{L}^{\prime\prime}_{{\mathcal{A}_{ij},ij}}(\overline{\theta})||_1$
%measures the ``influence" of the edge $(i,j)$ on other edges.  By
%giving an upper bound, more stringent sparsity on the network is
%imposed (note that, under C1, the $\ell_2$ norm is already
%bounded).
\end{itemize}
}

{\thm \label{thm:two} Suppose that conditions C0-C2 and D are
satisfied. Suppose further that $p=O(n^{\kappa})$ for some $\kappa
\geq 0$; $q_n \sim o(\sqrt{\frac{n}{\log n}}), \ \sqrt{
\frac{q_n\log n}{n}}=o(\lambda_n), \ \lambda_n \sqrt{\frac{n}{\log
n}} \rightarrow \infty$ and $\sqrt{q_n}\lambda_n \sim o(1)$, as $n
\rightarrow \infty$. Then for any $\eta>0$, for $n$ sufficiently
large, the solution of (\ref{eqn:restrict}) satisfies

$$
P_{(\overline{\theta},\overline{\sigma})}\left(\max_{(i,j)\in
\mathcal{A}^c}|L^{\prime}_{n,ij}(\widehat{\theta}^{\mathcal{A},\lambda_n},\widehat{\sigma},\mathbf{Y})|<\lambda_n\right)
\geq 1-O(n^{-\eta}),
$$
where $
L^{\prime}_{n,ij}:=\frac{\partial L_n}{\partial \rho^{ij}}.
$
}

{\thm \label{thm:three} Assume the same conditions of Theorem
\ref{thm:two}.  Then there exists a constant
$C(\overline{\theta})>0$, such that for any $\eta>0$ the following
events hold with probability at least $1-O(n^{-\eta})$:
 \begin{itemize}
 \item there exists a solution
$\widehat{\theta}^{\lambda_n}=\widehat{\theta}^{\lambda_n}(\widehat{\sigma})$
of the $\ell_1$ penalization problem
\begin{eqnarray}
\label{eqn:l1_whole} \min_{\theta} L_n(\theta,\widehat{\sigma},
\mathbf{Y})+\lambda_n||\theta||_1,
\end{eqnarray}
where the loss function $L_n$ is defined via (\ref{eqn:loss}).

\item \textit{(estimation consistency)}: any solution
$\widehat{\theta}^{\lambda_n}$ of (\ref{eqn:l1_whole}) satisfies:

$$ ||\widehat{\theta}^{\lambda_n}-\overline{\theta}||_2\leq
C(\overline{\theta})(\sqrt{q_n}\lambda_n). $$

 \item \textit{(Model selection consistency/sign consistency)}: $$
{\rm sign}(\widehat{\theta}^{\lambda_n}_{ij})={\rm
sign}(\overline{\theta}_{ij}), \ \hbox{for all $1\leq i<j \leq p$}.$$

\end{itemize}

}

Proofs of these theorems are given in the Supplemental Material.
Finally, due to exponential small tails of the probabilistic
bounds, model selection consistency can be easily extended when
the network consists of $N$ disjointed components with
$N=O(n^{\alpha})$ for some $\alpha \geq 0$, as long as the size
and the number of true edges of each component satisfy the
corresponding conditions in Theorem \ref{thm:two}.

{\remark The condition $\lambda_n \sqrt{\frac{n}{\log n}}
\rightarrow \infty$ is indeed implied by the condition $\sqrt{
\frac{q_n\log n}{n}}=o(\lambda_n)$ as long as $q_n$ does not go to zero. Moreover,  under the ``worst case" scenario, that is when
$q_n$ is almost in the order of  $\sqrt{\frac{n}{\log n}}$,
$\lambda_n$ needs to be nearly in the order of $n^{-1/4}$. On the
other hand, for the``best case" scenario, that is when $q_n=O(1)$
(for example, when the dimension $p$ is fixed), the order of
$\lambda_n$ can be nearly as small as  $n^{-1/2}$ (within a factor
of $\log n$). Consequently, the $\ell_2$-norm distance of the estimator from the true parameter is in the order of $\sqrt{\log n/n}$, with probability tending to one.}

%\textcolor{red}{when sigma is estimated; disjoint components}

%\subsection{Consistency results when $\overline{\sigma}$ is independently estimated}

%\subsection{Some extensions for structured networks}

%\textcolor{red}{when sigma is estimated; disjoint components}

%\subsection{Consistency results when $\overline{\sigma}$ is independently estimated}

%\subsection{Some extensions for structured networks}

{\centering \section{SUMMARY}}\label{Sec:Discussion}
In this paper, we propose a joint sparse regression model --
\texttt{space} -- for selecting non-zero partial correlations
under the high-dimension-low-sample-size setting. By controlling
the overall sparsity of the partial correlation matrix,
\texttt{space} is able to automatically adjust for different
neighborhood sizes and thus to utilize data more effectively. The
proposed method also explicitly employs the symmetry among the
partial correlations, which also helps to improve efficiency.
Moreover, this joint model makes it easy to incorporate prior
knowledge about network structure. We develop a fast algorithm
\texttt{active-shooting} to implement the proposed procedure,
which can be readily extended to solve some other penalized
optimization problems.  We
also propose a ``BIC-type" criterion for the selection of the
tuning parameter. With extensive simulation studies, we
demonstrate that this method achieves good power in non-zero
partial correlation selection as well as hub identification, and
also performs favorably compared to two existing methods. The
impact of the sample size and dimensionality has been examined on
simulation examples as well. We then apply this method on a
microarray data set of $1217$ genes from $244$ breast cancer tumor
samples, and find $11$ candidate hubs, of which five are known
breast cancer related regulators. In the end, we show consistency
(in terms of model selection and estimation) of the proposed
procedure under suitable regularity and sparsity conditions.

The R package
\textit{space} -- Sparse PArtial Correlation Estimation -- is
available on \texttt{cran}.

{\centering \section*{ACKNOWLEDGEMENT}}
A paper based on this report has been accepted for publication on Journal of the American Statistical Association (\texttt{http://www.amstat.org/publications/JASA/}).
We are grateful to two anonymous reviewers and an associate editor for their valuable comments.

%whose comments led to many improvements of
%the paper.

Peng is partially supported by grant DMS-0806128 from the National
Science Foundation and grant 1R01GM082802-01A1 from the National
Institute of General Medical Sciences. Wang is partially supported
by grant 1R01GM082802-01A1 from the National Institute of General
Medical Sciences. Zhou and Zhu are partially supported by grants
DMS-0505432 and DMS-0705532 from the National Science Foundation.
\bibliographystyle{chicago}

%\input{jsrm-2-28.bbl}

%\bibliography{pei_network,pei_others,pei_revision,tibs,pei_others_1_22_b,pei_CGH}

%\newpage
%\input{Appendix_Paper.tex}
%\input{Table_TechniqualReport.tex}

%%%%%%%%%
%\input{table-9-16.tex}

%\input{Graph_TechniqualReport.tex}
%%%%%%%%%
%\input{graph.color.tex}
\newpage

%% simulate data, hubnetwork

\begin{figure}[h]
\begin{center}
 \subfigure[Hub network: 500 nodes and 568 edges.
 15 nodes (in black) have degrees of around 15.]{
\label{Fig:HubNet}
\includegraphics[width=5in,
 height=3.3in]{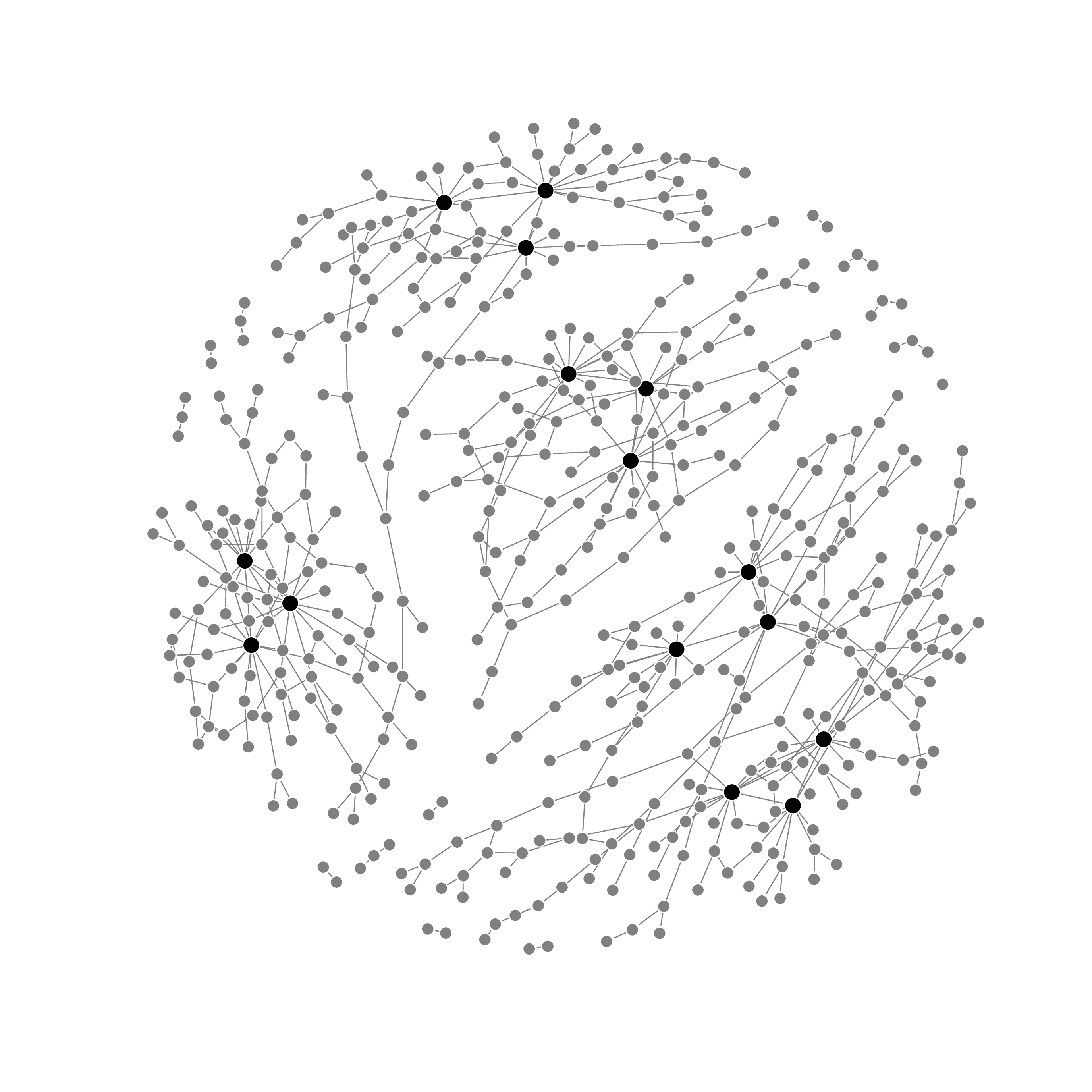}}
 \subfigure[Power-law network: 500 nodes and 495
 edges. 3 nodes (in black) have degrees at least 20.]{
\label{Fig:PowerLawNet}
 \includegraphics[width=5in,
 height=3.3in]{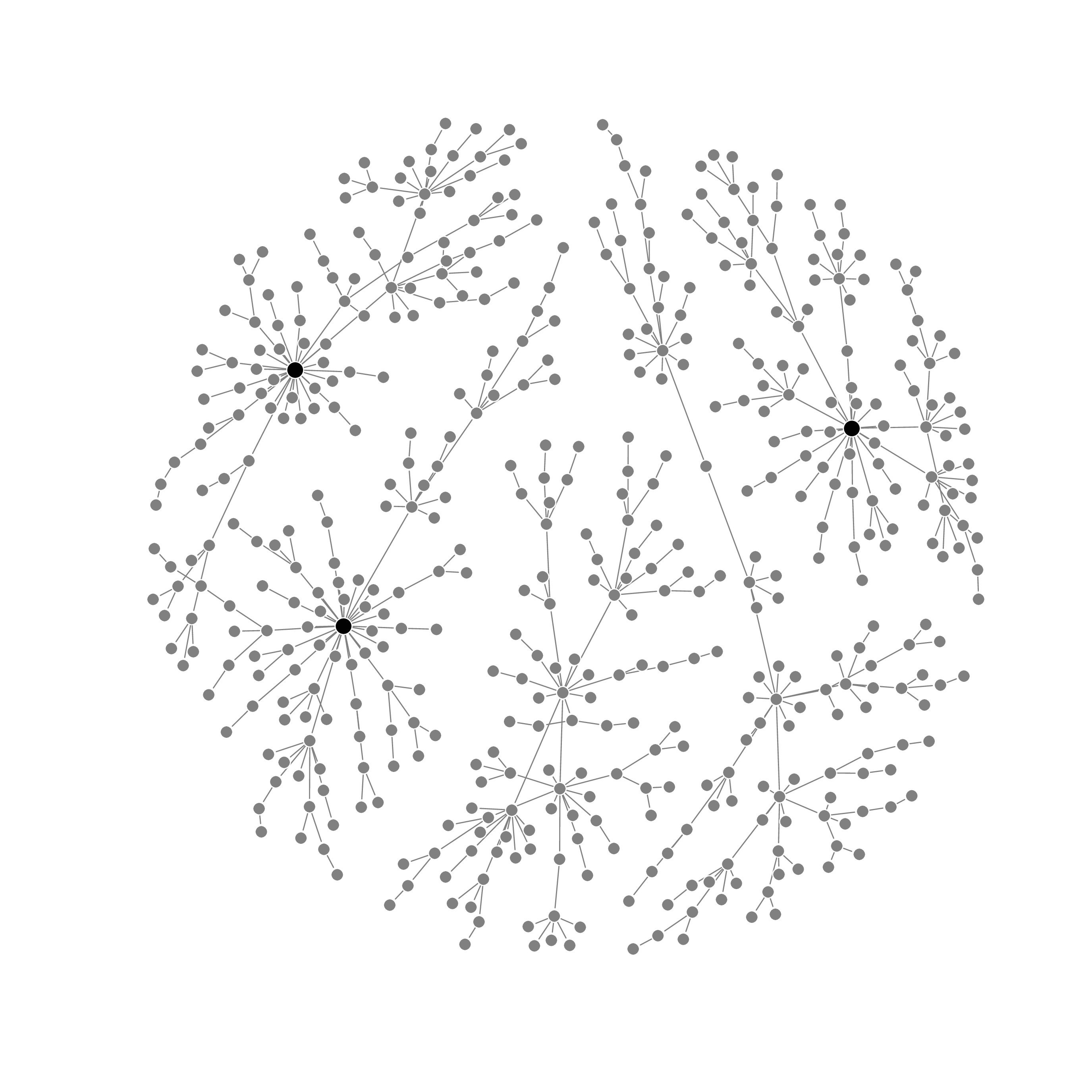}}
 \caption{Topology of simulated networks.}\label{Fig:Net}
\end{center}
\end{figure}

\begin{figure}[h]
\begin{center}
 \subfigure[\textit{x-axis}: the
 number of total detected edges(i.e., the total number of pairs
 $(i,j)$ with $\widehat{\rho}^{ij} \not=0$); \textit{y-axis}: the
 number of correctly identified edges. The vertical grey line
 corresponds to the number of true edges.]{
\label{Fig:HubDetectEdge}
 \includegraphics[width=10cm,
 angle=0]{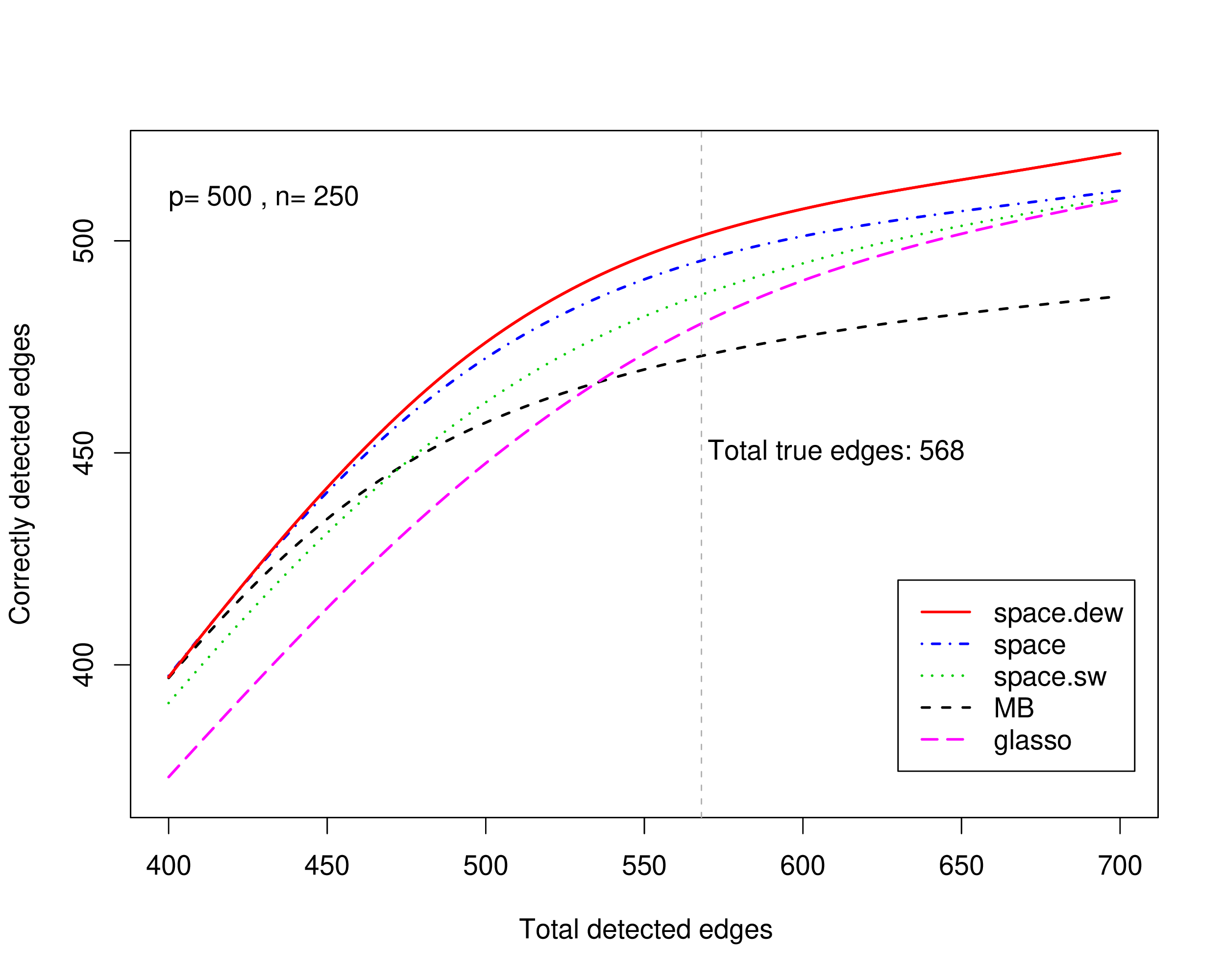}}
 \hspace{.0in}
 \subfigure[\textit{x-axis}: the number of total detected edges;
 \textit{y-axis}: the average rank of the estimated degrees of the 15
 true hub nodes.]{
 \label{Fig:HubRank}
 \includegraphics[width=10cm,
 angle=0]{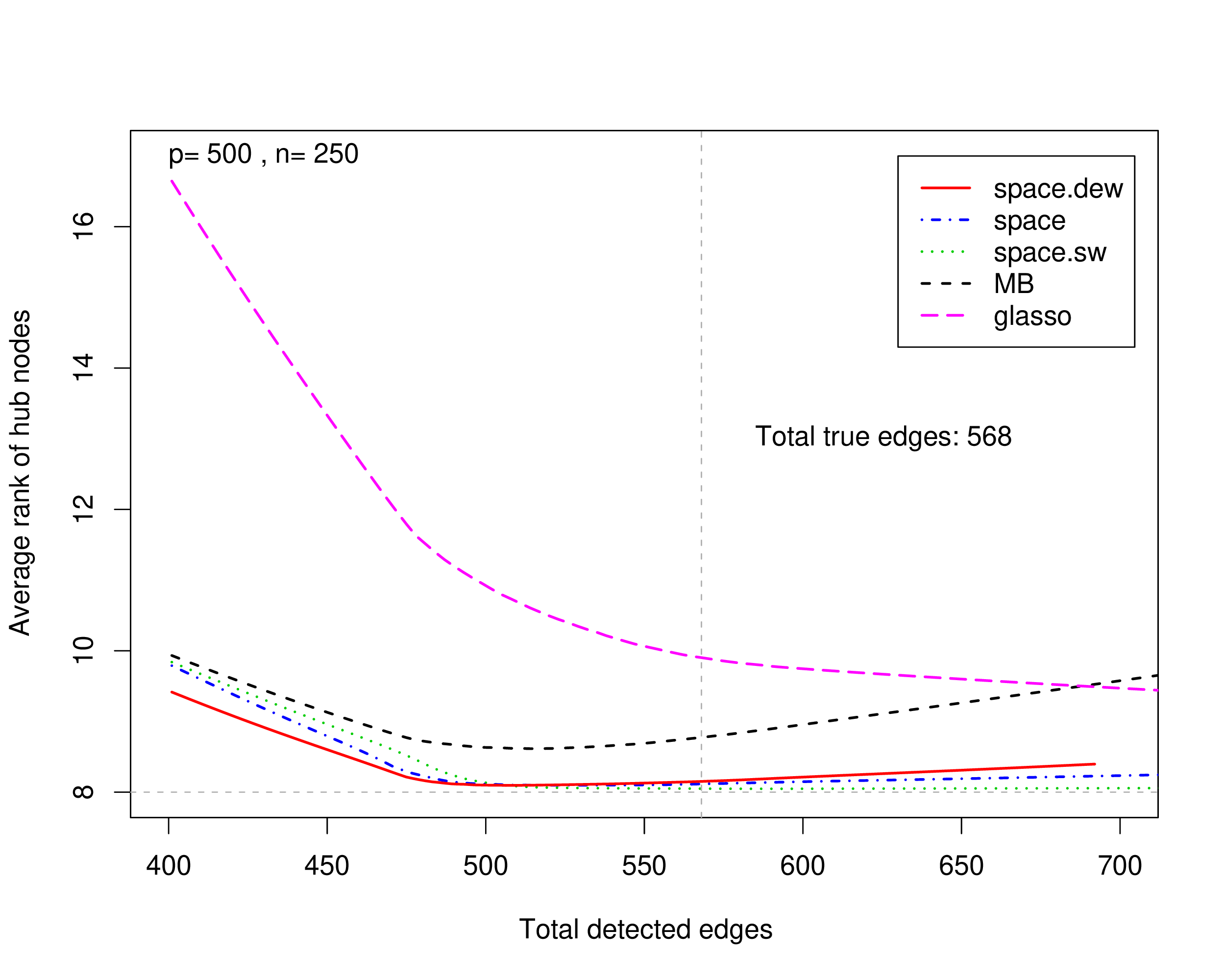}} \hspace{.0in}
\caption{Simulation results for Hub network. }
\label{Fig:HubResult}
\end{center}
\vspace{-23pt}
\end{figure}

\begin{figure}[h]
\begin{center}
 \includegraphics[width=11cm,
 angle=0]{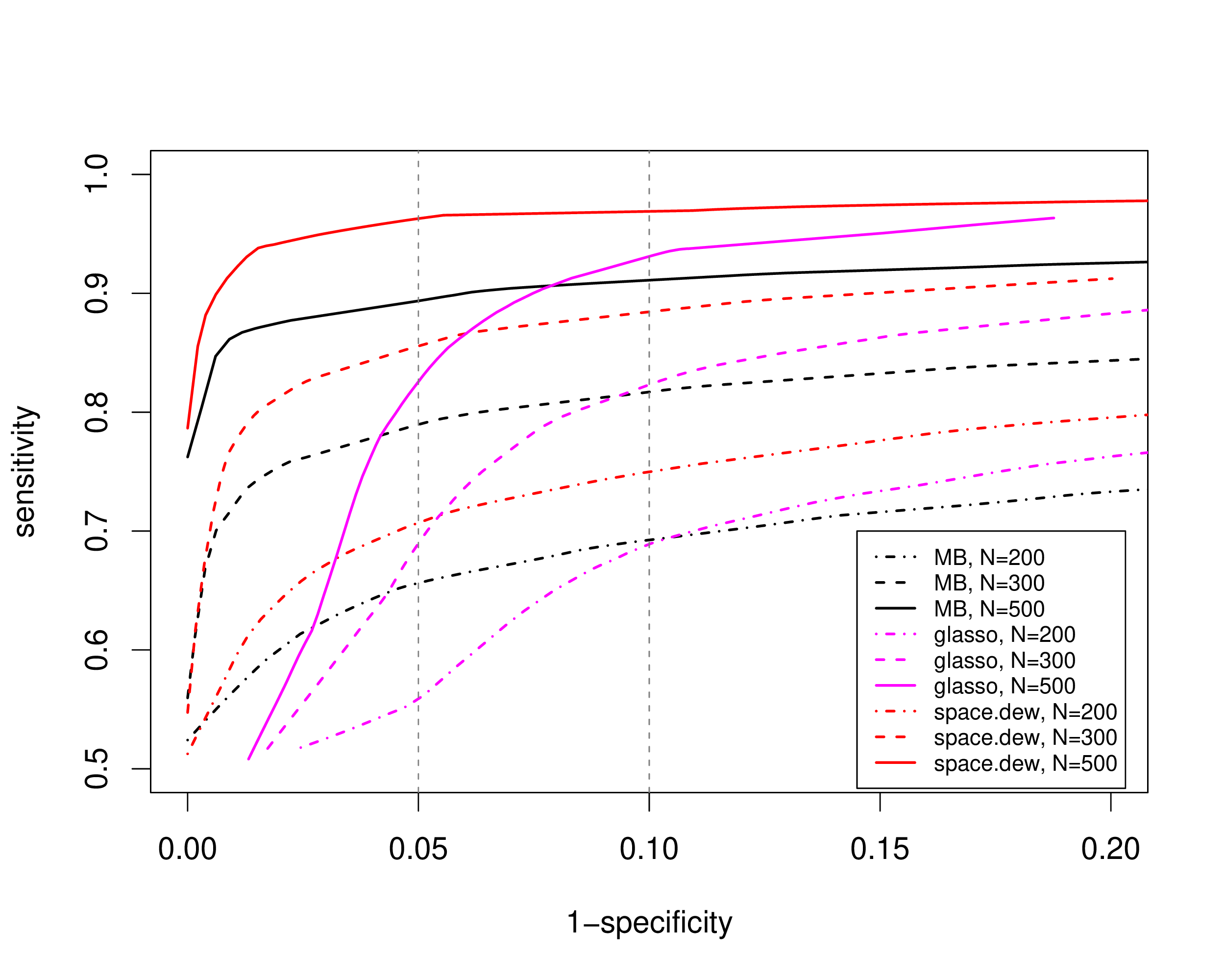}
\caption{Hub network: ROC curves for different samples sizes
($p=1000$).}\label{Fig:HubNet1000}
\end{center}
\vspace{-23pt}
\end{figure}

\begin{figure}[h]
\begin{center}
\includegraphics[width=11cm,
 angle=0]{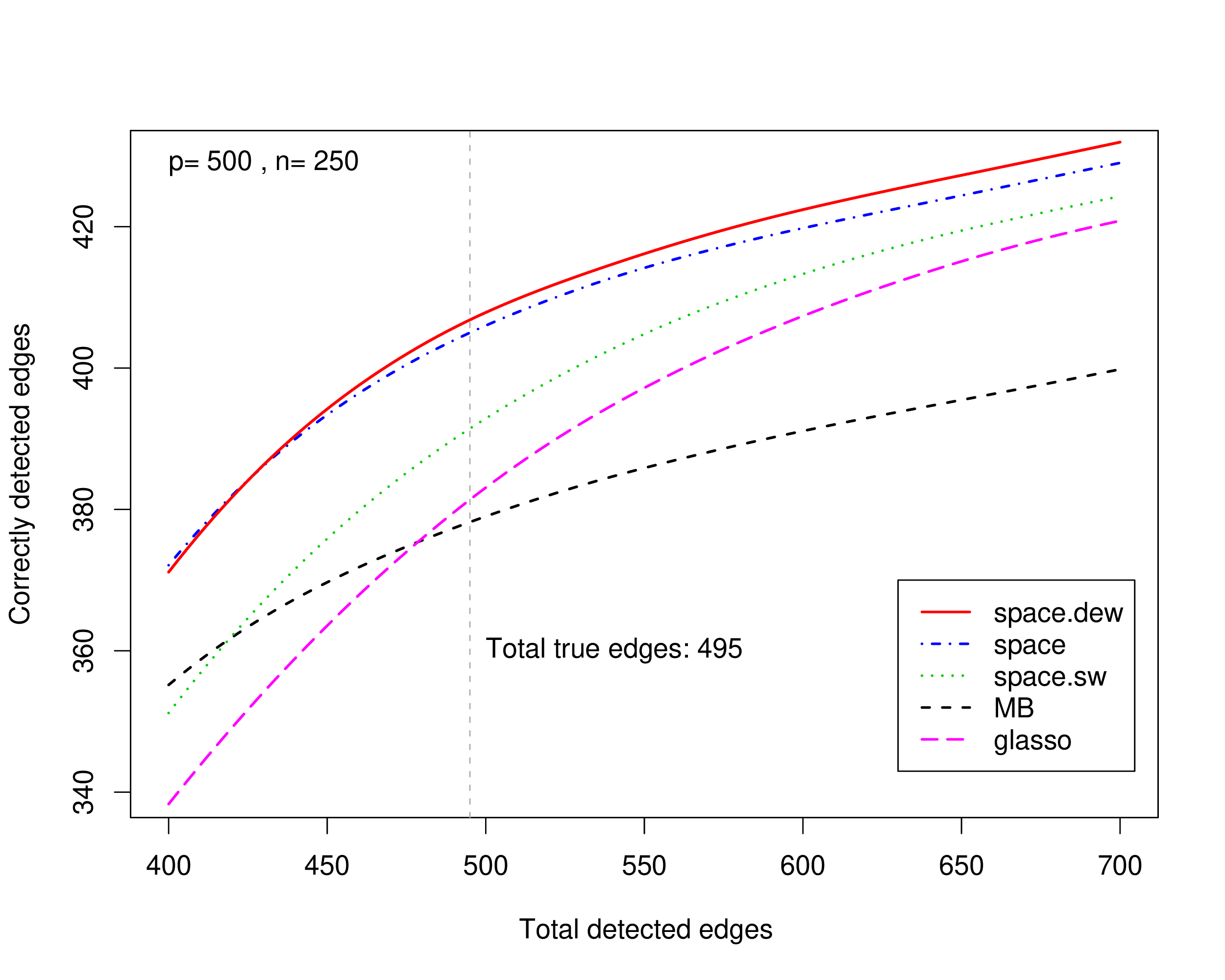}
\caption{Simulation results for Power-law network.
\textit{x-axis}:
 the number of total detected edges; \textit{y-axis}: the
 number of correctly identified edges. The vertical grey line
 corresponds to the number of true edges.}\label{Fig:PowerLawCurve}
\end{center}
\end{figure}

%% simulate data : powerlaw

%\begin{figure}[h]
%\begin{center}
 %\subfigure[Power-law network. Power law parameter $\alpha=2.3$.]{
 %\label{Fig:PowerLawDegree}
 %\includegraphics[width=3.5in,angle=0]{PowerLaw.degree.dist.ps}}
 %\subfigure[Empirical network. Power law parameter $\alpha=2.56$.]{
 %\label{Fig:RealDataDegree}
 %\includegraphics[width=3.5in]{DegreeDistribution.ps}}
%\includegraphics[width=2.5in,angle=0]{PowerLaw.ParCor.ps}
%\caption{The degree distributions of the Power-law network used in
%the simulation study and the empirical network based on the breast
%cancer expression data set.}\label{Fig:DegreeDist}
%\end{center}
%\end{figure}

%%Uniform

\begin{figure}[h]
\begin{center}
 \subfigure[Uniform network: 500 nodes and 447 edges.]{
 \label{Fig:Uniform}
 \includegraphics[width=5in,
 height=3.3in]{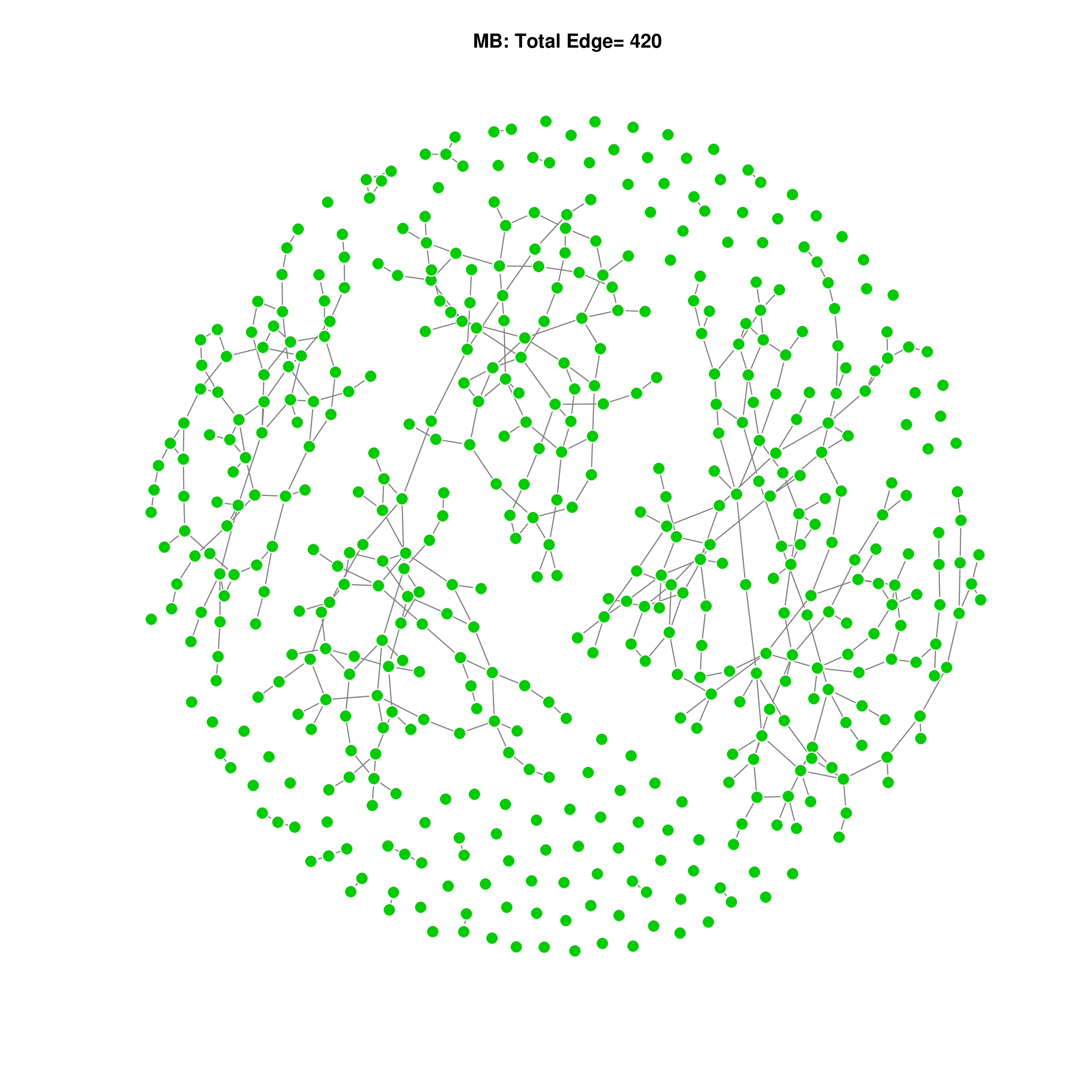}}
 \subfigure[Simulation results for Uniform network.
\textit{x-axis}:
 the total number of edges detected; \textit{y-axis}: the total
 number of correctly identified edges. The vertical grey line
 corresponds to the number of true edges.]{
 \label{Fig:Uniform-curve}
 \includegraphics[width=5in,
 height=3.3in]{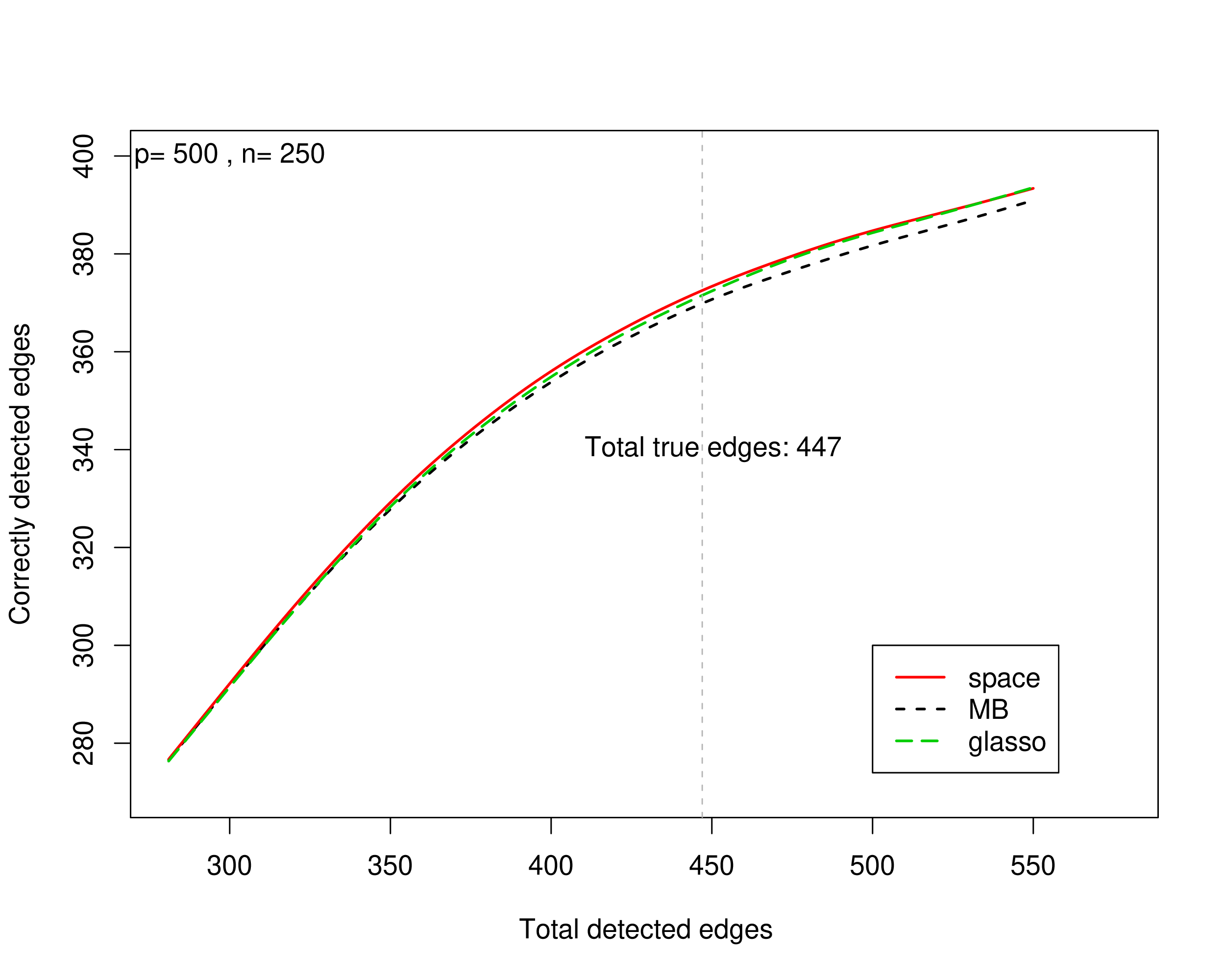}}
 \caption{Simulation results for Uniform networks.}\label{Fig:Uniform-result}
\end{center}
\end{figure}

%%AR

\begin{figure}[h]
\begin{center}
 \subfigure[AR network: 500 nodes and 499 edges.]{
 \label{Fig:AR}
 \includegraphics[width=5in,
 height=3.3in]{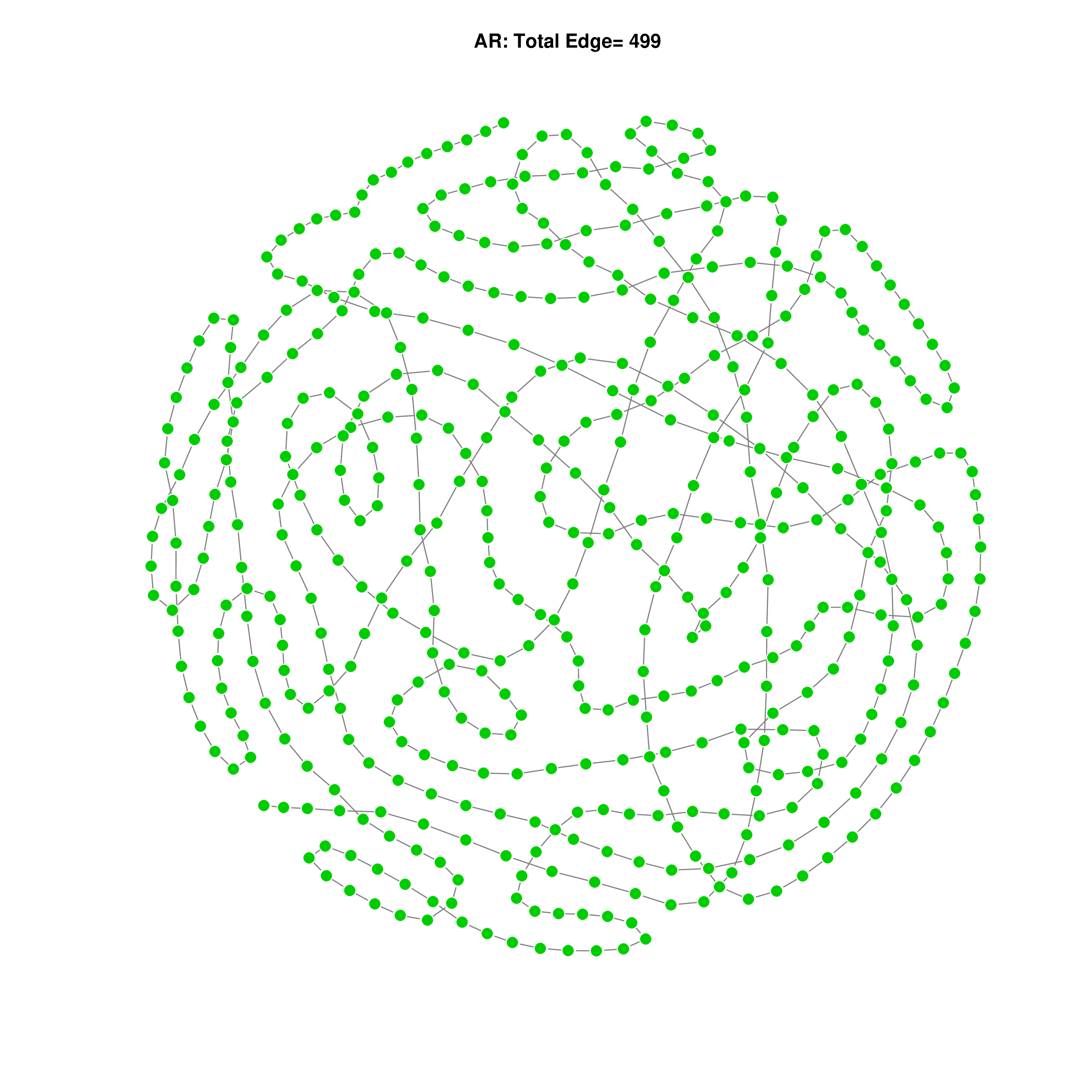}}

 \subfigure[Simulation results for AR network.
\textit{x-axis}:
 the total number of edges detected; \textit{y-axis}: the total
 number of correctly identified edges. The vertical grey line
 corresponds to the number of true edges.]{\label{Fig:AR-curve}
 \includegraphics[width=5in,
 height=3.3in]{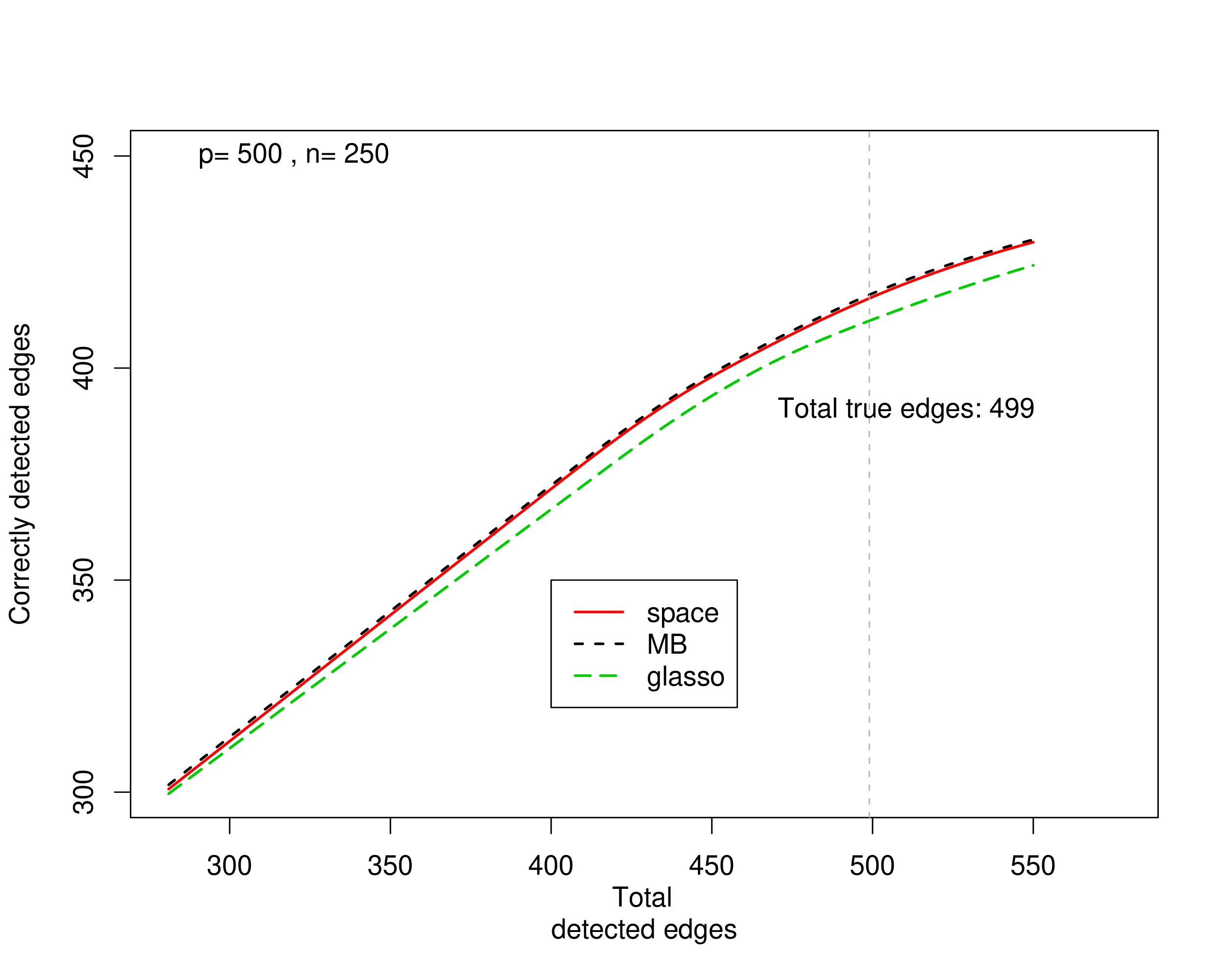}}
 \caption{Simulation results for AR networks.}\label{Fig:AR-result}
\end{center}
\end{figure}

%%circle

\begin{figure}[h]
\begin{center}
 \subfigure[Big-circle network: 500 nodes and 500 edges.]{ \label{Fig:Cicle}
 \includegraphics[width=5in,
 height=3.3in]{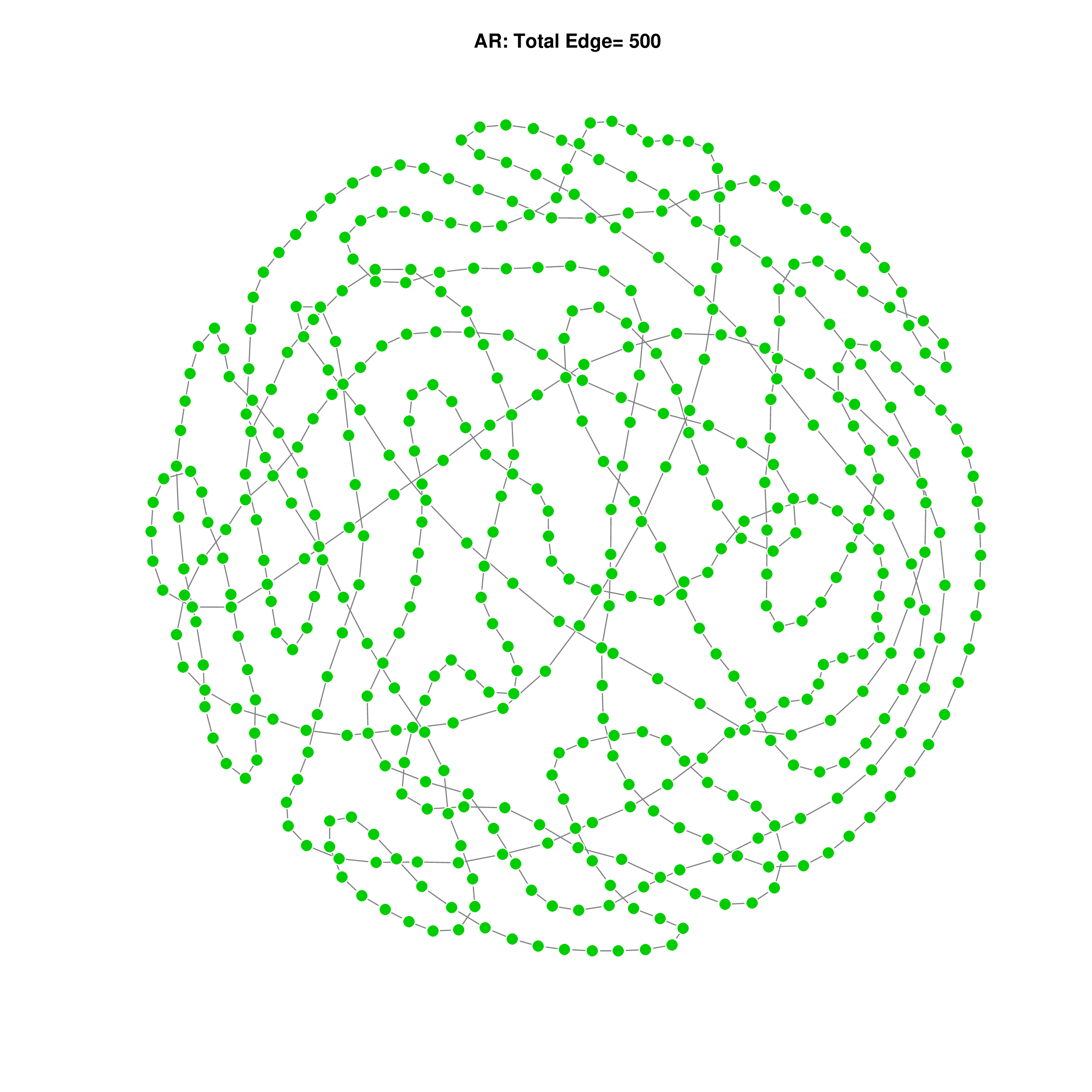}}

 \subfigure[Simulation results for Circle network.
\textit{x-axis}:
 the total number of edges detected; \textit{y-axis}: the total
 number of correctly identified edges. The vertical grey line
 corresponds to the number of true edges.]{\label{Fig:Cicle-result}
 \includegraphics[width=5in,
 height=3.3in]{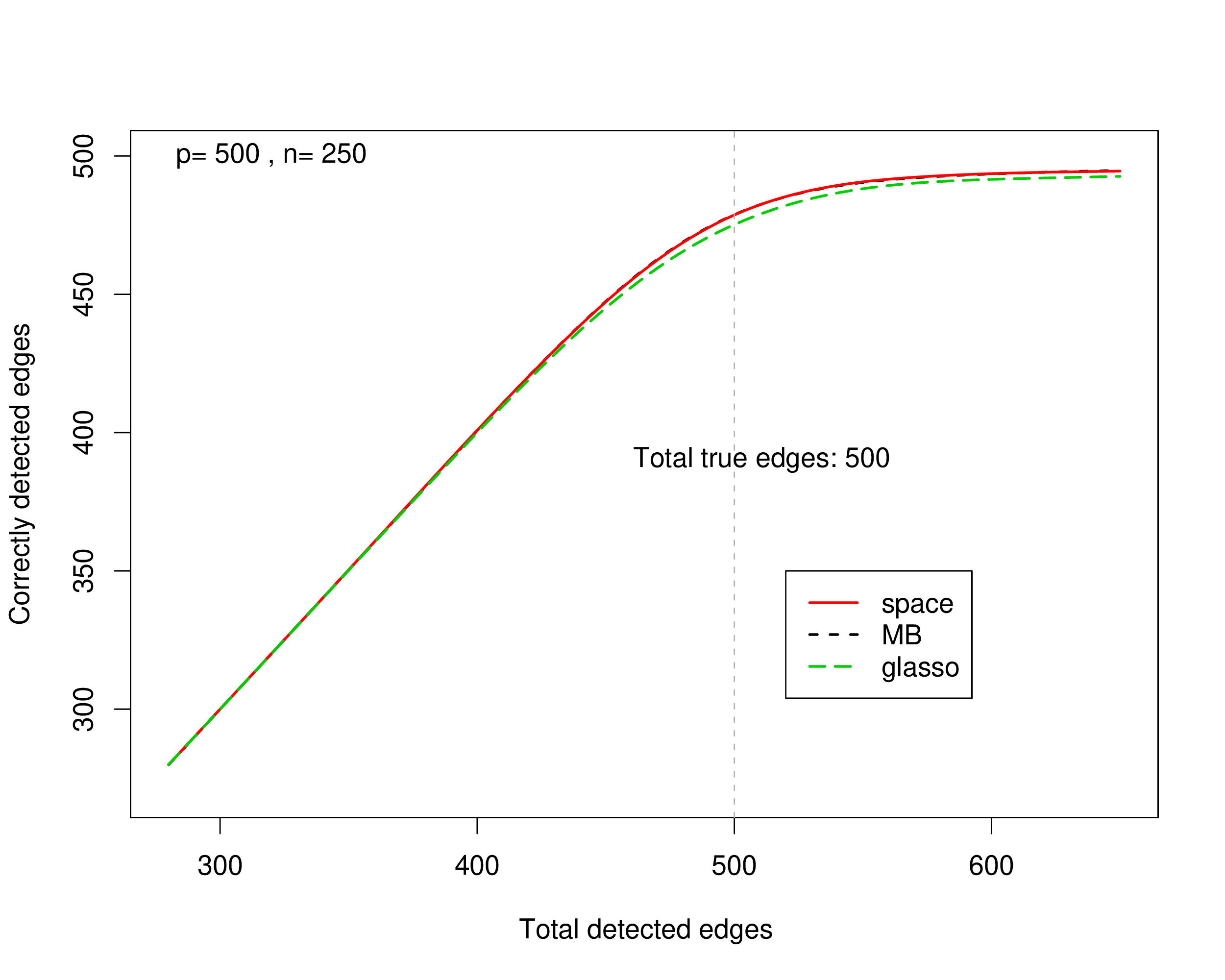}}
 \caption{Simulation results for Circle networks.}
  \label{Fig:Circle-curve}
\end{center}
\end{figure}

%% real data

\begin{figure}[h]
\begin{center}
 \subfigure[Network inferred from the real data (only showing components
 with at least three nodes). The gene annotation of the hub nodes
 (numbered) are given in Table 5.]{
\label{Fig:RealNet}
 \includegraphics[width=4in, height=3in]{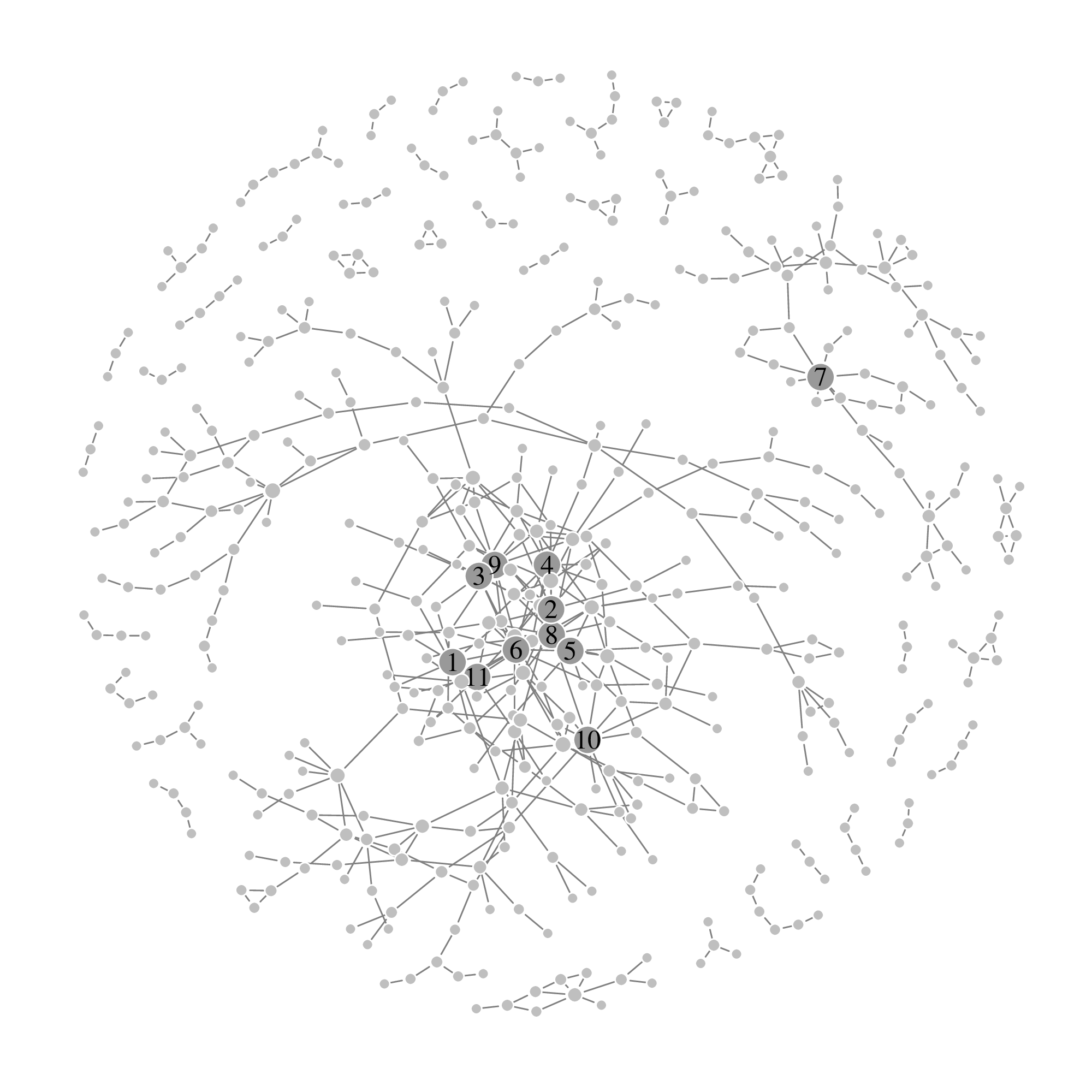}}

 \subfigure[Degree ranks (for the 100
 genes with highest degrees). Different circles represent different
 genes. \textit{Solid circles}: the 11 genes with highest degrees.
  \textit{Circles}: the other genes.
 The sd(rank) of the top 11 genes are
 all smaller than $4.62$ ($4.62$ is the $1\%$ quantile of sd(rank) among all the 1217
 genes), and thus are identified as hub nodes.]{
 \label{Fig:RealDegreeHub}
 \includegraphics[width=3.7in, height=3in]{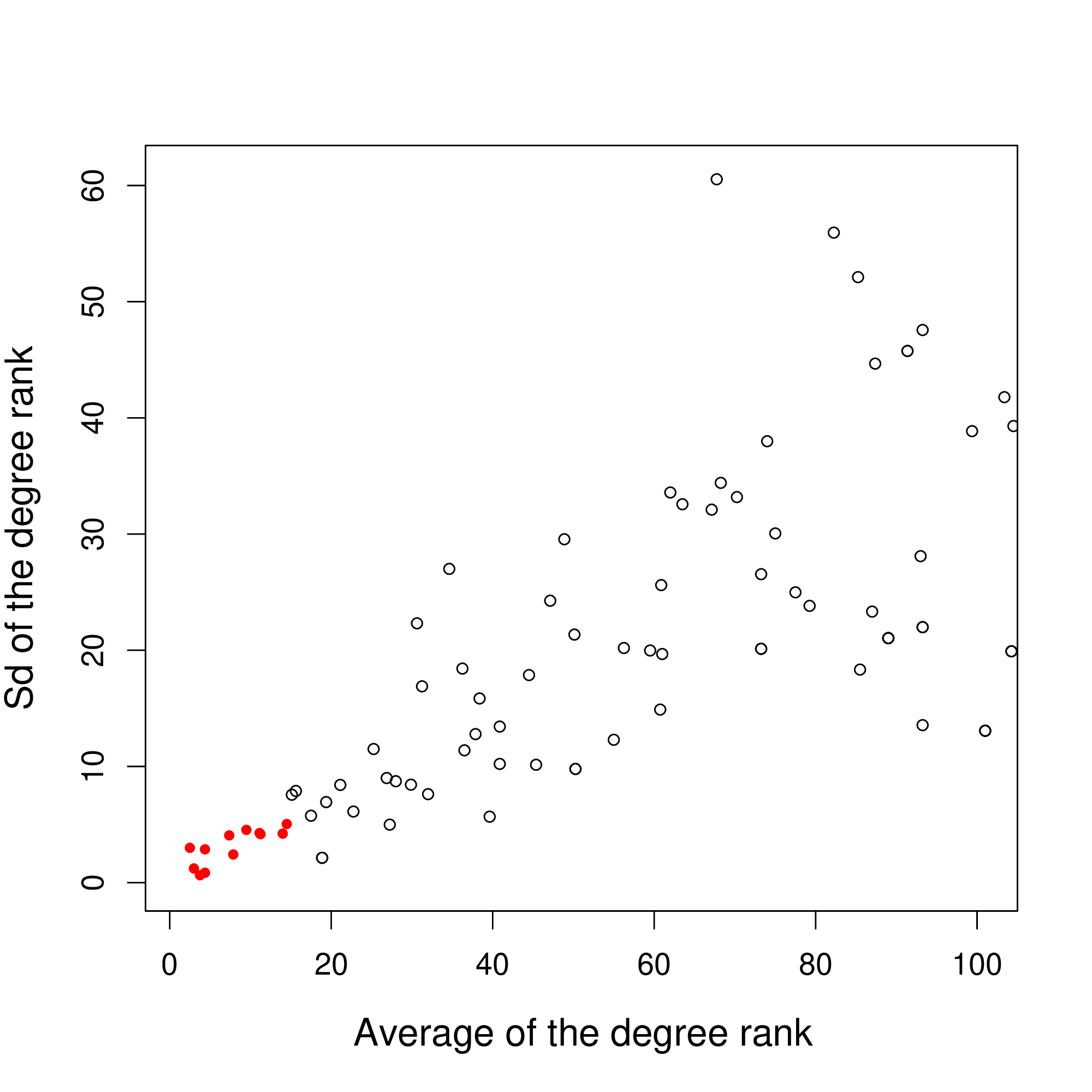}}

\caption{Results for the breast cancer expression data
set.}\label{Fig:RealNetResult}
\end{center}
\end{figure}

%
%\begin{figure}[h]
%\begin{center}
% \subfigure[Empirical network inferred from the real data (only showing components
% with at least three nodes). The gene annotation of the hub nodes
% (numbered) are given in Table 3.]{
% \label{Fig:RealNet}
% \includegraphics[width=4in, height=3.5in]{NKI.net.11.29.whole.new.ps}}
% \subfigure[Degree ranks (for the 100
% genes with highest degrees). Different symbols represent results from different
% $\lambda$. \textit{Solid line}: the average rank across different
% $\lambda$. \textit{Dashed lines}: the average rank $\pm
% 4.62$ ($4.62$ is the $1\%$ quantile of sd(rank) among all the 1217
% genes).
% The sd(rank) of the top 11 genes (left to the vertical grey line) are
% all smaller than $4.62$, and thus are identified as hub nodes.]{
% \label{Fig:RealDegreeHub}
%% \includegraphics[width=4in]{Hub.rank.11.29.small.final.ps}}
% \includegraphics[width=4in]{Hub.rank.ave.sd.ps}}
%\caption{Result for the breast cancer expression data
%set.}\label{Fig:RealNetResult}
%\end{center}
%\end{figure}
\clearpage
\setcounter{equation}{0}
\renewcommand{\theequation}{S-\arabic{equation}}
\setcounter{lemma}{0}
\renewcommand{\thelemma}{S-\arabic{lemma}}
\begin{center}
{\Large Supplemental Material}
\end{center}

\section*{Part I}

In this section, we list properties of the loss function:
\begin{eqnarray}
\label{eqn:loss_s}
 L(\theta,\sigma,Y)=\frac{1}{2}\sum_{i=1}^p w_i(y_i-\sum_{j \not=
 i}\sqrt{\sigma^{jj}/\sigma{ii}}\rho^{ij}y_j)^2=\frac{1}{2}\sum_{i=1}^p \tilde{w}_i(\tilde{y}_i-\sum_{j \not=i}\rho^{ij}\tilde{y}_j)^2,
\end{eqnarray}
where $Y=(y_1,\cdots,y_p)^T$ and $ \tilde{y}_i=\sqrt{\sigma^{ii}}y_i$,$\tilde{w}_i=w_i/\sigma^{ii}$.
These properties are used for the proof of the main results.
Note: throughout the supplementary material,  when evaluation is taken place
at $\sigma=\bar{\sigma}$, sometimes we omit the argument $\sigma$
in the notation for simplicity. Also we use $Y=(y_1,\cdots,y_p)^T$ to denote a generic sample and use $\mathbf{Y}$ to denote the $p \times n$ data matrix consisting of $n$ i.i.d. such samples: $\mathbf{Y^1},\cdots,\mathbf{Y^n}$, and define
\begin{eqnarray}
\label{eqn:loss_data}
L_n(\theta,\sigma,\mathbf{Y}):=\frac{1}{n}\sum_{k=1}^n
L(\theta,\sigma,\mathbf{Y}^k).
\end{eqnarray}

\begin{itemize}
\item [\bf A1:] for all $\theta, \sigma$ and $Y \in
\mathcal{R}^p$, $ L(\theta,\sigma, Y) \geq 0. $ \item [\bf A2:]
for any $Y \in \mathcal{R}^p$ and any $\sigma>0$,
$L(\cdot,\sigma,Y)$ is convex in $\theta$; and with probability
one, $L(\cdot,\sigma,Y)$ is strictly convex.

\item [\bf A3:] for $1 \leq i<j \leq p$
$$
\overline{L}^{\prime}_{ij}(\bar{\theta},\bar{\sigma}) :=
E_{(\bar{\theta},\bar{\sigma})}\left(\frac{\partial
L(\theta,\sigma,Y)}{\partial
\rho^{ij}}\Bigl|_{\theta=\bar{\theta},
\sigma=\bar{\sigma}}\right)=0.
$$

\item [\bf A4:] for  $1 \leq i<j \leq p$ and  $1 \leq k<l \leq p$,
$$
\overline{L}_{ij,kl}^{''}(\theta,\sigma) :=
E_{(\theta,\sigma)}\left(\frac{\partial^2
L(\theta,\sigma,Y)}{\partial \rho^{ij}
\rho^{kl}}\right)=\frac{\partial}{\partial
\rho^{kl}}\left[E_{(\theta,\sigma)}\left(\frac{\partial
L(\theta,\sigma,Y)}{\partial \rho^{ij}}\right)\right],
$$
and $\overline{L}^{''}(\bar{\theta},\bar{\sigma})$ is positive
semi-definite.
\end{itemize}

If assuming C0-C1, then we have
\begin{itemize}
\item [\bf B0]: There exist constants $0<\bar{\sigma}_0 \leq
\bar{\sigma}_{\infty}< \infty$ such that: $0<\bar{\sigma}_0 \leq
\min\{\bar{\sigma}^{ii}: 1 \leq i \leq p\}\leq
\max\{\bar{\sigma}^{ii}: 1 \leq i \leq p\} \leq
\bar{\sigma}_{\infty}$.

 \item [\bf B1]: There exist constants
$0<\Lambda^L_{\min}(\bar{\theta}) \leq
\Lambda^L_{\max}(\bar{\theta}) <\infty $, such that
$$
0<\Lambda^L_{\min}(\bar{\theta}) \leq
\lambda_{min}(\overline{L}^{\prime\prime}(\bar{\theta})) \leq
\lambda_{\max}(\overline{L}^{\prime\prime}(\bar{\theta})) \leq
\Lambda^L_{\max}(\bar{\theta}) < \infty
$$

\item [\bf B1.1]: There exists a constant
$K(\bar{\theta})<\infty$, such that for all $1\leq i <j \leq p$, $
\overline{L}_{ij,ij}^{\prime\prime}(\bar{\theta}) \leq
K(\bar{\theta}). $

\item [\bf B1.2]: There exist constants $M_1(\bar{\theta}),
M_2(\bar{\theta})  < \infty$, such that for any $1\leq i<j \leq p$
$$
{\rm Var}_{(\bar{\theta},
\bar{\sigma})}(L^{\prime}_{ij}(\bar{\theta},\bar{\sigma},Y)) \leq
M_1(\bar{\theta}),  \ {\rm Var}_{(\bar{\theta},
\bar{\sigma})}(L^{\prime\prime}_{ij,ij}(\bar{\theta},\bar{\sigma},Y))
\leq M_2(\bar{\theta}).
$$

\item [\bf B1.3]: There exists a constant
$0<g(\bar{\theta})<\infty$, such that for all $(i,j) \in
\mathcal{A}$
$$
\overline{L}^{\prime\prime}_{ij,ij}(\bar{\theta},\bar{\sigma})-\overline{L}^{\prime\prime}_{ij,\mathcal{A}_{ij}}(\bar{\theta},\bar{\sigma})\left[\overline{L}^{\prime\prime}_{\mathcal{A}_{ij},\mathcal{A}_{ij}}(\bar{\theta},\bar{\sigma})\right]^{-1}\overline{L}^{\prime\prime}_{\mathcal{A}_{ij},ij}(\bar{\theta},\bar{\sigma})
\geq g (\bar{\theta}),
$$
where $\mathcal{A}_{ij}=\mathcal{A}/\{(i,j)\}$.

\item [ \bf B1.4]: There exists a constant
$M(\bar{\theta})<\infty$, such that for any $(i,j) \in
\mathcal{A}^c$

$$
||\overline{L}^{\prime\prime}_{ij,\mathcal{A}}(\bar{\theta})[\overline{L}^{\prime\prime}_{\mathcal{A}\mathcal{A}}(\bar{\theta})]^{-1}||_2
\leq M(\bar{\theta}).
$$

 \item [\bf B2] There exists a constant $K_1(\bar{\theta}) < \infty$, such that for any $1 \leq i \leq j
\leq p$,
$||E_{\bar{\theta}}(\tilde{y}_i\tilde{y}_j\tilde{y}\tilde{y}^T)||
\leq K_1(\bar{\theta})$, where $
\tilde{y}=(\tilde{y}_1,\cdots,\tilde{y}_p)^{T}$.

\item [\bf B3] If we further assume that condition $D$ holds for
$\widehat{\sigma}$ and $q_n \sim o(\frac{n}{\log n})$, we have:
for any $\eta>0$, there exist constants $C_{1,\eta},C_{2,\eta}>0$,
such that for sufficiently large $n$
$$
\max _{1 \leq i<k \leq p}\left|L^{\prime}_{n,ik}(
\bar{\theta},\bar{\sigma},\mathbf{Y})- L^{\prime}_{n,ik}(
\bar{\theta},\widehat{\sigma},\mathbf{Y})\right|\leq
C_{1,\eta}(\sqrt{\frac{\log n}{n}}),
$$
$$
\max _{1 \leq i<k \leq p, 1 \leq t<s\leq
p}\left|L^{\prime\prime}_{n,ik,ts}(
\bar{\theta},\bar{\sigma},\mathbf{Y})- L^{\prime\prime}_{n,ik,ts}(
\bar{\theta},\widehat{\sigma},\mathbf{Y})\right|\leq
C_{2,\eta}(\sqrt{\frac{\log n}{n}}),
$$
hold with probability at least $1-O(n^{-\eta})$.

\end{itemize}

B0 follows from C1 immediately. B1.1--B1.4 are direct consequences
of B1. B2 follows from B1 and Gaussianity. B3 follows from
conditions C0-C1 and D.

\medskip
\noindent \underline{\textit{proof of A1}}: obvious.\\
\noindent \underline{\textit{proof of A2}}: obvious.\\
\noindent \underline{\textit{proof of A3}}: denote the residual
for the ith term by
$$
e_i(\theta,\sigma)=\tilde{y}_i-\sum_{j
\not=i}\rho^{ij}\tilde{y}_j.
$$
Then evaluated at the true parameter values
$(\bar{\theta},\bar{\sigma})$, we have
$e_i(\bar{\theta},\bar{\sigma})$ uncorrelated with
$\tilde{y}_{(-i)}$ and
$E_{(\bar{\theta},\bar{\sigma})}(e_i(\bar{\theta},\bar{\sigma}))=0$.
It is easy to show
$$
\frac{\partial L(\theta,\sigma,Y)}{\partial
\rho^{ij}}=-\tilde{w}_ie_i(\theta,\sigma)\tilde{y}_j-\tilde{w}_je_{j}(\theta,\sigma)\tilde{y}_i.
$$
This proves A3.\\
\noindent \underline{\textit{proof of A4}}: see the proof of B1.

\noindent \underline{\textit{proof of B1}}: Denote $
\tilde{y}=(\tilde{y}_1,\cdots,\tilde{y}_p)^{T}, $ and
$\tilde{x}=(\tilde{x}_{(1,2)}, \tilde{x}_{(1,3)},\cdots,
\tilde{x}_{(p-1,p)})$ with $
\tilde{x}_{(i,j)}=(0,\cdots,0,\tilde{y}_j,\cdots,\tilde{y}_i,0,\cdots,0)^T.
$ Then the loss function (\ref{eqn:loss_s}) can be written as $
L(\theta,\sigma,Y)=\frac{1}{2}||\tilde{w}(\tilde{y}-\tilde{x}\theta)||^2_2,$
with
$\tilde{w}=diag(\sqrt{\tilde{w}}_1,\cdots,\sqrt{\tilde{w}}_p)$.
Thus $
\overline{L}^{\prime\prime}(\theta,\sigma)=E_{(\theta,\sigma)}\left[\tilde{x}^{T}\tilde{w}^2\tilde{x}\right]$
(this proves A4). Let $d=p(p-1)/2$, then $\tilde{x}$ is a $p$ by
$d$ matrix. Denote its $i$th row by $x_i^T$ ($1 \leq i  \leq p$).
Then for any $a \in \mathcal{R}^d$, with $||a||_2=1$, we have
\begin{eqnarray*}
a^T\overline{L}^{\prime\prime}(\bar{\theta})a=E_{\bar{\theta}}(a^T\tilde{x}^T\tilde{w}^2\tilde{x}a)=E_{\bar{\theta}}\left(\sum_{i=1}^p
\tilde{w}_i(x_i^Ta)^2\right).
\end{eqnarray*}
Index the elements of  $a$ by $
a=(a_{(1,2)},a_{(1,3)},\cdots,a_{(p-1,p)})^T, $ and for each $1
\leq i \leq p$, define $a_i \in \mathcal{R}^p$ by $
a_i=(a_{(1,i)},\cdots,a_{(i-1,i)},0,a_{(i,i+1)},\cdots,a_{(i,p)})^T.$
Then by definition $ x_i^Ta=\tilde{y}^Ta_i.$ Also note that $
\sum_{i=1}^p ||a_i||^2_2=2||a||^2_2=2.$ This is because, for $i
\not= j$, the $jth$ entry of $a_i$ appears exactly twice in $a$.
Therefore
$$
a^T\overline{L}^{\prime\prime}(\bar{\theta})a=\sum_{i=1}^p
\tilde{w}_i
E_{\bar{\theta}}\left(a_i^T\tilde{y}\tilde{y}^Ta_i\right)=\sum_{i=1}^p
\tilde{w}_i a_i^T \tilde{\mathbf{\Sigma}} a_i \geq \sum_{i=1}^p
\tilde{w}_i\lambda_{\min}(\tilde{\mathbf{\Sigma}})||a_i||^2_2 \geq
2\tilde{w}_0 \lambda_{\min}(\tilde{\mathbf{\Sigma}}),
$$
where $\tilde{\mathbf{\Sigma}}={\rm Var}(\tilde{y})$ and
$\tilde{w}_0=w_0/\bar{\sigma}_{\infty}$. Similarly $
a^T\overline{L}^{\prime\prime}(\bar{\theta})a \leq 2
\tilde{w}_{\infty} \lambda_{\max}(\tilde{\mathbf{\Sigma}}), $ with
$\tilde{w}_{\infty}=w_{\infty}/\bar{\sigma}_0$. By C1,
$\tilde{\mathbf{\Sigma}}$ has bounded eigenvalues, thus B1 is proved.\\

\noindent \underline{\textit{proof of B1.1:}} obvious.\\

\noindent \underline{\textit{proof of B1.2:}} note that ${\rm
Var}_{(\bar{\theta},\bar{\sigma})}(e_i(\bar{\theta},\bar{\sigma}))=1/\bar{\sigma}^{ii}$
and ${\rm
Var}_{(\bar{\theta},\bar{\sigma})}(\tilde{y}_i)=\bar{\sigma}^{ii}$.
Then for any $1 \leq i<j \leq p$, by Cauchy-Schwartz
\begin{eqnarray*}
 {\rm Var}_{(\bar{\theta},
\bar{\sigma})}(L^{\prime}_{n,ij}(\bar{\theta},\bar{\sigma},Y))&=&{\rm
Var}_{(\bar{\theta},
\bar{\sigma})}(-\tilde{w}_ie_i(\bar{\theta},\bar{\sigma})\tilde{y}_j-\tilde{w}_je_j(\bar{\theta},\bar{\sigma})\tilde{y}_i)\\
&\leq& E_{(\bar{\theta},
\bar{\sigma})}(\tilde{w}_i^2e_i^2(\bar{\theta},\bar{\sigma})\tilde{y}_j^2)+E_{(\bar{\theta},
\bar{\sigma})}(\tilde{w}_j^2e_j^2(\bar{\theta},\bar{\sigma})\tilde{y}_i^2)\\
&+&2\sqrt{\tilde{w}_i^2\tilde{w}_j^2E_{(\bar{\theta},
\bar{\sigma})}(e_i^2(\bar{\theta},\bar{\sigma})\tilde{y}_j^2)E_{(\bar{\theta},
\bar{\sigma})}(e_j^2(\bar{\theta},\bar{\sigma})\tilde{y}_i^2)}\\
&=&\frac{w_i^2\bar{\sigma}^{jj}}{(\bar{\sigma}^{ii})^3}+\frac{w_j^2\bar{\sigma}^{ii}}{(\bar{\sigma}^{jj})^3}+2\frac{w_iw_j}{\bar{\sigma}^{ii}\bar{\sigma}^{jj}}.
\end{eqnarray*}
The right hand side is bounded because of  C0 and B0.\\

\noindent \underline{\textit{proof of B1.3:}} for $(i,j) \in
\mathcal{A}$, denote
$$
D:=\overline{L}^{\prime\prime}_{ij,ij}(\bar{\theta},\bar{\sigma})-\overline{L}^{\prime\prime}_{ij,\mathcal{A}_{ij}}(\bar{\theta},\bar{\sigma})\left[\overline{L}^{\prime\prime}_{\mathcal{A}_{ij},\mathcal{A}_{ij}}(\bar{\theta},\bar{\sigma})\right]^{-1}\overline{L}^{\prime\prime}_{\mathcal{A}_{ij},ij}(\bar{\theta},\bar{\sigma}).
$$
Then $D^{-1}$ is the $(ij,ij)$ entry in
$\left[\overline{L}_{\mathcal{A},\mathcal{A}}^{\prime\prime}(\bar{\theta})\right]^{-1}$.
Thus by B1, $D^{-1}$ is positive and bounded from above, so $D$ is
bounded away from zero.\\

\noindent \underline{\textit{proof of B1.4:}} note that
$||\overline{L}^{\prime\prime}_{ij,\mathcal{A}}(\bar{\theta})
[\overline{L}^{\prime\prime}_{\mathcal{A}\mathcal{A}}(\bar{\theta})]^{-1}||^2_2
\leq
||\overline{L}^{\prime\prime}_{ij,\mathcal{A}}(\bar{\theta})||^2_2
\lambda_{\max}([\overline{L}^{\prime\prime}_{\mathcal{A}\mathcal{A}}(\bar{\theta})]^{-2})$.
By B1,
$\lambda_{\max}([\overline{L}^{\prime\prime}_{\mathcal{A}\mathcal{A}}(\bar{\theta})]^{-2})$
is bounded from above, thus it suffices to show that
$||\overline{L}^{\prime\prime}_{ij,\mathcal{A}}(\bar{\theta})||^2_2$
is bounded. Since $(i,j) \in \mathcal{A}^c$, define $\mathcal{A}^+
:=(i,j) \cup \mathcal{A}$. Then
$\overline{L}^{\prime\prime}_{ij,ij}(\bar{\theta})-\overline{L}^{\prime\prime}_{ij,\mathcal{A}}(\bar{\theta})
[\overline{L}^{\prime\prime}_{\mathcal{A}\mathcal{A}}(\bar{\theta})]^{-1}
\overline{L}^{\prime\prime}_{\mathcal{A},ij}(\bar{\theta})$ is the
inverse of the $(1,1)$ entry of
$\overline{L}^{\prime\prime}_{\mathcal{A}^+,\mathcal{A}^+}(\bar{\theta})$.
Thus by B1, it is bounded away from zero. Therefore by B1.1,
$\overline{L}^{\prime\prime}_{ij,\mathcal{A}}(\bar{\theta})
[\overline{L}^{\prime\prime}_{\mathcal{A}\mathcal{A}}(\bar{\theta})]^{-1}
\overline{L}^{\prime\prime}_{\mathcal{A},ij}(\bar{\theta})$ is
bounded from above. Since
$\overline{L}^{\prime\prime}_{ij,\mathcal{A}}(\bar{\theta})
[\overline{L}^{\prime\prime}_{\mathcal{A}\mathcal{A}}(\bar{\theta})]^{-1}
\overline{L}^{\prime\prime}_{\mathcal{A},ij}(\bar{\theta}) \geq
||\overline{L}^{\prime\prime}_{ij,\mathcal{A}}(\bar{\theta})||^2_2
\lambda_{\min}([\overline{L}^{\prime\prime}_{\mathcal{A}\mathcal{A}}(\bar{\theta})]^{-1})$,
and by B1,
$\lambda_{\min}([\overline{L}^{\prime\prime}_{\mathcal{A}\mathcal{A}}(\bar{\theta})]^{-1})$
is bounded away from zero, we have
$||\overline{L}^{\prime\prime}_{ij,\mathcal{A}}(\bar{\theta})||^2_2$
 bounded from above.

\medskip\noindent \underline{\textit{proof of B2}}: the $(k,l)$-th entry of the matrix
$\tilde y_i \tilde y_j\tilde y \tilde y^T$ is $\tilde y_i \tilde
y_j\tilde y_k \tilde y_l$, for $1 \leq  k<l \leq p$. Thus, the
$(k,l)$-th entry of the matrix $\mathbb{E}[\tilde y_i \tilde
y_j\tilde y \tilde y^T]$ is $\mathbb{E}[\tilde y_i \tilde
y_j\tilde y_k \tilde y_l] = \tilde \sigma_{ij} \tilde \sigma_{kl}
+ \tilde \sigma_{ik} \tilde \sigma_{jl} + \tilde \sigma_{il}
\tilde \sigma_{jk}$. Thus, we can write
\begin{equation}\label{eq:expect_D_expr}
\mathbb{E}[\tilde y_i \tilde y_j\tilde y \tilde y^T] = \tilde
\sigma_{ij} \tilde{\mathbf{\Sigma}} + \tilde \sigma_{i\cdot} \tilde
\sigma_{j\cdot}^T + \tilde \sigma_{j\cdot} \tilde
\sigma_{i\cdot}^T,
\end{equation}
where $\tilde \sigma_{i\cdot}$ is the $p\times 1$ vector
$(\tilde\sigma_{ik})_{k=1}^p$. From (\ref{eq:expect_D_expr}), we
have
\begin{eqnarray}\label{eq:expect_D_norm}
\parallel \mathbb{E}[\tilde y_i \tilde y_j\tilde y \tilde y^T]
\parallel &\leq& |\tilde \sigma_{ij}| \parallel \tilde{\mathbf {\Sigma}}
\parallel + 2 \parallel \tilde \sigma_{i\cdot} \parallel_2 \parallel \tilde
\sigma_{j\cdot} \parallel_2,
\end{eqnarray}
where $||\cdot||$ is the operator norm. By C0-C1, the first term
on the right hand side is uniformly bounded. Now, we also have,
\begin{equation}\label{eq:inverse_term}
\tilde \sigma_{ii} - \tilde \sigma_{i\cdot}^T \tilde
{\mathbf{\Sigma}}_{(-i)}^{-1} \tilde \sigma_{i\cdot} > 0
\end{equation}
where $\tilde {\mathbf{\Sigma}}_{(-i)}$ is the submatrix of $\tilde {\mathbf{\Sigma}}$
removing $i$-th row and column. From this, it follows that
\begin{eqnarray}\label{eq:sigma_i_bound}
\parallel \tilde \sigma_{i\cdot} \parallel_2 &=& \parallel
\tilde {\mathbf{\Sigma}}_{(-i)}^{1/2} \tilde {\mathbf{\Sigma}}_{(-i)}^{-1/2} \tilde
\sigma_{i\cdot} \parallel_2 \nonumber\\
&\leq& \parallel \tilde {\mathbf{\Sigma}}_{(-i)}^{1/2} \parallel ~\parallel
\tilde {\mathbf{\Sigma}}_{(-i)}^{-1/2} \tilde \sigma_{i\cdot} \parallel_2
\nonumber\\
&\leq& \sqrt{\parallel \tilde {\mathbf{\Sigma}} \parallel} \sqrt{\tilde
\sigma_{ii}},
\end{eqnarray}
where the last inequality follows from (\ref{eq:inverse_term}),
and the fact that $\tilde {\mathbf{\Sigma}}_{(-i)}$ is a principal submatrix
of $\tilde {\mathbf{\Sigma}}$. Thus the result follows by applying
(\ref{eq:sigma_i_bound}) to bound the last term in
(\ref{eq:expect_D_norm}).

\medskip\noindent \underline{\textit{proof of B3}}:
\begin{eqnarray*}
L^{\prime}_{n, ik}(\bar{\theta},\sigma, \mathbf{Y})&=&\frac{1}{n}
\sum_{l=1}^n -w_i\left(y^l_i-\sum_{j \neq
i}\sqrt{\frac{\sigma^{jj}}{\sigma^{ii}}}
\rho^{ij}y^l_j\right)\sqrt{\frac{\sigma^{kk}}{\sigma^{ii}}}y^l_k\\
&&-w_k \left(y^l_k-\sum_{j \neq
k}\sqrt{\frac{\sigma^{jj}}{\sigma^{kk}}}
\rho^{kj}y^l_j\right)\sqrt{\frac{\sigma^{ii}}{\sigma^{kk}}}y^l_i.
\end{eqnarray*}
Thus,
\begin{eqnarray*}
&&L^{\prime}_{n, ik}(\bar{\theta},\bar{\sigma},
\mathbf{Y})-L^{\prime}_{n, ik}(\bar{\theta},\widehat{\sigma},
\mathbf{Y})\\
&=&-w_i\left[\overline{y_iy_k}
\left(\sqrt{\frac{\sigma^{kk}}{\sigma^{ii}}}-\sqrt{\frac{\widehat{\sigma}^{kk}}{\widehat{\sigma}^{ii}}}\right)
-\sum_{j \neq i}
\overline{y_jy_k}\rho^{ij}\left(\frac{\sqrt{\sigma^{jj}\sigma^{kk}}}{\sigma^{ii}}-\frac{\sqrt{\widehat{\sigma}^{jj}\widehat{\sigma}^{kk}}}{\widehat{\sigma}^{ii}}
\right)\right]\\
&&-w_k\left[\overline{y_iy_k}
\left(\sqrt{\frac{\sigma^{ii}}{\sigma^{kk}}}-\sqrt{\frac{\widehat{\sigma}^{ii}}{\widehat{\sigma}^{kk}}}\right)
-\sum_{j \neq k}
\overline{y_jy_i}\rho^{kj}\left(\frac{\sqrt{\sigma^{jj}\sigma^{ii}}}{\sigma^{kk}}-\frac{\sqrt{\widehat{\sigma}^{jj}\widehat{\sigma}^{ii}}}{\widehat{\sigma}^{kk}}
\right)\right],
\end{eqnarray*}
where for $1 \leq i, j \leq p$,
$\overline{y_iy_j}:=\frac{1}{n}\sum_{l=1}^n y^l_iy^l_j$. Let
$\sigma_{ij}$ denote the $(i,j)$-th element of the true covariance
matrix $\overline{\Sigma}$. By C1, $\{\sigma_{ij}: 1 \leq i,j \leq
p\}$ are bounded from below and above, thus
$$\max_{1 \leq i,j \leq
p}|\overline{y_iy_j}-\sigma_{ij}|=O_p(\sqrt{\frac{\log n}{n}}).$$
(Throughout the proof, $O_p(\cdot)$ means that for any $\eta>0$,
for sufficiently large $n$, the left hand side is bounded by the
order within $O_p(\cdot)$  with probability at least
$1-O(n^{-\eta})$.) Therefore
$$
\sum_{j \neq i} |\overline{y_jy_k}-\sigma_{jk}||\rho^{ij}| \leq
(\sum_{j \neq i} |\rho^{ij}|) \max_{1 \leq i,j \leq
p}|\overline{y_iy_j}-\sigma_{ij}| \leq (\sqrt{q_n \sum_{j \neq i}
(\rho^{ij})^2)} \max_{1 \leq i,j \leq
p}|\overline{y_iy_j}-\sigma_{ij}|=o(1),
$$
where the last inequality is by Cauchy-Schwartz and the fact that,
for fixed $i$, there are at most $q_n$ non-zero $\rho^{ij}$. The
last equality is due to the assumption $q_n \sim o(\frac{n}{\log
n})$, and the fact that $\sum_{j \neq i} (\rho^{ij})^2$ is bounded
which is in turn implied by condition C1. Therefore,
\begin{eqnarray*}
&&|L^{\prime}_{n, ik}(\bar{\theta},\bar{\sigma},
\mathbf{Y})-L^{\prime}_{n, ik}(\bar{\theta},\widehat{\sigma},
\mathbf{Y})|\\
&\leq& (w_i |\sigma_{ik}|+w_k |\sigma_{ik}|)
\max_{i,k}\left|\sqrt{\frac{\sigma^{kk}}{\sigma^{ii}}}-\sqrt{\frac{\widehat{\sigma}^{kk}}{\widehat{\sigma}^{ii}}}\right|
+(w_i\tau_{ki}+w_k\tau_{ik})\max_{i,j,k}\left|\frac{\sqrt{\sigma^{jj}\sigma^{kk}}}{\sigma^{ii}}-\frac{\sqrt{\widehat{\sigma}^{jj}\widehat{\sigma}^{kk}}}{\widehat{\sigma}^{ii}}
\right|+R_n,
\end{eqnarray*}
where $\tau_{ki}:=\sum_{j \neq i} |\sigma_{jk}\rho^{ij}|$, and the
reminder term $R_n$ is of smaller order of the leading terms.
Since C1 implies B0, thus together with condition D, we have
$$
\max_{1 \leq i,k \leq
p}\left|\sqrt{\frac{\sigma^{ii}}{\sigma^{kk}}}-
\sqrt{\frac{\widehat{\sigma}^{ii}}{\widehat{\sigma}^{kk}}}\right|=O_p(\sqrt{\frac{\log
n}{n}}),
$$

$$
\max_{1 \leq i,j,k \leq p}\left|
\frac{\sqrt{\sigma^{jj}\sigma^{ii}}}{\sigma^{kk}}-\frac{\sqrt{\widehat{\sigma}^{jj}\widehat{\sigma}^{ii}}}{\widehat{\sigma}^{kk}}\right|=O_p(\sqrt{\frac{\log
n}{n}}).
$$
Moreover, by Cauchy-Schwartz
$$
\tau_{ki} \leq \sqrt{\sum_{j} (\rho^{ij})^2} \sqrt{\sum_{j}
(\sigma_{jk})^2},
$$
and the right hand side is uniformly bounded (over $(i,k)$) due to
condition C1. Thus by C0,C1 and D, we have showed
$$
\max_{i,k}|L^{\prime}_{n, ik}(\bar{\theta},\bar{\sigma},
\mathbf{Y})-L^{\prime}_{n, ik}(\bar{\theta},\widehat{\sigma},
\mathbf{Y})|=O_p(\sqrt{\frac{\log n}{n}}).
$$

Observe that, for $1 \leq i<k \leq p, 1 \leq t <s \leq p$
\begin{eqnarray*}
L^{\prime\prime}_{n,ik,ts}=\left\{\begin{array}{ccc}
\frac{1}{n}\sum_{l=1}^n w_i\frac{\sigma^{kk}}{\sigma^{ii}}
y^l_k+ w_k\frac{\sigma^{ii}}{\sigma^{kk}}y^l_i&, if &
(i,k)=(t,s)\\
\frac{1}{n}\sum_{l=1}^n
w_i\frac{\sqrt{\sigma^{kk}\sigma^{ss}}}{\sigma^{ii}}y^l_sy^l_k,
&if& i=t, k \neq s\\
\frac{1}{n}\sum_{l=1}^n
w_k\frac{\sqrt{\sigma^{tt}\sigma^{ii}}}{\sigma^{kk}}y^l_ty^l_i,
&if& i \neq t, k=s\\
 0& if & otherwise.
\end{array}
\right.
\end{eqnarray*}
Thus by similar arguments as in the above, it is easy to proof the
claim.

\section*{Part II}
In this section, we proof the main results (Theorems 1--3). We
first give a few lemmas.

 {\lemma \label{lemma:kkt}
(Karush-Kuhn-Tucker condition) $\widehat{\theta}$ is a solution of
the optimization problem
$$
{\rm
arg}\min_{\theta: \theta_{\mathcal{S}^c}=0} L_n(\theta,\widehat{\sigma},
\mathbf{Y})+\lambda_n||\theta||_1,
$$
where $\mathcal{S}$ is a subset of $\mathcal{T}:=\{(i,j): 1 \leq i<j \leq p\}$,
if and only if
$$
\begin{array}{ccccc}
L^{\prime}_{n,ij}(\widehat{\theta},
\widehat{\sigma},\mathbf{Y})&=&\lambda_n {\rm
sign}(\widehat{\theta}_{ij}), &{\rm if} &\widehat{\theta}_{ij}
\not= 0\\
|L^{\prime}_{n,ij}(\widehat{\theta},\widehat{\sigma},\mathbf{Y})|
&\leq& \lambda_n ,& {\rm if} &\widehat{\theta}_{ij} = 0,
\end{array}
$$
for $(i,j) \in \mathcal{S}$. Moreover, if the solution is not unique,
$|L^{\prime}_{n,ij}(\tilde{\theta},\widehat{\sigma},\mathbf{Y})| <
\lambda_n$ for some specific solution $\tilde{\theta}$ and
$L^{\prime}_{n,ij}(\theta,\widehat{\sigma},\mathbf{Y})$ being
continuous in $\theta$ imply that $\widehat{\theta}_{ij}=0$ for
all solutions $\widehat{\theta}$. (Note that optimization problem (9) corresponds to $\mathcal{S}=\mathcal{T}$ and the restricted optimization problem (11) corresponds to $\mathcal{S}=\mathcal{A}$.) }

{\lemma \label{lemma1} For the loss function defined by (\ref{eqn:loss_data}), if conditions C0-C1 hold and condition D
holds for $\widehat{\sigma}$ and if $q_n \sim o(\frac{n}{\log
n})$, then for any $\eta>0$, there exist constants
$c_{0,\eta},c_{1,\eta},c_{2,\eta},c_{3,\eta}>0$,
 such that for any $u \in R^{q_n}$ the following hold with probability as least $1-O(n^{-\eta})$ for sufficiently large
$n$:

\begin{eqnarray*}
&&||L^{\prime}_{n,\mathcal{A}}(\overline{\theta},\widehat{\sigma},\mathbf{Y})||_2
\leq c_{0,\eta}\sqrt{\frac{q_n \log n}{n}}\\
&&|u^TL^{\prime}_{n,\mathcal{A}}(\overline{\theta},\widehat{\sigma},\mathbf{Y})| \leq c_{1,\eta}||u||_2(\sqrt{\frac{q_n\log n}{n}})\\
&&|u^TL^{\prime\prime}_{n,\mathcal{A}\mathcal{A}}(\overline{\theta},\widehat{\sigma},\mathbf{Y})u-u^T\overline{L}^{\prime\prime}_{\mathcal{A}\mathcal{A}}(\overline{\theta})u|
\leq c_{2,\eta}||u||^2_2(q_n\sqrt{\frac{\log n}{n}})\\
&&||L^{\prime\prime}_{n,\mathcal{A}\mathcal{A}}(\overline{\theta},\widehat{\sigma},\mathbf{Y})u-\overline{L}^{\prime\prime}_{\mathcal{A}\mathcal{A}}(\overline{\theta})u||_2
\leq c_{3,\eta}||u||_2(q_n\sqrt{\frac{\log n}{n}})
\end{eqnarray*}
}

 \noindent{\underline{\it proof of Lemma \ref{lemma1}}:}
If we replace $\widehat{\sigma}$ by $\bar{\sigma}$ on the left
hand side, then the above results follow easily from
Cauchy-Schwartz and Bernstein's inequalities by using B1.2.
Further observe that,
$$
||L^{\prime}_{n,\mathcal{A}}(\overline{\theta},\widehat{\sigma},\mathbf{Y})||_2
\leq
||L^{\prime}_{n,\mathcal{A}}(\overline{\theta},\bar{\sigma},\mathbf{Y})||_2+||L^{\prime}_{n,\mathcal{A}}(\overline{\theta},\bar{\sigma},\mathbf{Y})-L^{\prime}_{n,\mathcal{A}}(\overline{\theta},\widehat{\sigma},\mathbf{Y})||_2,
$$
and the second term on the right hand side has order $\sqrt{
\frac{q_n \log n}{n}}$, since there are $q_n$ terms and  by B3,
they are uniformly bounded by $\sqrt{\frac{\log n}{n}}$. The rest
of the lemma can be proved by similar arguments.

The following two lemmas are used for proving Theorem 1.

{\lemma \label{lemma2} Assuming the same conditions of Theorem 1. Then there exists a constant
$C_1(\overline{\theta})>0$, such that for any $\eta>0$, the
probability that there exists a local minima of the restricted
problem (11) within the disc:
\begin{eqnarray*}
\{\theta: ||\theta-\overline{\theta}||_2 \leq
C_1(\overline{\theta})\sqrt{q_n}\lambda_n\}.
\end{eqnarray*}
is at least $1-O(n^{-\eta})$ for sufficiently large $n$. }

\noindent{\underline{\it proof of Lemma \ref{lemma2}}:} Let $
\alpha_n=\sqrt{q_n}\lambda_n,$ and $
Q_n(\theta,\widehat{\sigma},\mathbf{Y},\lambda_n)=L_n(\theta,\widehat{\sigma},\mathbf{Y})+\lambda_n||\theta||_1.$
Then for any given constant $C>0$ and any vector $u \in R^p$ such
that $u_{\mathcal{A}^c}=0$ and $||u||_2=C$, by the triangle
inequality and Cauchy-Schwartz inequality, we have
$$
||\overline{\theta}||_1-||\overline{\theta}+\alpha_n u||_1 \leq
\alpha_n||u||_1 \leq C \alpha_n\sqrt{q_n}.
$$
Thus
\begin{eqnarray*}
&&Q_n(\overline{\theta}+\alpha_nu,\widehat{\sigma},\mathbf{Y},\lambda_n)-Q_n(\overline{\theta},\widehat{\sigma},\mathbf{Y},
\lambda_n)\\
&=&\{L_n(\overline{\theta}+\alpha_nu,\widehat{\sigma},\mathbf{Y})-L_n(\overline{\theta},\widehat{\sigma},\mathbf{Y})\}-\lambda_n\{||\overline{\theta}||_1-||\overline{\theta}+\alpha_n
u||_1 \}\\
&\geq&
\{L_n(\overline{\theta}+\alpha_nu,\widehat{\sigma},\mathbf{Y})-L_n(\overline{\theta},\widehat{\sigma},\mathbf{Y})\}-C\alpha_n\sqrt{q_n}\lambda_n\\
&=&\{L_n(\overline{\theta}+\alpha_nu,\widehat{\sigma},\mathbf{Y})-L_n(\overline{\theta},\widehat{\sigma},\mathbf{Y})\}-C\alpha_n^2.
\end{eqnarray*}
Thus for any $\eta>0$, there exists $c_{1,\eta}, c_{2,\eta}>0$,
such that, with probability at least $1-O(n^{-\eta})$
\begin{eqnarray*}
&&L_n(\overline{\theta}+\alpha_nu,\widehat{\sigma},\mathbf{Y})-L_n(\overline{\theta},\widehat{\sigma},\mathbf{Y})=\alpha_n
u_{\mathcal{A}}^TL^{\prime}_{n,\mathcal{A}}(\overline{\theta},\widehat{\sigma},\mathbf{Y})
+\frac{1}{2}\alpha_n^2u_{\mathcal{A}}^TL^{\prime\prime}_{n,\mathcal{A}\mathcal{A}}(\overline{\theta},\widehat{\sigma},\mathbf{Y})u_{\mathcal{A}}\\
&=&\frac{1}{2}\alpha_n^2u_{\mathcal{A}}^T\overline{L}^{\prime\prime}_{\mathcal{A}\mathcal{A}}(\overline{\theta})u_{\mathcal{A}}
+\alpha_n
u_{\mathcal{A}}^TL^{\prime}_{n,\mathcal{A}}(\overline{\theta},\widehat{\sigma},\mathbf{Y})
+\frac{1}{2}\alpha_n^2u_{\mathcal{A}}^T\left(L^{\prime\prime}_{n,\mathcal{A}\mathcal{A}}(\overline{\theta},\widehat{\sigma},\mathbf{Y})-\overline{L}^{\prime\prime}_{\mathcal{A}\mathcal{A}}(\overline{\theta})\right)u_{\mathcal{A}}\\
&\geq&\frac{1}{2}\alpha_n^2u_{\mathcal{A}}^T\overline{L}^{\prime\prime}_{\mathcal{A}\mathcal{A}}(\overline{\theta})u_{\mathcal{A}}
-c_{1,\eta}(\alpha_nq_n^{1/2}n^{-1/2}\sqrt{\log
n})-c_{2,\eta}(\alpha_n^2q_nn^{-1/2}\sqrt{\log n}).
\end{eqnarray*}
In the above, the first equation is because the loss function
$L(\theta,\sigma,Y)$ is quadratic in $\theta$ and
$u_{\mathcal{A}^c}=0$. The inequality is due to Lemma \ref{lemma1}
and the union bound. By the {\bf assumption $\lambda_n
\sqrt{\frac{n}{\log n}} \rightarrow \infty$}, we have $ \alpha_n
q_n^{1/2}n^{-1/2}\sqrt{\log n}=o(\alpha_n
\sqrt{q_n}\lambda_n)=o(\alpha_n^2)$. Also by the {\bf assumption
that $q_n \sim o(\sqrt{n/\log n})$}, we have
$\alpha_n^2q_nn^{-1/2}\sqrt{\log n}=o(\alpha_n^2)$. Thus, with $n$
sufficiently large
$$
Q_n(\overline{\theta}+\alpha_nu,\widehat{\sigma},\mathbf{Y},\lambda_n)-Q_n(\overline{\theta},\widehat{\sigma},\mathbf{Y},
\lambda_n) \geq
\frac{1}{4}\alpha_n^2u_{\mathcal{A}}^T\overline{L}^{\prime\prime}_{\mathcal{A}\mathcal{A}}(\overline{\theta})u_{\mathcal{A}}
-C\alpha_n^2
$$
with probability at least $1-O(n^{-\eta})$. By B1, $
u_{\mathcal{A}}^T\overline{L}^{\prime\prime}_{\mathcal{A}\mathcal{A}}(\overline{\theta})u_{\mathcal{A}}
\geq
\Lambda^L_{\min}(\bar{\theta})||u_{\mathcal{A}}||_2^2=\Lambda^L_{\min}(\bar{\theta})C^2.
$ Thus, if we choose
$C=4/\Lambda^L_{\min}(\bar{\theta})+\epsilon$, then for any
$\eta>0$, for sufficiently large $n$, the following holds
$$
\inf_{u:u_{\mathcal{A}^c}=0,
||u||_2=C}Q_n(\overline{\theta}+\alpha_nu,\widehat{\sigma},\mathbf{Y},\lambda_n)>Q_n(\overline{\theta},\widehat{\sigma},\mathbf{Y},
\lambda_n),
$$
with probability at least $1-O(n^{-\eta})$. This means that a
local minima exists within the disc $ \{\theta:
||\theta-\overline{\theta}||_2 \leq
C\alpha_n=C\sqrt{q_n}\lambda_n\}$ with probability at least
$1-O(n^{-\eta})$.

{\lemma \label{lemma3} Assuming the same conditions of Theorem 1. Then there exists a constant
$C_2(\overline{\theta})>0$, such that for any $\eta>0$, for
sufficiently large $n$, the following holds with probability at
least $1-O(n^{-\eta})$: for any $\theta$ belongs to the set
$S=\{\theta: ||\theta-\overline{\theta}||_2 \geq
C_2(\overline{\theta})\sqrt{q_n}\lambda_n,
\theta_{\mathcal{A}^c}=0\},$ it has $
||L^{\prime}_{n,\mathcal{A}}(\theta,\widehat{\sigma},\mathbf{Y})||_2
> \sqrt{q_n}\lambda_n$. }

\noindent{\underline{\it proof of Lemma \ref{lemma3}}:} Let $
\alpha_n=\sqrt{q_n}\lambda_n$. Any $\theta$ belongs to $S$ can be
written as: $\theta=\overline{\theta}+\alpha_n u$, with
$u_{\mathcal{A}^c}=0$ and $||u||_2 \geq C_2(\bar{\theta})$. Note
that
\begin{eqnarray*}
L^{\prime}_{n,\mathcal{A}}(\theta,\widehat{\sigma},\mathbf{Y})&=&L^{\prime}_{n,\mathcal{A}}(\overline{\theta},\widehat{\sigma},\mathbf{Y})+\alpha_nL^{\prime\prime}_{n,\mathcal{A}\mathcal{A}}(\overline{\theta},\widehat{\sigma},\mathbf{Y})u\\
&=&L^{\prime}_{n,\mathcal{A}}(\overline{\theta},\widehat{\sigma},\mathbf{Y})
+\alpha_n(L^{\prime\prime}_{n,\mathcal{A}\mathcal{A}}(\overline{\theta},\widehat{\sigma},\mathbf{Y})-\overline{L}^{\prime\prime}_{\mathcal{A}\mathcal{A}}(\overline{\theta}))u
+\alpha_n\overline{L}^{\prime\prime}_{\mathcal{A}\mathcal{A}}(\overline{\theta}))u.
\end{eqnarray*}
By the triangle inequality and Lemma \ref{lemma1}, for any
$\eta>0$, there exists constants $c_{0,\eta},c_{3,\eta}>0$, such
that
$$
||L^{\prime}_{n,\mathcal{A}}(\theta,\widehat{\sigma},\mathbf{Y})||_2
\geq
\alpha_n||\overline{L}^{\prime\prime}_{\mathcal{A}\mathcal{A}}(\overline{\theta}))u||_2
-c_{0,\eta}(q_n^{1/2}n^{-1/2}\sqrt{\log n})
-c_{3,\eta}||u||_2(\alpha_n q_nn^{-1/2}\sqrt{\log n})
$$
with probability at least $1-O(n^{-\eta})$. Thus, similar as in
Lemma \ref{lemma2}, for $n$ sufficiently large,
$||L^{\prime}_{n,\mathcal{A}}(\theta,\widehat{\sigma},\mathbf{Y})||_2
\geq
\frac{1}{2}\alpha_n||\overline{L}^{\prime\prime}_{\mathcal{A}\mathcal{A}}(\overline{\theta}))u||_2$
with probability at least $1-O(n^{-\eta})$. By B1, $
||\overline{L}^{\prime\prime}_{\mathcal{A}\mathcal{A}}(\overline{\theta}))u||_2
\geq \Lambda^L_{\min}(\bar{\theta})||u||_2. $ Therefore
$C_2(\overline{\theta})$ can be taken as
$2/\Lambda^L_{\min}(\bar{\theta})+\epsilon$.

The following lemma is used in proving Theorem 2.
{\lemma \label{lemma4} Assuming conditions C0-C1. Let
$D_{\mathcal{A}\mathcal{A}}(\bar{\theta},Y)
=L^{\prime\prime}_{1,\mathcal{A}\mathcal{A}}(\bar{\theta},Y)-\overline{L}^{\prime\prime}_{\mathcal{A}\mathcal{A}}(\bar{\theta})$.
Then there exists a constant $K_2(\bar{\theta})<\infty$, such that
for any $(k,l) \in \mathcal{A}$, $\lambda_{\max}({\rm
Var}_{\bar{\theta}}(D_{\mathcal{A},kl}(\bar{\theta},Y))) \leq
K_2(\bar{\theta})$. }

\noindent \underline{\it proof of Lemma \ref{lemma4}}: $ {\rm
Var}_{\bar{\theta}}(D_{\mathcal{A},kl}(\bar{\theta},Y))=E_{\bar{\theta}}(L^{\prime\prime}_{1,\mathcal{A},kl}
(\bar{\theta},Y)L^{\prime\prime}_{1,\mathcal{A},kl}(\bar{\theta},Y)^T)
-\overline{L}^{\prime\prime}_{\mathcal{A},kl}(\bar{\theta})\overline{L}^{\prime\prime}_{\mathcal{A},kl}(\bar{\theta})^T.$
Thus it suffices to show that, there exists a constant
$K_2(\bar{\theta})>0$, such that for all $(k,l)$
\begin{eqnarray*}
\lambda_{\max}(E_{\bar{\theta}}(L^{\prime\prime}_{1,\mathcal{A},kl}(\bar{\theta},Y)
L^{\prime\prime}_{1,\mathcal{A},kl}(\bar{\theta},Y)^T)) \leq
K_2(\bar{\theta}).
\end{eqnarray*}
Use the same notations as in the proof of B1. Note that $
L_{1,\mathcal{A},kl}''(\overline{\theta},Y) = \tilde x^T \tilde
w^2 \tilde{x}_{(k,l)}= \tilde w_k \tilde y_l x_k + \tilde w_l
\tilde y_k x_l. $ Thus
\begin{eqnarray*}
E_{\bar{\theta}}(L^{\prime\prime}_{1,\mathcal{A},kl}(\bar{\theta},Y)L^{\prime\prime}_{1,\mathcal{A},kl}(\bar{\theta},Y)^T)=
\tilde w_k^2 \mathbb{E}[\tilde y_l^2 x_k x_k^T] + \tilde w_l^2
\mathbb{E}[\tilde y_k^2 x_l x_l^T] + \tilde w_k \tilde
w_l\mathbb{E}[\tilde y_k \tilde y_l (x_k x_l^T + x_l x_k^T)],
\end{eqnarray*}
and for $a \in \mathcal{R}^{p(p-1)/2}$
\begin{eqnarray*}
&&a^TE_{\bar{\theta}}(L^{\prime\prime}_{1,\,\mathcal{A},kl}(\bar{\theta},Y)L^{\prime\prime}_{1,\mathcal{A},kl}(\bar{\theta},Y)^T)a\\
&=&\tilde w_k^2 a_k^T \mathbb{E}[\tilde y_l^2 \tilde y \tilde y^T]
a_k + \tilde w_l^2 a_l^T \mathbb{E}[\tilde y_k^2 \tilde y \tilde
y^T] a_l + 2 \tilde w_k \tilde w_l a_k^T \mathbb{E}[\tilde y_k
\tilde y_l\tilde y \tilde y^T] a_l.
\end{eqnarray*}
Since $\sum_{k=1}^p ||a_k||_2^2=2||a||_2^2$, and by B2:
$\lambda_{\max}(\mathbb{E}[\tilde y_i \tilde y_j\tilde y \tilde
y^T]) \leq K_1(\bar{\theta})$ for any $1 \leq i \leq j \leq p$,
the conclusion follows.

\medskip
\noindent \underline{{\it proof of Theorem 1}}: The
existence of a solution of (11) follows from Lemma
\ref{lemma2}. By the Karush-Kuhn-Tucker condition (Lemma
\ref{lemma:kkt}), for any solution $\widehat{\theta}$ of
(11), it has $
||L^{\prime}_{n,\mathcal{A}}(\widehat{\theta},\widehat{\sigma},\mathbf{Y})||_{\infty}
\leq \lambda_n.$ Thus $
||L^{\prime}_{n,\mathcal{A}}(\widehat{\theta},\widehat{\sigma},\mathbf{Y})||_2
\leq \sqrt{q_n}
||L^{\prime}_{n,\mathcal{A}}(\widehat{\theta},\widehat{\sigma},\mathbf{Y})||_{\infty}
\leq \sqrt{q_n}\lambda_n$. Thus by Lemma \ref{lemma3}, for any
$\eta>0$, for $n$ sufficiently large with probability at least
$1-O(n^{-\eta})$, all solutions of (11) are inside
the disc $\{\theta: ||\theta-\overline{\theta}||_2 \leq
C_2(\overline{\theta})\sqrt{q_n}\lambda_n\}$. Since
$\frac{s_n}{\sqrt{q_n}\lambda_n} \rightarrow \infty$, for
sufficiently large $n$ and $(i,j) \in \mathcal{A}$: $
\overline{\theta}_{ij} \geq s_n >2
C_2(\overline{\theta})\sqrt{q_n} \lambda_n.$ Thus
\begin{eqnarray*}
1-O(n^{-\eta})
&\leq&P_{\overline{\theta}}\left(||\widehat{\theta}^{\mathcal{A},\lambda_n}-\overline{\theta}_{\mathcal{A}}||_2
\leq C_2(\overline{\theta})\sqrt{q_n} \lambda_n, \bar{\theta}_{ij}
>2 C_2(\overline{\theta})\sqrt{q_n} \lambda_n, \ \hbox{for all} (i,j) \in
\mathcal{A}\right)\\ & \leq &P_{\bar{\theta}}\left({\rm
sign}(\widehat{\theta}^{\mathcal{A},\lambda_n}_{ij})={\rm
sign}(\overline{\theta}_{ij}), \ \hbox{for all} (i,j) \in
\mathcal{A}\right). \end{eqnarray*}

\medskip
\noindent \underline{{\it proof of Theorem 2}}: For
any given $\eta>0$, let $\eta'=\eta+\kappa$. Let
$\mathcal{E}_n=\{{\rm
sign}(\widehat{\theta}^{\mathcal{A},\lambda_n})={\rm
sign}(\bar{\theta})\}$. Then by Theorem 1,
$P_{\bar{\theta}}(\mathcal{E}_n) \geq 1-O(n^{-\eta'})$ for
sufficiently large $n$. On $\mathcal{E}_n$, by the
Karush-Kuhn-Tucker condition and the expansion of
$L^{\prime}_{n,\mathcal{A}}(\widehat{\theta}^{\mathcal{A},\lambda_n},\widehat{\sigma},\mathbf{Y})$
at $\bar{\theta}$
\begin{eqnarray*}
-\lambda_n{\rm sign}(\bar{\theta}_{\mathcal{A}})
&=&L^{\prime}_{n,\mathcal{A}}(\widehat{\theta}^{\mathcal{A},\lambda_n},\widehat{\sigma},\mathbf{Y})=L^{\prime}_{n,\mathcal{A}}(\bar{\theta},\widehat{\sigma},\mathbf{Y})+L^{\prime\prime}_{n,\mathcal{A}\mathcal{A}}(\bar{\theta},\widehat{\sigma},\mathbf{Y})\nu_n\\
&=&\overline{L}^{\prime\prime}_{\mathcal{A}\mathcal{A}}(\bar{\theta})\nu_n
+L^{\prime}_{n,\mathcal{A}}(\bar{\theta},\widehat{\sigma},\mathbf{Y})+\left(L^{\prime\prime}_{n,\mathcal{A}\mathcal{A}}(\bar{\theta},\widehat{\sigma},\mathbf{Y})-\overline{L}^{\prime\prime}_{\mathcal{A}\mathcal{A}}(\bar{\theta})\right)\nu_n,
\end{eqnarray*}
where $\nu_n
:=\widehat{\theta}^{\mathcal{A},\lambda_n}_{\mathcal{A}}-\bar{\theta}_{\mathcal{A}}$.
By the above expression
\begin{eqnarray}
\label{eqn:nu_n}
\nu_n=-\lambda_n[\overline{L}^{\prime\prime}_{\mathcal{A}\mathcal{A}}(\bar{\theta})]^{-1}{\rm
sign}(\bar{\theta}_{\mathcal{A}})-[\overline{L}^{\prime\prime}_{\mathcal{A}\mathcal{A}}(\bar{\theta})]^{-1}[L^{\prime}_{n,\mathcal{A}}(\bar{\theta},\widehat{\sigma},\mathbf{Y})
+D_{n,\mathcal{A}\mathcal{A}}(\bar{\theta},\widehat{\sigma},\mathbf{Y})\nu_n],
\end{eqnarray}
where
$D_{n,\mathcal{A}\mathcal{A}}(\bar{\theta},\widehat{\sigma},\mathbf{Y})=L^{\prime\prime}_{n,\mathcal{A}\mathcal{A}}(\bar{\theta},\widehat{\sigma},\mathbf{Y})-\overline{L}^{\prime\prime}_{\mathcal{A}\mathcal{A}}(\bar{\theta})$.
Next, fix $(i,j) \in \mathcal{A}^{c}$, and consider the expansion
of
$L^{\prime}_{n,ij}(\widehat{\theta}^{\mathcal{A},\lambda_n},\widehat{\sigma},\mathbf{Y})$
around $\bar{\theta}$:
\begin{eqnarray}
\label{eqn:A_c}
L^{\prime}_{n,ij}(\widehat{\theta}^{\mathcal{A},\lambda_n},\widehat{\sigma},\mathbf{Y})
&=&L^{\prime}_{n,ij}(\bar{\theta},\widehat{\sigma},\mathbf{Y})+L^{\prime\prime}_{n,ij,\mathcal{A}}(\bar{\theta},\widehat{\sigma},\mathbf{Y})\nu_n.
\end{eqnarray}
Then plug in (\ref{eqn:nu_n}) into (\ref{eqn:A_c}), we get
\begin{eqnarray}
\label{eqn:25}
&&L^{\prime}_{n,ij}(\widehat{\theta}^{\mathcal{A},\lambda_n},\widehat{\sigma},\mathbf{Y})
=-\lambda_n\overline{L}^{\prime\prime}_{ij,\mathcal{A}}(\bar{\theta})[\overline{L}^{\prime\prime}_{\mathcal{A}\mathcal{A}}(\bar{\theta})]^{-1}
{\rm
sign}(\bar{\theta}_{\mathcal{A}})-\overline{L}^{\prime\prime}_{ij,\mathcal{A}}(\bar{\theta})[\overline{L}^{\prime\prime}_{\mathcal{A}\mathcal{A}}(\bar{\theta})]^{-1}L^{\prime}_{n,\mathcal{A}}(\bar{\theta},\widehat{\sigma},\mathbf{Y})\nonumber \\
&+&L^{\prime}_{n,ij}(\bar{\theta},\widehat{\sigma},\mathbf{Y})+\left[D_{n,ij,\mathcal{A}}(\bar{\theta},\widehat{\sigma},\mathbf{Y})-\overline{L}^{\prime\prime}_{ij,\mathcal{A}}(\bar{\theta})[\overline{L}^{\prime\prime}_{\mathcal{A}\mathcal{A}}(\bar{\theta})]^{-1}D_{n,\mathcal{A}\mathcal{A}}(\bar{\theta},\widehat{\sigma},\mathbf{Y})
\right]\nu_n.
\end{eqnarray}
By condition C2, for any $(i,j) \in \mathcal{A}^c$: $
|\overline{L}^{\prime\prime}_{ij,\mathcal{A}}(\bar{\theta})[\overline{L}^{\prime\prime}_{\mathcal{A}\mathcal{A}}(\bar{\theta})]^{-1}
{\rm sign}(\bar{\theta}_{\mathcal{A}})|\leq \delta <1$. Thus it
suffices to prove that the remaining terms in (\ref{eqn:25}) are
all $o(\lambda_n)$ with probability at least $1-O(n^{-\eta'})$
(uniformly for all $(i,j) \in \mathcal{A}^c$). Then since
$|\mathcal{A}^c| \leq p \sim O(n^{\kappa})$, by the union bound,
the event $ \max_{(i,j)\in
\mathcal{A}^c}|L^{\prime}_{n,ij}(\widehat{\theta}^{\mathcal{A},\lambda_n},\widehat{\sigma},\mathbf{Y})|<\lambda_n
$ holds with probability at least
$1-O(n^{\kappa-\eta'})=1-O(n^{-\eta})$, when $n$ is sufficiently
large.

By B1.4, for any $(i,j) \in \mathcal{A}^c$:
$||\overline{L}^{\prime\prime}_{ij,\mathcal{A}}(\bar{\theta})[\overline{L}^{\prime\prime}_{\mathcal{A}\mathcal{A}}(\bar{\theta})]^{-1}||_2
\leq M(\bar{\theta})$. Therefore by Lemma \ref{lemma1}, for any
$\eta>0$, there exists a constant $C_{1,\eta}>0$, such that
$$\max_{(i,j) \in
\mathcal{A}^c}|\overline{L}^{\prime\prime}_{ij,\mathcal{A}}(\bar{\theta})
[\overline{L}^{\prime\prime}_{\mathcal{A}\mathcal{A}}(\bar{\theta})]^{-1}L^{\prime}_{n,\mathcal{A}}(\bar{\theta},\widehat{\sigma},\mathbf{Y})|
\leq C_{1,\eta}(\sqrt{\frac{q_n\log n}{n}})=(o(\lambda_n))$$
with
probability at least $1-O(n^{-\eta})$. The claim follows by the
{\bf assumption $\sqrt{\frac{q_n\log n}{n}} \sim o(\lambda_n)$}.

By B1.2, $||{\rm
Var}_{\bar{\theta}}(L^{\prime}_{ij}(\bar{\theta},\bar{\sigma},\mathbf{Y}))||_2
\leq M_1(\bar{\theta})$. Then similarly as in Lemma \ref{lemma1}, for any
$\eta>0$, there exists a constant $C_{2,\eta}>0$, such that $
\max_{i,j}|L^{\prime}_{n,ij}(\bar{\theta},\widehat{\sigma},\mathbf{Y})|
\leq C_{2,\eta}(\sqrt{\frac{\log n}{n}})=(o(\lambda_n))$, with
probability at least $1-O(n^{-\eta})$. The claims follows by the
{\bf assumption that $\lambda_n \sqrt{\frac{n}{\log n}}
\rightarrow \infty$.}
%holds uniformly for $(i,j) \in \mathcal{A}^c$.

Note that by Theorem 1, for any $\eta>0$, $||\nu_n||_2
\leq C(\bar{\theta})\sqrt{q_n}\lambda_n$ with probability at least
$1-O(n^{-\eta})$ for large enough $n$. Thus, similarly as in Lemma
\ref{lemma1}, for any $\eta>0$, there exists a constant
$C_{3,\eta}$, such $
|D_{n,ij,\mathcal{A}}(\bar{\theta},\widehat{\sigma},\mathbf{Y})\nu_n|
\leq C_{3,\eta}(\sqrt{\frac{q_n \log
n}{n}}\sqrt{q_n}\lambda_n)(=o(\lambda_n))$, with probability at
least $1-O(n^{-\eta})$. The claims follows from {\bf the
assumption $q_n \sim o(\sqrt{\frac{n}{\log n}})$}.
%holds uniformly for $(i,j) \in \mathcal{A}^c$.

Finally, let
$b^T=|\overline{L}^{\prime\prime}_{ij,\mathcal{A}}(\bar{\theta})
[\overline{L}^{\prime\prime}_{\mathcal{A}\mathcal{A}}(\bar{\theta})]^{-1}$.
By Cauchy-Schwartz inequality
$$
|b^T
D_{n,\mathcal{A}\mathcal{A}}(\bar{\theta},\bar{\sigma},\mathbf{Y})
\nu_n| \leq ||b^T
D_{n,\mathcal{A}\mathcal{A}}(\bar{\theta},\bar{\sigma},\mathbf{Y})||_2
||\nu_n||_2  \leq q_n \lambda_n \max_{(k,l) \in \mathcal{A}} |b^T
D_{n,\mathcal{A}, kl}(\bar{\theta},\bar{\sigma},\mathbf{Y})|.
$$
In order to show the right hand side is $o(\lambda_n)$ with
probability at least $1-O(n^{-\eta})$, it suffices to show
$\max_{(k,l) \in \mathcal{A}} |b^T D_{n,\mathcal{A},
kl}(\bar{\theta},\bar{\sigma},\mathbf{Y})|=O(\sqrt{\frac{\log
n}{n}})$ with probability at least $1-O(n^{-\eta})$, because of
the {\bf the assumption $q_n \sim o(\sqrt{\frac{n}{\log n}})$}.
This is implied by
$$
E_{\bar{\theta}}(|b^T D_{\mathcal{A},
kl}(\bar{\theta},\bar{\sigma},Y)|^2) \leq ||b||^2_2
\lambda_{\max}({\rm
Var}_{\bar{\theta}}(D_{\mathcal{A},kl}(\bar{\theta},\bar{\sigma},Y)))
$$
being bounded, which follows immediately from B1.4 and Lemma
\ref{lemma4}. Finally, similarly as in Lemma \ref{lemma1},
$$
|b^T
D_{n,\mathcal{A}\mathcal{A}}(\bar{\theta},\widehat{\sigma},\mathbf{Y})
\nu_n| \leq |b^T
D_{n,\mathcal{A}\mathcal{A}}(\bar{\theta},\bar{\sigma},\mathbf{Y})
\nu_n|+|b^T
(D_{n,\mathcal{A}\mathcal{A}}(\bar{\theta},\bar{\sigma},\mathbf{Y})-D_{n,\mathcal{A}\mathcal{A}}(\bar{\theta},\widehat{\sigma},\mathbf{Y}))
\nu_n|,
$$
where by B3, the second term on the right hand side is bounded by
$O_p(\sqrt{\frac{\log n}{n}}) ||b||_2 ||\nu_n||_2$. Note that
$||b||_2 \sim \sqrt{q_n}$, thus the second term is also of order
$o(\lambda_n)$ by {\bf the assumption $q_n \sim
o(\sqrt{\frac{n}{\log n}})$}. This completes the proof.

\medskip
\noindent \underline{\it proof of Theorem 3}: By Theorems 1 and 2
and the Karush-Kuhn-Tucker condition, for any $\eta>0$, with
probability at least $1-O(n^{-\eta})$, a solution of the
restricted problem is also a solution of the original problem. On
the other hand, by Theorem 2 and the Karush-Kuhn-Tucker condition,
with high probability, any solution of the original problem is a
solution of the restricted problem. Therefore, by Theorem 1, the
conclusion follows.

\section*{Part III}
In this section, we provide details for the implementation of
\texttt{space} which takes advantage of the sparse structure of
$\mathcal{X}$. Denote the target loss function as
\begin{equation}
f(\theta)=\frac{1}{2}\left\|\mathcal{Y}-\mathcal{X}\theta\right\|^2+\lambda_1\sum_{i<
j}|\rho^{ij}|.
%+\lambda_2\sum_{i< j}|\rho^{ij}|^2.
\end{equation}
Our goal is to find
$\widehat{\theta}=\textrm{argmin}_{\theta}f(\theta)$ for a given
$\lambda_1$.
% and $\lambda_2$.
We will employ \texttt{active-shooting}
algorithm (Section 2.3) to solve this optimization problem.

Without loss of generality, we assume
$\textrm{mean}(\mathbf{Y}_i)=1/n\sum_{k=1}^n y^k_i=0$ for $i=1, \ldots, p$.
Denote $\xi_i=\mathbf{Y}_i^{T}\mathbf{Y}_i$. We have
$$
\mathcal{X}_{(i,j)}^{T}\mathcal{X}_{(i,j)}=\xi_j\frac{\sigma^{jj}}{\sigma^{ii}}+\xi_i\frac{\sigma^{ii}}{\sigma^{jj}};
$$
$$\mathcal{Y}^{T}\mathcal{X}_{(i,j)}=\sqrt{\frac{\sigma^{jj}}{\sigma^{ii}}}\mathbf{Y}_i^T\mathbf{Y}_j
+\sqrt{\frac{\sigma^{ii}}{\sigma^{jj}}}\mathbf{Y}_j^T\mathbf{Y}_i.$$
Denote
$\rho^{ij}=\rho_{(i,j)}$. We now present details of the initialization step and
the updating steps in the \texttt{active-shooting} algorithm.
\vspace{5pt}

\noindent{\textbf{1. Initialization}}\\
Let
\begin{equation}
\begin{array}{rl}
\rho_{(i,j)}^{(0)} & =
%\frac{1}{1+\lambda_2}
\frac{\left(|\mathcal{Y}^T\mathcal{X}_{(i,j)}|-\lambda_1\right)_+
\cdot \textrm{sign}(\mathcal{Y}^T\mathcal{X}_{(i,j)})}
                  {\mathcal{X}_{(i,j)}^T\mathcal{X}_{(i,j)}}\\
              &=
%\frac{1}{1+\lambda_2}
\frac{\left(\left|\sqrt{\frac{\sigma^{jj}}{\sigma^{ii}}}\mathbf{Y}_i^T\mathbf{Y}_j
+\sqrt{\frac{\sigma^{ii}}{\sigma^{jj}}}\mathbf{Y}_j^T\mathbf{Y}_i\right|-\lambda_1\right)_+
\cdot \textrm{sign}(\mathbf{Y}_i^T\mathbf{Y}_j)}
                  {\xi_j \frac{\sigma^{jj}}{\sigma^{ii}} + \xi_i
            \frac{\sigma^{ii}}{\sigma^{jj}}}.
\end{array}
\end{equation}
For $j=1, \ldots, p$, compute
\begin{equation}
\widehat{\mathbf{Y}}_j^{(0)}=\left(\sqrt{\frac{\sigma^{11}}{\sigma^{jj}}}\mathbf{Y}_1,
..., \sqrt{\frac{\sigma^{pp}}{\sigma^{jj}}}\mathbf{Y}_p\right)\cdot \left(
\begin{array}{c}
\rho_{(1,j)}^{(0)}\\
\vdots\\
\rho_{(p,j)}^{(0)}
\end{array}
\right),
\end{equation}
and
\begin{equation}
E^{(0)}=\mathcal{Y}-\widehat{\mathcal{Y}}^{(0)}=\left((E^{(0)}_1)^T,
..., (E^{(0)}_p)^T\right),
\end{equation}
where $E^{(0)}_{j}=\mathbf{Y}_j-\widehat{\mathbf{Y}}_j^{(0)}$, for $1 \leq j \leq p$.
\vspace{5pt}

\noindent{\textbf{2. Update $\rho_{(i,j)}^{(0)}\longrightarrow
\rho_{(i,j)}^{(1)}$}}\\
Let
\begin{equation}
A_{(i,j)}=(E^{(0)}_{j})^T \cdot
\sqrt{\frac{\sigma^{ii}}{\sigma^{jj}}}\mathbf{Y}_i,
\end{equation}
\begin{equation}
A_{(j,i)}=(E^{(0)}_{i})^T \cdot
\sqrt{\frac{\sigma^{jj}}{\sigma^{ii}}}\mathbf{Y}_j.
\end{equation}

We have
\begin{equation}
\begin{array}{rl}
(E^{(0)})^T \mathcal{X}_{(i,j)}&=(E^{(0)}_i)^T \cdot
\sqrt{\frac{\sigma^{jj}}{\sigma^{ii}}}\mathbf{Y}_j+(E^{(0)}_j)^T
\cdot \sqrt{\frac{\sigma^{ii}}{\sigma^{jj}}}\mathbf{Y}_i\\
&=A_{(j,i)}+A_{(i,j)}.
\end{array}
\end{equation}
It follows
\begin{equation}
\begin{array}{rl}
\rho^{(1)}_{(i,j)}&=
%\frac{1}{1+\lambda_2}
\textrm{sign} \left( \frac{(E^{(0)})^T
  \mathcal{X}_{_(i,j)}}{\mathcal{X}^T_{(i,j)}
  \mathcal{X}_{(i,j)}} + \rho^{(0)}_{(i,j)}\right)
\left(\left|\frac{(E^{(0)})^T
  \mathcal{X}_{_(i,j)}}{\mathcal{X}^T_{(i,j)}
  \mathcal{X}_{(i,j)}} + \rho^{(0)}_{(i,j)}\right|
-\frac{\lambda_1}{\mathcal{X}^T_{(i,j)}\mathcal{X}_{(i,j)}}\right)_+\\
&=
%\frac{1}{1+\lambda_2}
\textrm{sign} \left(\frac{A_{(j,i)} + A_{(i,j)}}{\xi_j
  \frac{\sigma^{jj}}{\sigma^{ii}} + \xi_i\frac{\sigma^{ii}}{\sigma^{jj}}}
+\rho^{(0)}_{(i,j)}\right)
\left(\left| \frac{A_{(j,i)} + A_{(i,j)}}{\xi_j
  \frac{\sigma^{jj}}{\sigma^{ii}} +
  \xi_i\frac{\sigma^{ii}}{\sigma^{jj}}} + \rho^{(0)}_{(i,j)}\right|
-\frac{\lambda_1}{\xi_j \frac{\sigma^{jj}}{\sigma^{ii}} + \xi_i
  \frac{\sigma^{ii}}{\sigma^{jj}}} \right)_+.
\end{array}
\end{equation}
\vspace{5pt}

\noindent{\textbf{3. Update $\rho^{(t)}\longrightarrow
\rho^{(t+1)}$}}\\ From the previous iteration, we have
\begin{itemize}
\item $E^{(t-1)}$: residual in the previous iteration ($np\times
1$ vector). \item $(i_0, j_0)$: index of coefficient that is
updated in the previous iteration. \item $
\rho^{(t)}_{(i,j)}=\left\{
\begin{array}{ll}
\rho^{(t-1)}_{(i,j)} & \textrm{ if } (i,j)\neq (i_0, j_0),
\textrm{ nor
}(j_0, i_0)\\
\rho^{(t-1)}_{(i,j)}- \Delta& \textrm{ if } (i,j)= (i_0, j_0),
\textrm{ or }(j_0, i_0)
\end{array}
\right. $
\end{itemize}
Then,
\begin{equation}
\begin{array}{rl}
E^{(t)}_{k} &=E^{(t-1)}_{k} \textrm{ for }k\neq i_0, j_0;\\
E^{(t)}_{j_0}&=E^{(t-1)}_{j_0}+\widehat{\mathbf{Y}}^{(t-1)}_{j_0}-\widehat{\mathbf{Y}}^{(t)}_{j_0}\\
             &=E^{(t-1)}_{j_0}+\sum_{i=1}^{p}
             \sqrt{\frac{\sigma^{ii}}{\sigma^{j_0j_0}}}\mathbf{Y}_i
             (\rho^{(t-1)}_{(i,j_0)}-\rho^{(t)}_{(i,j_0)})\\
             &=E^{(t-1)}_{j_0}+\sqrt{\frac{\sigma^{i_0i_0}}{\sigma^{j_0j_0}}}\mathbf{Y}_{i_0}\cdot
             \Delta;\\
E^{(t)}_{i_0}&=E^{(t-1)}_{i_0}+\sqrt{\frac{\sigma^{j_0j_0}}{\sigma^{i_0i_0}}}\mathbf{Y}_{j_0}\cdot
             \Delta.
\end{array}
\end{equation}
Suppose the index of the coefficient we would like to update in this
iteration is $(i_1, j_1)$, then let
$$
 A_{(i_1,j_1)}=(E^{(t)}_{j_1})^T
\cdot \sqrt{\frac{\sigma^{i_1i_1}}{\sigma^{j_1j_1}}}\mathbf{Y}_{i_1},
$$
$$
 A_{(j_1,i_1)}=(E^{(t)}_{i_1})^T
\cdot \sqrt{\frac{\sigma^{j_1j_1}}{\sigma^{i_1i_1}}}\mathbf{Y}_{j_1}.
$$
We have
\begin{equation}
\begin{array}{ll}
\rho^{(t+1)}_{(i,j)}
=&
%\frac{1}{1+\lambda_2}
\textrm{sign} \left( \frac{A_{(j_1,i_1)} + A_{(i_1,j_1)}}{\xi_j
  \frac{\sigma^{j_1j_1}}{\sigma^{i_1i_1}}
+\xi_{i_1}\frac{\sigma^{i_1i_1}}{\sigma^{j_1j_1}}}
+\rho^{(t)}_{(i_1,j_1)}\right)\\
&\times
\left(\left| \frac{A_{(j_1,i_1)} + A_{(i_1,j_1)}}{\xi_j
  \frac{\sigma^{j_1j_1}}{\sigma^{i_1i_1}}
+\xi_{i_1}\frac{\sigma^{i_1i_1}}{\sigma^{j_1j_1}}}
+\rho^{(t)}_{(i_1,j_1)}\right|
-\frac{\lambda_1}{\xi_j \frac{\sigma^{jj}}{\sigma^{ii}} +
  \xi_i\frac{\sigma^{ii}}{\sigma^{jj}}} \right)_+.
\end{array}
\end{equation}
\vspace{5pt}

Using the above steps 1--3, we have implemented the
\texttt{active-shooting} algorithm in \texttt{c}, and the
corresponding \texttt{R} package \texttt{space} to fit the
\texttt{space} model is available on \texttt{cran}.

\end{document}